\documentclass[aas_macros]{aa}

\usepackage{txfonts}
\usepackage{xcolor}
\usepackage{rotating}
\usepackage{natbib}
\usepackage{aas_macros}

\usepackage{graphicx}	
\usepackage{amsmath}	
\usepackage{amssymb}	
\usepackage{array}
\DeclareMathAlphabet\mathbfcal{OMS}{cmsy}{b}{n}

\begin{document}
\renewcommand{\arraystretch}{1.2}

\title{Understanding and improving the timing of PSR~J0737$-$3039B}


\author{
A.~Noutsos\inst{1}\thanks{anoutsos@mpifr-bonn.mpg.de}
\and G.~Desvignes\inst{1}\fnmsep\inst{10}
\and M.~Kramer\inst{1}\fnmsep\inst{7}
\and N.~Wex\inst{1}
\and P.~C.~C.~Freire\inst{1}
\and I.~H.~Stairs\inst{2}
\and M.~A.~McLaughlin\inst{3}\fnmsep\inst{8}\fnmsep\inst{9}
\and R.~N.~Manchester\inst{4}
\and A.~Possenti\inst{5}\fnmsep\inst{12}
\and M.~Burgay\inst{5}
\and A.~G.~Lyne\inst{6}
\and R.~P.~Breton\inst{7}
\and B.~B.~P.~Perera\inst{11}
\and R.~D.~Ferdman\inst{13}
}

\institute{
Max-Planck-Institut f\"ur Radioastronomie, Auf dem H\"ugel 69, D-53121 Bonn, Germany
\and Department of Physics and Astronomy, University of British Columbia, 6224 Agricultural Road, Vancouver, British  Columbia V6T 1Z1, Canada
\and Department of Physics, West Virginia University, Morgantown, WV 26506, USA
\and Australia Telescope National Facility, Commonwealth Scientific and Industrial Research Organisation (CSIRO), P.O. Box 76, Epping, New South Wales 1710, Australia
\and INAF-Osservatorio Astronomico di Cagliari, via della Scienza 5, I-09047 Selargius (CA), Italy
\and University of Manchester, Jodrell Bank Observatory, Macclesfield SK11 9DL, UK
\and University of Manchester, School of Physics and Astronomy, Jodrell Bank Center for Astrophysics, Alan Turing Building, Manchester, M13 9PL, UK
\and Alfred P. Sloan Research Fellow
\and National Radio Astronomy Observatory, Green Bank, WV 24944, USA
\and LESIA, Observatoire de Paris, Universit\'e PSL, CNRS, Sorbonne Universit\'e, Universit\'e de Paris, 5 place Jules Janssen, 92195 Meudon, France
\and Arecibo Observatory, HC3 Box 53995, Arecibo, PR 00612, USA
\and Universit\`a  di Cagliari, Dip di Fisica, S.P. Monserrato-Sestu Km 0,700 - 09042 Monserrato - Italy
\and School of Chemistry, University of East Anglia, Norwich, NR4 7TJ, United Kingdom
}

\date{Accepted XXX. Received YYY; in original form ZZZ}

\abstract
{The double pulsar (PSR J0737$-$3039A/B) provides some of the most stringent tests of general relativity (GR) and its alternatives. The success of this system in tests of GR is largely due to the high-precision, long-term timing of its recycled-pulsar member, pulsar A. On the other hand, pulsar B is a young pulsar that exhibits significant short-term and long-term timing variations due to the electromagnetic-wind interaction with its companion and geodetic precession. Improving pulsar B's timing precision is a key step towards improving the precision in a number of GR tests with PSR J0737$-$3039A/B. In this paper, red noise signatures in the timing of pulsar B are investigated using roughly a four-year time span, from 2004 to 2008, beyond which time the pulsar's radio beam precessed out of view. In particular, we discuss the profile variations seen on timescales ranging from minutes --- during the so-called `bright' orbital phases --- to hours --- during its full 2.5\,h orbit --- to years, as geodetic precession displaces the pulsar's beam with respect to our line of sight. Also, we present our efforts to model the orbit-wide, harmonic modulation that has been previously seen in the timing residuals of pulsar B, using simple geometry and the impact of a radial electromagnetic wind originating from pulsar A. Our model successfully accounts for the long-term precessional changes in the amplitude of the timing residuals but does not attempt to describe the fast profile changes observed during each of the bright phases, nor is it able to reproduce the lack of observable emission between phases. Using a nested sampling analysis, our simple analytical model allowed us to extract information about the general properties of pulsar B's emission beam, such as its approximate shape and intensity, as well as the magnitude of the deflection of that beam, caused by pulsar A's wind. We also determined for the first time that the most likely sense of rotation of pulsar B, consistent with our model, is prograde with respect to its orbital motion. Finally, we discuss the potential of combining our model with future timing of pulsar B, when it becomes visible again, towards improving the precision of tests of GR with the double pulsar. The timing of pulsar B presented in this paper depends on the size of the pulsar's orbit, which was calculated from GR, in order to precisely account for orbital timing delays. Consequently, our timing cannot directly be used to test theories of gravity. However, our modelling of the beam shape and radial wind of pulsar B can indirectly aid future efforts to time this pulsar by constraining part of the additional red noise observed on top of the orbital delays. As such, we conclude that, in the idealised case of zero covariance between our model's parameters and those of the timing model, our model can bring about a factor 2.6 improvement on the measurement precision of the mass ratio, $R=m_{\rm A}/m_{\rm B}$, between the two pulsars: a theory-independent parameter, which is pivotal in tests of GR.}


\maketitle

\clearpage

\section{Introduction}
The PSR J0737$-$3039 system is commonly known as the double pulsar and is composed of an old, recycled pulsar (pulsar A), with a spin period of $P_{\rm A} \approx 22.7$\,ms, and a young pulsar (pulsar B), with a spin period of $P_{\rm B} \approx 2.77$\,s (Burgay et al.~2003\nocite{bdp+03}; Lyne et al.~2004\nocite{lbk+04}). The two pulsars orbit each other in tight, low-eccentricity orbits: the orbital period and eccentricity of the orbits is $P_{\rm b} \approx 2.45$\,h and $e\approx 0.088$, respectively. The double pulsar exhibits a plethora of physical effects, many of them due to the intense gravitational interaction between the two neutron stars (NS), which are never separated by more than $\approx 3$\,light-s. Moreover, the system is viewed almost perfectly edge-on. These properties have allowed for very precise tests of general relativity (GR; Kramer et al.~2006\nocite{ksm+06}).

\section{Previous work}
\subsection{Tests of GR with the double pulsar}
Amongst its many properties, the double pulsar is unique in that it is the only double-NS system we know to date, where we detect pulsed emission from both its members, thus allowing us to directly measure the length of its semi-major axes through pulsar timing. Pulsar A is a very stable rotator, exhibiting little timing noise. Solely from timing pulsar A, several orbital (Keplerian and post-Keplerian) parameters have been measured with high precision for this system; and this pulsar's timing precision has served in testing theories of gravity (Kramer et al.~2006\nocite{ksm+06}; Kramer \& Wex~2009\nocite{kw09}; Kramer et al.~{\em in preparation}). The timing of pulsar B, on the other hand, exhibits significant amounts of systematic noise, over a variety of timescales (see section below). To date, the precision of a number of tests of GR with the Double Pulsar has been limited by the significantly less precise timing of pulsar B, which dominates the uncertainty of the ratio between the two pulsars' semi-major axes, $x_{\rm B}/x_{\rm A}$: in a very large set of gravity theories, this ratio is equal to the mass ratio, $R=m_{\rm A}/m_{\rm B}$, an important theory-independent parameter in those tests.

\subsection{Timing pulsar B}
Within an orbital period, pulsar B exhibits a periodic modulation of its brightness. More specifically, during two orbital phase ranges, called bright phase 1 (hereafter BP1) and bright phase 2 (hereafter BP2), the pulsar appears much brighter than anywhere else (Lyne et al.~2004\nocite{lbk+04}). These two phases correspond to orbital phase (from the ascending node) of $\approx 210^\circ$ and $\approx 280^\circ$, although it has been observed that their location and extent gradually change with time, at the rate of a few degrees per year (Burgay et al.~2005\nocite{bpm+05}; Perera et al.~2010\nocite{pmk+10}; hereafter PMK+10). During the rest of its orbit, for roughly 85\% of the orbital period, pulsar B is barely detectable, which has granted this phase the title, weak phase (hereafter WP). Furthermore, during  each of BP1 and BP2, the profile of pulsar B evolves dramatically as a function of orbital phase, in contrast to the WP, during which the profile evolution is significantly less. This significant profile evolution during each of the bright phases (BPs) was already noticed soon after its discovery (Lyne et al.~2004\nocite{lbk+04}). The strong profile evolution during the BPs, coupled with the intermittent visibility during the orbit, has limited the amount of timing precision that can be achieved for this pulsar. 

Kramer et al.~(2006; hereafter KSM+06)\nocite{ksm+06} presented a detailed timing analysis of pulsar B from observations between MJD\:52760 and MJD\:53736. The authors divided the orbit into five intervals (hereafter also `orbital-phase windows') and generated a timing template per interval to time the pulsar. A different set of five templates was used for every subsequent three-month period, thus accounting for secular pulse-shape changes due to geodetic precession (see Section~\ref{subsec:geoprec}). They published pulsar B's spin parameters, based only on timing data during the WP, having a much more stable profile than the BPs, while excluding BP1 and BP2 from their timing analysis on the grounds of significant profile variation during those phases.

\subsection{The geometry of pulsar B}
The inclination of the double pulsar's orbit is $i\approx 89^\circ$, resulting in periodic, 30s-long eclipses of pulsar A's emission by the magnetosphere of pulsar B, as the latter pulsar moves in front of the former, at conjunction, during their orbit. Breton et al.~(2008; hereafter BKK+08)\nocite{bkk+08} successfully modelled the flux-density modulation of pulsar A during those eclipses, with a simple geometric model describing the relative orientation of the two pulsars. In that work, the authors were able to determine a number of parameters of pulsar B's geometry with high precision: for example, the magnetic inclination, which is the angle between the spin and the magnetic axis, $\alpha=70\fdg 9(4)$, and the angle between the spin axis and the orbital angular momentum, $\delta=50\fdg 0(4)$. It must be noted that in the geometry of BKK+08, the angle that was constrained was the `colatitude' of the spin axis, $\theta=180^\circ-\delta$, which is supplementary to the angle $\delta$ defined in Damour \& Taylor (1992)\nocite{dt92}. Interestingly, the work of BKK+08 could not distinguish between $\delta$ (prograde spin) and $180^\circ-\delta$ (retrograde spin), both of these solutions being degenerate, and hence the sense of the pulsar's rotation could not be uniquely determined.

\subsection{Geodetic precession}
\label{subsec:geoprec}
The problem of timing pulsar B is exacerbated by the long-term modulation of the effects mentioned above, due to the large degree of geodetic precession that this pulsar exhibits. The misalignment of pulsar B's spin axis with the total angular momentum of the binary system, the latter being approximately equal to the orbital angular momentum, $\boldsymbol{L}$, results in the geodetic precession of pulsar B's spin around $\boldsymbol{L}$. BKK+08, through their modelling of pulsar A's eclipses, measured the rate of geodetic precession to be $\Omega_{\rm SO}=4\fdg7(7)$\,yr$^{-1}$, which is consistent with the prediction of GR, meaning $\Omega_{\rm SO}^{\rm GR}=5\fdg0734(7)$\,yr$^{-1}$ (Barker \& O'Connell 1975\nocite{bo75b}). Also, the precession phase, which is the angle between the plane containing $\boldsymbol{L}$ and the spin axis and that containing $\boldsymbol{L}$ and the line of sight (LOS), was determined to be $\phi_{\rm SO}({\rm MJD}\:53857)=51\fdg 2(8)$. Geodetic precession causes the gradual shift of the trace of our LOS across the emission region of pulsar B, as a function of time, which in turn causes the observed pulse profile to change significantly over timescales of years (Burgay et al.~2005\nocite{bpm+05}; PMK+10\nocite{pmk+10}). Ultimately, the emission beam of pulsar B precessed out of view in March 2008 (MJD\:54552), hence limiting the total span of available pulsar B data, from its discovery to its disappearance, to $\approx 4.3$ years (i.e.~MJD\:52997--MJD\:54552). It should be noted that pulsar A does not exhibit geodetic precession, as its spin axis is closely aligned with $\boldsymbol{L}$ (Ferdman et al.~2013\nocite{fsk+13}).

\subsection{Theoretical modelling}
The nature of the profile-shape changes is poorly understood; however, it was proposed early on that the distortion of pulsar B's emission is caused by the pressure of pulsar A's relativistic wind (McLaughlin et al.~2004\nocite{mkl+04}; Arons et al.~2005\nocite{absk05}). In that context, there have been a number of attempts for a physical description: for example, Lyutikov (2005)\nocite{lyu05} proposed a model based on distorted Euler potentials of pulsar B's magnetic dipole in order to explain the observed profile and intensity modulation during the orbit. Although the model predicts that the phase of the pulsed emission drifts by $\sim 15$\,ms across each of the BPs, those drifts progress in opposite directions for BP1 and BP2, according to the model; in contrast, observations show that the pulse drifts towards later pulse phases during both BP1 and BP2 (e.g.~compare Fig.~3 of Ransom et al.~2005\nocite{rdk+05} to Fig.~5 of Lyutikov~2005\nocite{lyu05}). 

More recently, notable efforts were made by Perera et al.~(2012) and Lomiashvili \& Lyutikov (2014\nocite{ll14}; hereafter LL14) to model the shape and location of the emission region of Pulsar B. The former authors used all available pulse profiles at 820\,MHz, from the pulsar's discovery to its disappearance, to calculate the geometry of the bow shock that is formed at the equilibrium points between the wind pressure of pulsar A and the magnetic pressure of pulsar B. By tracing the last closed field lines of pulsar B's confined magnetosphere, those authors were able to estimate emission heights, assuming they are produced at the locations where the tangent to the field lines coincide with the observer's LOS. The estimated heights ranged from $\approx$ 100 to 400\,km. Furthermore, that work tried to model the beam shape of pulsar B's emission, assuming it follows the topology of concentric, elliptic hollow cones of diminishing intensity towards the edge of the beam. Using solely the observed pulse-profile widths during BP1, the authors performed a maximum likelihood analysis to determine the beam ellipticity: the beam's semi-major to semi-minor axis ratio was determined to be $2.6_{-0.6}^{+0.4}$. In addition, their analysis constrained the values of $\alpha$ and $\delta$ to be $\alpha=61^\circ\,^{+8^\circ}_{-2^\circ}$ and $\delta=139^\circ\,^{+5^\circ}_{-4^\circ}$, which the authors claimed were both consistent within $2\sigma$ with those derived by BKK+08. We note, however, that in the geometry of Perera et al.~(2012), the angle $\delta$ (the angle between the orbital momentum and the spin) is identified as $\theta$, which is in fact supplementary to the angle $\theta$ defined in BKK+08. In that sense, the constraint placed on $\delta$ is only consistent with the degenerate solution of BKK+08 where the pulsar spin is retrograde with respect to the orbital motion. Finally, Perera et al.~used the magneto-hydrodynamic confinement model of Tsyganenko (2002a,b\nocite{tsy02a}\nocite{tsy02b}) to set an upper limit on the emission height, based on the maximum deflection angle of the polar magnetic-field lines. The latter was $14\fdg3$, which is equal to the full width, along the semi-major axis, at 10\% of the maximum intensity of their calculated elliptical beam. The upper limit on the emission height was $\sim 25,000$\,km, which is compatible with the values derived by the same authors from field-line tracing, although less constraining.  

Additionally, the work of LL14 modelled the orbital and secular variations of pulsar B, using the MHD model of Tsyganenko (2002a,b\nocite{tsy02a}\nocite{tsy02b}) and a Dungey-type model (Dungey~1961\nocite{dun61}) to describe the deformation of pulsar B's magnetosphere under the strong wind pressure of pulsar A. Motivated by the work of PMK+10\nocite{pmk+10}, LL14 assumed a smooth, analytic shape for the emission beam (at rest) of pulsar B, whose intensity distribution resembles a `horseshoe' convolved with a 3D Gaussian profile. In particular, this shape is consistent with the secular evolution of the average BP1 and BP2 profiles of pulsar B (i.e.~averaged over each of the BPs), which change from predominantly single peaked to double peaked, over the span of $\approx 4$ years (PMK+10\nocite{pmk+10}). LL14 determined the best beam-shape and axial-orientation parameters of pulsar B, as well as the height of radio emission, by fitting the model's predictions of the peak pulse intensity in a pulse period to the observed intensity variations, as a function of orbital and precession phase. It is worth noting that their model did not consider the pulse-phase delays with respect to an inertial system, and hence pulsar timing was not attempted in that work.

\subsection{Paper layout}
The present paper is an attempt to model the periodic and secular variations observed in the pulse profiles of pulsar B, using a geometric model of the interaction with its companion (pulsar A) and a synthetic model of the structure of its emission beam. The ultimate goal of this paper is to provide a semi-analytic description of those variations, as a function of time and orbital phase, which can then be used to improve the timing of pulsar B and hence the precision of tests of gravity with the double-pulsar system. Towards that purpose, this work is organised as follows. In Section~\ref{sec:dataredux}, we describe the data we have used and present an initial timing analysis of pulsar B in order to obtain a qualitative characterisation of the systematic timing variations, which we attempt to model in the following section. In addition, we present our measurements of physical properties such as the dependence of pulsar B's average flux density on orbital phase and observation epoch. We also provide an estimate of the spectral index of pulsar B's flux-density spectrum. In Section~\ref{sec:simwindeff}, we first employ an empirical, toy-model description of the observed systematic variations, based only on the initial timing analysis: this contributes to quantitative estimates of the magnitude and functional dependence of the systematic variations. Following that, we lay out the full 3D geometry of our main model, including the parametrisation of pulsar B's emission beam. We then provide a description of our method of parameter estimation, based on a nested sampling algorithm and the observed flux-density profiles of pulsar B. Finally, we present the resulting most likely configuration of the pulsar's emission geometry, as well as the most likely magnitude of the deflection of pulsar B's beam by pulsar A's wind. In Section~\ref{sec:discussion}, we discuss the magnitude of the timing improvements resulting from the use of our model in timing pulsar B. We also discuss the prospects of future modelling improvements that would give rise to measurements of post-Keplerian parameters, such as beam aberration. At the end of the section, we present tantalising evidence for a systematic displacement of pulsar B's BPs, likely caused by geodetic precession, using a simple model of the secular changes in their orbital locations. The paper concludes with Section~\ref{sec:summaryfuture}, wherein we summarise and discuss the main results.

\begin{table*}
	\centering
	\caption{Properties of the observations and data products used in the analysis of this paper, listed per 100-day MJD bin. Columns 3--6 show the telescope gain ($G$), the observing frequency ($\nu$), the observation bandwidth ($\delta\nu$) and the typical system-noise temperature ($T_{\rm sys}$) of each observational setup, respectively. Column 8 shows the integration time ($t_{\rm sub}$) of each sub-integration in the original data. The second-to-last column shows the final number of sub-integrations contained in each MJD bin, after omitting those that were plagued by RFI and applying a cut-off to the distribution of uncertainties in our subsequent timing analysis (see Section~\ref{sec:timing}). The last column shows the number of observing epochs (separated by at least one day) corresponding to the data in each MJD bin.}
	\label{tab:paperdata}
	\begin{tabular}{lccccccccc} 
		\hline
		MJD & Telescope & $G$  & $\nu$ & $\delta\nu$ & $T_{\rm sys}$ &  Back end & $t_{\rm sub}$  & $N_{\rm sub}$ & $N_{\rm epoch}$\\
	        &            & (K Jy$^{-1}$) & (MHz) & (MHz) & (K) &  & (s) &  & \\
		\hline
		53000--53100  & Parkes & 0.66 & 685  & 64  & 45 & Wide-band correlator & 60 & 33 & 3 \\
		53100--53200  & Parkes & 0.74 & 1400 & 256 & 22 & Wide-band correlator & 60 & 22 & 3 \\
		53200--53300  & GBT    & 2.00 & 820  & 48  & 35 & BCPM                 & 180 & 47 & 4 \\
		53300--53400  & GBT    & 2.00 & 820  & 48  & 35 & BCPM/SPIGOT          & 180/20 & 187 & 2 \\
		53400--53500  & GBT    & 2.00 & 820  & 48  & 35 & BCPM/SPIGOT          & 20 & 376 & 2 \\  
		53500--53600  & GBT    & 2.00 & 820  & 48  & 35 & BCPM/SPIGOT          & 20 & 869 & 9 \\  
		53600--53700  & GBT    & 2.00 & 820  & 48  & 35 & BCPM/SPIGOT          & 20 & 374 & 3 \\  
		53700--53800  & GBT    & 2.00 & 820  & 48  & 35 & BCPM/SPIGOT          & 20 & 951 & 9 \\  
		53800--53900  & GBT    & 2.00 & 820  & 48  & 35 & SPIGOT               & 20 & 567 & 7 \\  
		53900--54000  & GBT    & 2.00 & 820  & 48  & 35 & SPIGOT               & 20 & 65 & 4 \\  
		54000--54100  & GBT    & 2.00 & 820  & 48  & 35 & SPIGOT               & 20 & 285 & 6\\  
		54100--54200  & GBT    & 2.00 & 820  & 48  & 35 & SPIGOT               & 20 & 284 & 5 \\  
		54200--54300  & --     & --   & --   & --  & -- & --                   & -- & 0 & 0 \\  
		54300--54400  & GBT    & 2.00 & 820  & 48  & 35 & SPIGOT               & 20 & 5 & 1 \\  
		54400--54500  & GBT    & 2.00 & 820  & 48  & 35 & SPIGOT               & 20 & 50 & 6 \\  
		\hline
	\end{tabular}
\end{table*}

\begin{table*}
\centering
\caption{The orbital phases of the centroids of BP1, BP2 and the IP, $\phi_{\rm asc}^{\rm \{BP1,BP2,IP\}}$, and the corresponding $3\sigma$ widths, $W_{\rm 3\sigma}^{\rm \{BP1,BP2,IP\}}$, as a function of MJD, estimated via Gaussian fits to the distribution of S/N of the observed profiles (see Fig.~\ref{fig:bp_fits}). All orbital phases are shown as a fraction of the orbit, measured from the ascending node. The $1\sigma$ errors in parentheses correspond to the last significant digit. The second column shows the mean MJD, weighted by the number of observations.}
\label{tab:bplocations}
\begin{tabular}{llllllll} 
\hline
MJD   &  $\langle {\rm MJD}\rangle$     & $\phi_{\rm asc}^{\rm BP1}[{\rm rad}/2\pi]$ & $W_{3\sigma}^{\rm BP1}[{\rm rad}/2\pi]$ & $\phi_{\rm asc}^{\rm BP2}[{\rm rad}/2\pi]$ & $W_{3\sigma}^{\rm BP2}[{\rm rad}/2\pi]$ & $\phi_{\rm asc}^{\rm IP}[{\rm rad}/2\pi]$ & $W_{3\sigma}^{\rm IP}[{\rm rad}/2\pi]$ \\
\hline
53000--53100 & 53018.1  & 0.573(1)   & 0.154(9) & 0.782(3)   & 0.12(2)   & --       &  --  \\
53100--53200 & 53144.7  & 0.579(1)   & 0.140(6) & 0.779(3)   & 0.14(2)   & 0.051(2) &  0.07(1)  \\
53200--53300 & 53231.8  & 0.5753(9)  & 0.128(6) & 0.792(2)   & 0.11(1)   & 0.03(1)  &  0.25(7)  \\
53300--53400 & 53373.1  & 0.5786(4)  & 0.109(3) & 0.795(1)   & 0.106(7)  & 0.02(1)  &  0.25(7)  \\
53400--53500 & 53464.5  & 0.5781(4)  & 0.097(2) & 0.7956(8)  & 0.109(5)  & 0.04(1)  &  0.22(4)  \\
53500--53600 & 53528.0  & 0.5805(2)  & 0.104(1) & 0.7988(5)  & 0.119(3)  & 0.034(4) &  0.22(2)  \\
53600--53700 & 53646.2  & 0.5817(4)  & 0.110(2) & 0.7999(7)  & 0.115(4)  & 0.04(1)  &  0.28(6)  \\
53700--53800 & 53701.9  & 0.5813(2)  & 0.107(1) & 0.8015(4)  & 0.115(2)  & 0.040(7) &  0.23(4)  \\
53800--53900 & 53849.6  & 0.5832(3)  & 0.098(2) & 0.8054(3)  & 0.107(2)  & 0.04(1)  &  0.31(7)  \\
53900--54000 & 53944.4  & 0.589(1)   & 0.110(7) & 0.805(2)   & 0.131(9)  & --       &  --  \\
54000--54100 & 54053.9  & 0.5899(4)  & 0.097(2) & 0.8069(4)  & 0.112(3)  & --       &  --  \\
54100--54200 & 54186.3  & 0.5936(4)  & 0.095(3) & 0.8072(5)  & 0.106(3)  & --       &  --   \\
54300--54400 & 54373.6  & 0.600(5)   & 0.15(3)  & 0.808(1)   & 0.079(8)  & --       &  --  \\
54400--54500 & 54433.3  & 0.6027(8)  & 0.095(5) & 0.8088(7)  & 0.096(4)  & --       &  --   \\
\hline
\end{tabular}
\end{table*}

\section{Data reduction}
\label{sec:dataredux}
The data available for this work came from archival observations with the Parkes and GBT telescopes, at 685 and 1400\,MHz, for Parkes, and 820\,MHz, for GBT. A description of the observations can be found in KSM+06\nocite{ksm+06} and Kramer et al.~({\em in preparation}). The total span of the data was MJD\:53004 to MJD\:54496. The original data products of the GBT observations comprised sub-integrations of 20\,s or 180\,s in length, which contained de-dispersed and averaged profiles of pulsar B across a 48\,MHz band centred at 820\,MHz. The Parkes observations during MJD\:53004--53035 and MJD\:53102--53192 comprised sub-integrations of 60\,s in length, across a 64\,MHz band centred at 685\,MHz, and across a 256\,MHz band centred at 1375\,MHz, respectively. In total, the original data contained roughly 29,500 sub-integrations. The data contained no polarisation information. In a complete reprocessing of the data, the GBT observations were converted from the proprietary format of the original BCPM data to PSRFITS archives, for further processing with PSRCHIVE (Hotan et al.~2004\nocite{hvm04}). To maintain a uniform temporal resolution across all our profiles, we set the pulse-phase resolution to $N_{\rm bin}=512$\,bins, down-sampling where necessary: the corresponding temporal resolution is $\approx 5.5$\,ms. Following visual inspection of the data, we excised those sub-integrations and frequency channels that were plagued by RFI. Subsequently, the remaining frequency channels in each archive were averaged together, leaving only the sub-integrations in the data.

In order to determine the long-term temporal evolution of the emission properties of pulsar B, we uniformly grouped our data into 14 100-day intervals (hereafter MJD bins). Table~\ref{tab:paperdata} shows the instruments and observation properties corresponding to the data products within each MJD bin. Apart from the poorly sampled interval MJD\:54300--54400, and the interval MJD\:54200--54300, where no data were available, all other bins contain at least 20 sub-integrations. Our data set provides a nearly continuous coverage of pulsar B's profile evolution, with a 100-day resolution, of approximately four years.

The chosen length of the MJD bins was optimised to retain a useful amount of data in each MJD bin (see last column of Table~\ref{tab:paperdata}), while at the same time limiting the profile variations across an MJD bin due to geodetic precession. Theoretically, across the MJD range covered by our data, geodetic precession causes a drift of pulsar B's magnetic axis with respect to our LOS at the rate of $d\beta/dt\sim\sin\phi_{\rm SO}\sin(\delta-\alpha)\Omega_{\rm SO}\lesssim 1^\circ$ per MJD bin, where $\beta$ is the impact angle, that is, the angle between the LOS and the magnetic axis at the spin phase of the closest approach. Observationally, PMK+10\nocite{pmk+10} determined the rate of broadening pulsar B's profile to be $\approx 2\fdg 6$ yr$^{-1}$, for ca.~MJD\:53900--54500. Given that the profile's FWHM in BP1 and ca.~MJD\:54000 is $\approx 6^\circ$, the expected intra-bin smearing across an MJD bin is less than 15\%.

\begin{figure}
	\includegraphics[width=\columnwidth]{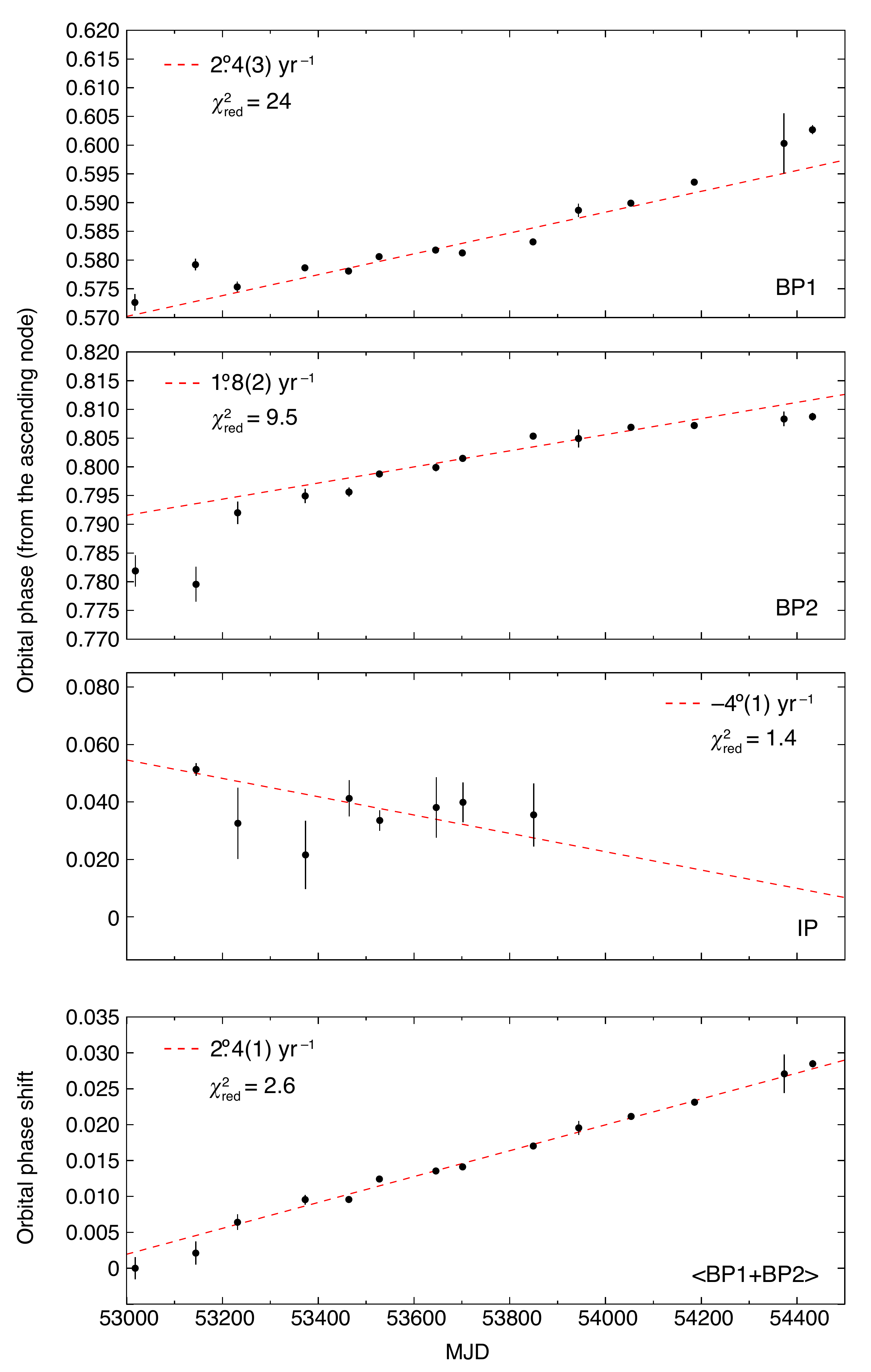}
    \caption{BP1,BP2 and IP: Orbital-phase evolution of the centroids of BP1, BP2, and the IP as a function of MJD epoch. The abscissa values correspond to the mean MJD across each bin (see Table~\ref{tab:bplocations}). The dashed red lines are the best linear fits to the data; the slope and reduced chi-squared of each fit is shown in the legend. <BP1+BP2>: The average orbital-phase shift of the centroids of BP1 and BP2 as a function of MJD epoch; the best linear fit is shown with a dashed red line.}
    \label{fig:bp_evol}
\end{figure}

As has been reported in previous studies, pulsar B is very bright during BP1 and BP2, while it is barely detectable elsewhere, particularly during the WP. Moreover, for at least 800 days, from MJD\:53100 to MJD\:53900, pulsar B appears to have an orbital-phase window of intermediate brightness, roughly an order of magnitude less than that of BP1, which we call the intermediate phase (hereafter IP). To determine the locations and extents of BP1, BP2, and the IP across our data, first we calculated the signal-to-noise ratio (hereafter S/N) of the profiles contained in each MJD bin. Then, we determined the orbital phase corresponding to each profile, from its time stamp and the orbital parameters of pulsar A, appropriately adding 180$^\circ$ to the longitude of periastron (Kramer et al.~{\em in preparation}). More specifically, we converted the site arrival times (SATs) to barycentric arrival times (BATs) using TEMPO2 and calculated for each SAT the phase from the ascending node ($\phi_{\rm asc}$) by solving Kepler's equations, at each step correcting for the periastron advance, $\dot{\omega}\approx 16\fdg9 {\rm \ yr}^{-1}$. The distribution of S/N as a function of $\phi_{\rm asc}$, for each MJD bin, is shown in Fig.~\ref{fig:bp_fits}, where it can be seen that in discrete segments of the orbit the distribution exhibits a roughly symmetric rise and fall of the S/N, with relatively sharp maxima. We approximated those features with a normal distribution, and fitted a Gaussian function to the S/N distribution of each of the BPs, and, depending on whether it was visible, that of the IP. The centroids and $3\sigma$ widths of BP1, BP2, and the IP for each MJD bin, corresponding to the maxima and standard deviations of the fitted Gaussians, are shown in Table~\ref{tab:bplocations}. In this work, the WP was defined as the orbital-phase interval between the upper $3\sigma$ bound of the IP and the lower 3$\sigma$ bound of BP1, which corresponds to roughly $\phi_{\rm asc}/2\pi\in [0.15,0.53]$. As can be seen in Fig.~\ref{fig:bp_evol}, the location of BP1 and BP2 changes significantly with time, whereas the orbital-phase evolution of the IP is more uncertain.

\begin{figure}
	\includegraphics[width=1\columnwidth]{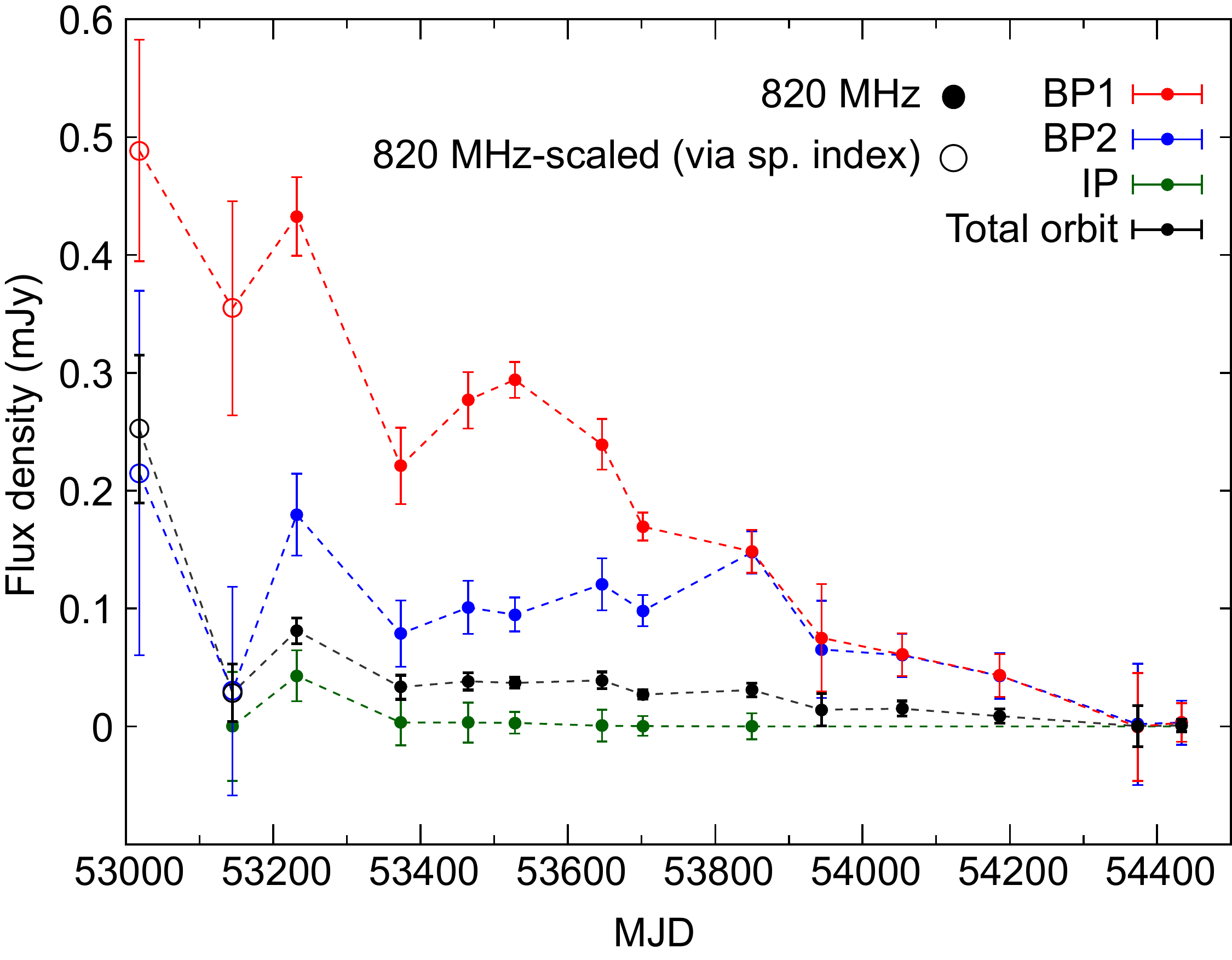}
    \caption{Pulse-averaged flux density of pulsar B at 820\,MHz (filled circles), as a function of MJD, averaged over BP1 (red), BP2 (blue), the IP (green), and over the entire orbit (black). The flux-density calculation of the individual pulse profiles is based on the radiometer equation, assuming a duty cycle equal to the full pulse width at 10\% of the pulse maximum. For the intervals MJD\:53000--53100 and MJD\:53100--53200, the flux densities (empty circles) were scaled to 820\,MHz, using the spectral indices reported in Section~\ref{subsec:spindx}, from observations at 685 and 1400\,MHz, respectively. In the interval MJD\:54200--54300, there were no available data. The dashed lines have been used as guides only.}
    \label{fig:avgfluxes}
\end{figure}

\begin{figure*}
	\includegraphics[width=1\textwidth]{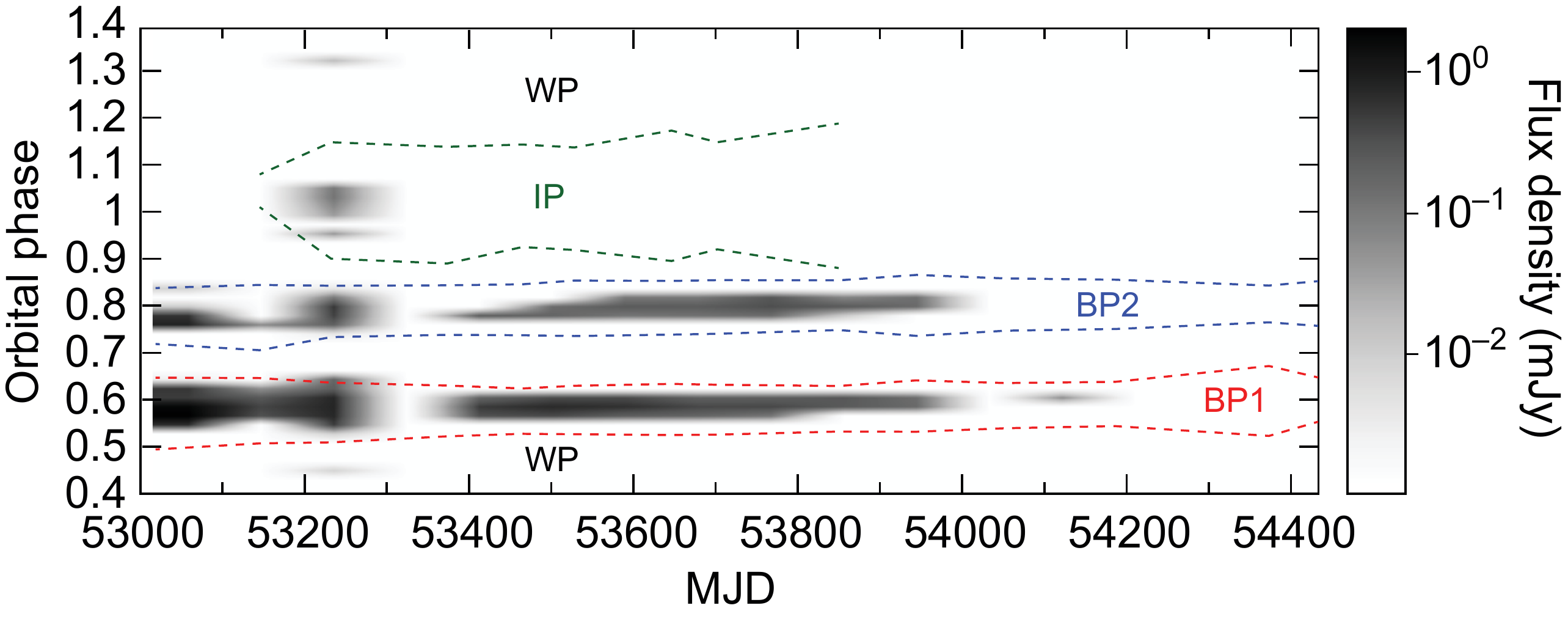}
    \caption{Greyscale map of pulse-averaged flux density of pulsar B, as in Fig.~\ref{fig:avgfluxes}, as a function of orbital phase and MJD. Bilinear interpolation between bins has been applied. The red, blue and green dashed lines delineate the orbital-phase regions of BP1, BP2, and IP as a function of MJD, assuming the corresponding 3$\sigma$ widths shown in Table~\ref{tab:bplocations}. We note that the IP is undetectable before MJD\:53100 and after MJD\:53900. The WP is defined as the orbital-phase region between the upper bound of IP and lower bound of BP1.}
    \label{fig:avgfluxes2}
\end{figure*}

In addition to the phase shifts of BP1 and BP2, there is significant brightness evolution of BP1, BP2, and the IP. We used the radiometer equation to calculate the mean flux density of the observed profiles, based on their S/N, integration length, bandwidth, and pulse width at 10\% of the maximum. In particular, for the 820\,MHz observations with the GBT we used $T_{\rm sys}=35$\,K (based on PMK+10\nocite{pmk+10}) and $G=2$\,K\,Jy$^{-1}$. In the MJD intervals 53000--53100 and 53100--53200, there were no available observations at 820\,MHz, so we decided to use the 685\,MHz and 1400\,MHz data, respectively, from Parkes observations. For those observations, we used $T_{\rm sys}=45$\,K and $G=0.66$\,K\,Jy$^{-1}$ at 685\,MHz, and $T_{\rm sys}=22$\,K and $G=0.74$\,K\,Jy$^{-1}$ at 1400\,MHz. Figure~\ref{fig:avgfluxes} shows the pulse-averaged flux density of pulsar B, averaged over BP1, BP2, the IP, and the total orbit, as a function of MJD. In Fig.~\ref{fig:avgfluxes2}, the pulse-averaged flux density as a function of orbital phase and MJD is presented in a greyscale plot.

\begin{figure}
	\includegraphics[width=\columnwidth]{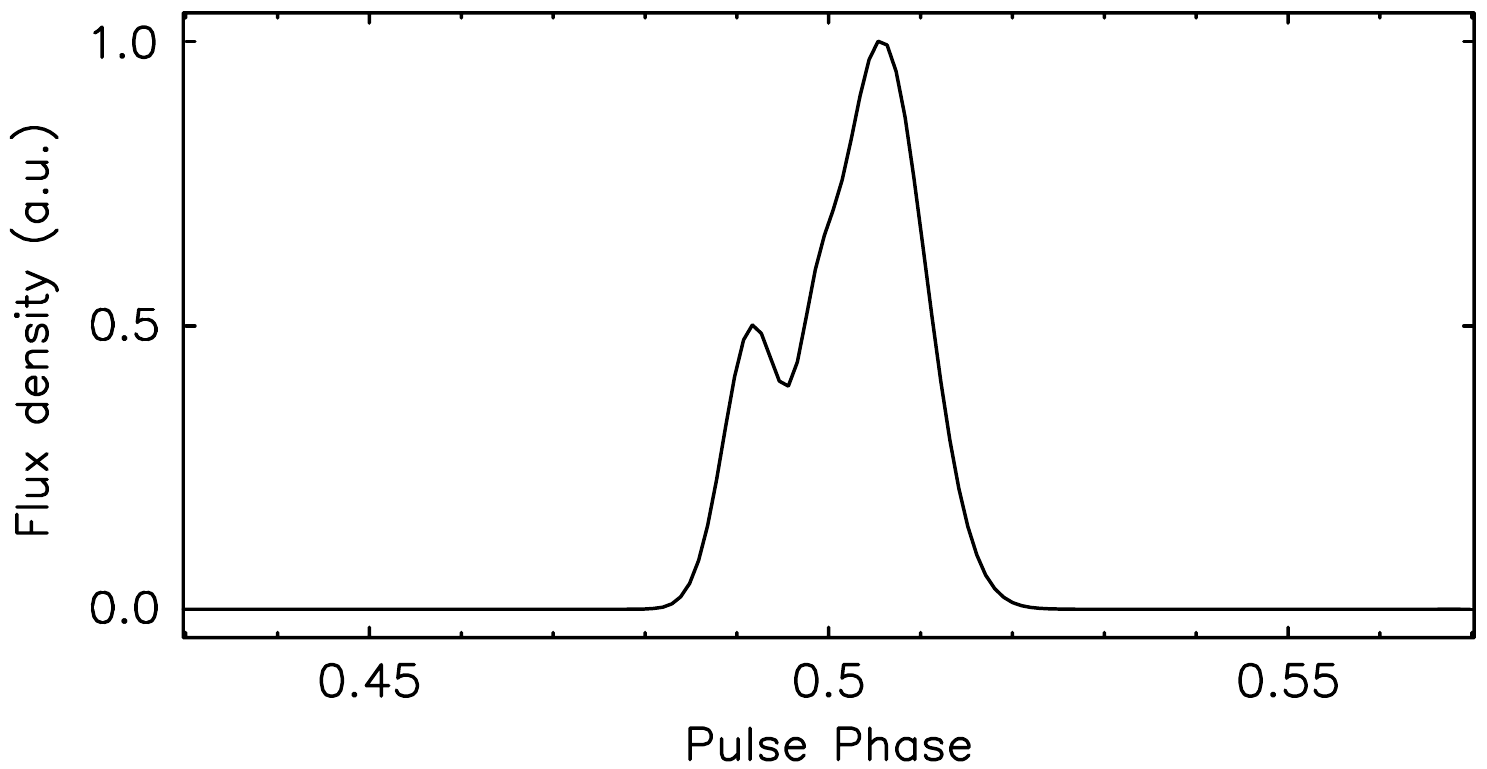}
    \caption{Analytic template constructed from averaging pulsar B observations during the WP ($\phi_{\rm asc}/2\pi\approx 0.15$--$0.53$), between MJD\:53200 and MJD\:53700.}
    \label{fig:wp_template}
\end{figure}

\subsection{Timing analysis}
\label{sec:timing}
\begin{figure}
	\includegraphics[width=\columnwidth]{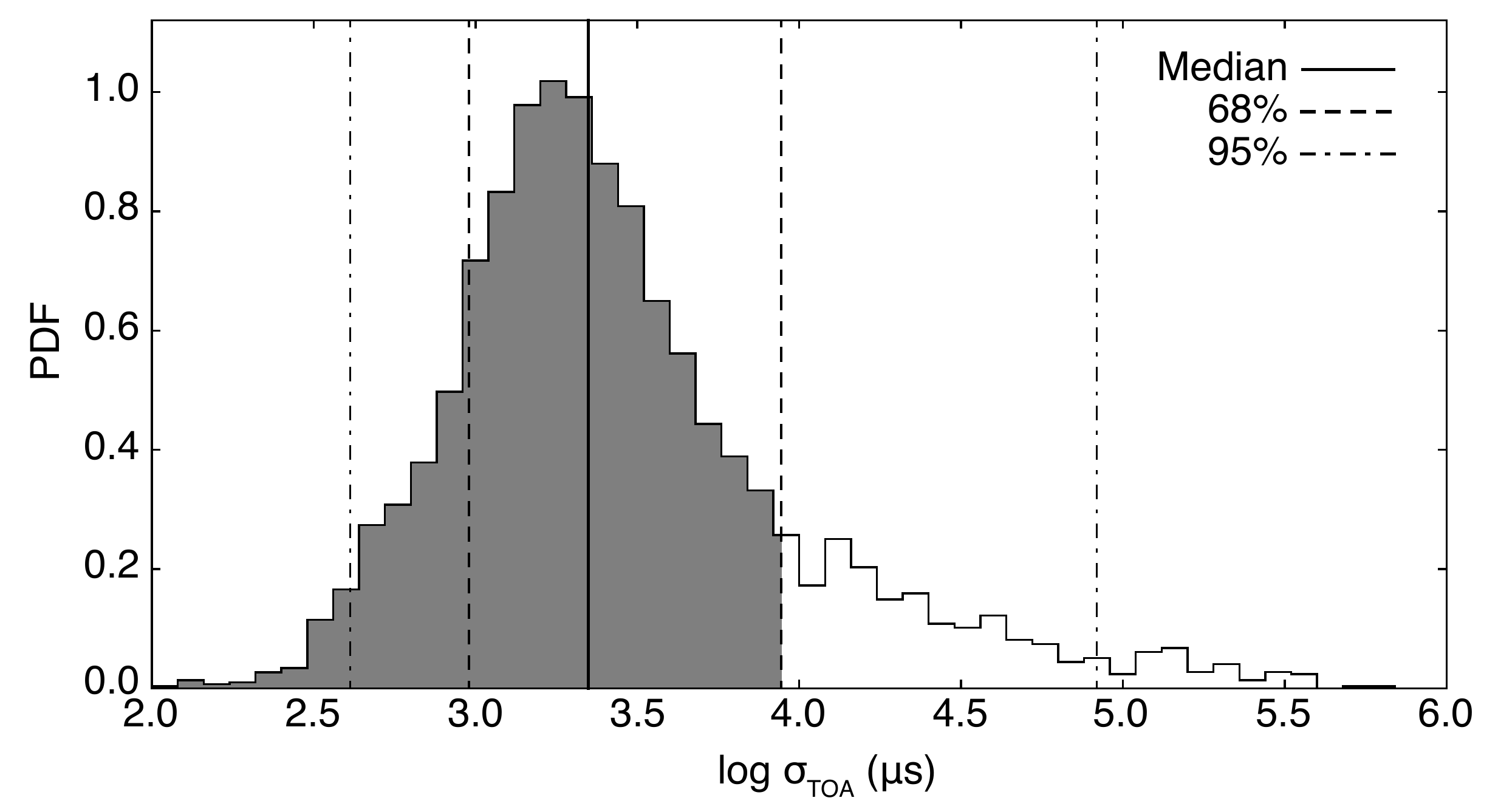}
    \caption{Probability density function (PDF) of the logarithm of TOA uncertainties, $\sigma_{\rm TOA}$ (in $\mu{\rm s}$), from the cross-correlation of all of the original data profiles (after RFI excision) with the WP template profile of Fig.~\ref{fig:wp_template}. The solid, dashed, and dashed-dotted vertical lines correspond to the median, the 68\%, and 95\% confidence intervals of the distribution, respectively. The unshaded area of the histogram corresponds to the range of uncertainties ($\sigma_{\rm TOA}>9.3$\,ms) corresponding to TOAs that were excluded from further analysis.}
    \label{fig:log10toaunc}
\end{figure}

In KSM+06\nocite{ksm+06}, the BPs were excluded from the timing analysis of pulsar B, as the pulse profile evolves dramatically as a function of orbital phase, during each of the orbital-phase intervals corresponding to BP1 and BP2. On the other hand, although much weaker, the profile during the WP appeared to be more stable. Based on those findings, we created an analytic template by fitting a three-Gaussian-component model to the average WP profile of pulsar B (see Fig.~\ref{fig:wp_template}): the latter was generated by averaging all available WP data in our data set, over the range of epochs where the WP emission was detectable. More specifically, we averaged all 820\,MHz BCPM data that were taken between MJD\:53200 and MJD\:53700, noting that although the WP was also detectable during the 1400\,MHz Parkes observations, between MJD\:53100 and MJD\:53200, we wanted to avoid mixing profiles at different frequencies, causing possible profile smearing due to frequency evolution. Subsequently, we generated the pulse times of arrival (TOAs) by cross-correlating the WP template profile of Fig.~\ref{fig:wp_template} with the data profiles of all sub-integrations. The cross-correlation was performed within PSRCHIVE using the Fourier domain fitting algorithm of Taylor~(1992)\nocite{tay92} and Markov chain Monte Carlo (MCMC) sampling to estimate the uncertainties. We note here that in most observations the S/N of the sub-integrations during the IP and the WP is very low. In order to obtain useful fits of the WP template to the profiles of those phases, we increased the S/N by averaging over a number of sub-integrations depending on the amount of signal in the observation. At that point, many of the TOAs still came from random fits to noise and are unwanted. To further excise data that are unrelated to the pulsar's emission, we rejected all TOAs with uncertainties ($\sigma_{\rm TOA}$) exceeding the upper 68\% confidence limit of the $\log(\sigma_{\rm TOA})$ distribution: $\sigma_{\rm TOA}>9.3$\,ms (see Fig.~\ref{fig:log10toaunc}). Finally, we visually inspected all profiles corresponding to the remaining TOAs and removed those that were produced by impulsive RFI. The final data set contained 4,115 TOAs.

Using the spin parameters for pulsar B, published by KSM+06\nocite{ksm+06}, and the orbital parameters, calculated from the ephemeris of pulsar A (Kramer et al.~{\em in preparation}) and assuming GR, we used TEMPO2 (Hotan et al.~2006\nocite{hem06}) and the DDGR model (Taylor \& Weisberg 1989\nocite{tw89}) to calculate the residuals of pulsar B for each MJD bin. We would like to clarify at this point that although it was also possible to use the orbital parameters from KSM+06, which were derived from directly timing pulsar B, this would have introduced systematics due to the noisy timing behaviour of this pulsar --- which is the subject of this paper's investigation. Using the DDGR model and pulsar A's ephemeris provides much higher precision and avoids the systematics of pulsar B. In particular, the DDGR model requires only the Keplerian parameters and the two pulsar masses (determined from pulsar A's timing to a high precision) to calculate all the relativistic effects affecting pulsar B's timing.  Equations~(\ref{eqs:ephempars})--(\ref{eqs:ephempars2}), below, show the definitions of the ephemeris parameters we used to calculate the orbital parameters of pulsar B from those of pulsar A.
\begin{align}
\label{eqs:ephempars}
&M=\left[\frac{\dot{\omega}(1-e^2)}{3T_{\odot}^{2/3}n^{5/3}}\right]^{3/2}\\
&M_{\rm B}=x_{\rm A}\frac{(Mn)^{2/3}}{T_{\odot}^{1/3}\sin i}\\
&M_{\rm A}=M-M_{\rm B}\\
&x_{\rm B}=x_{\rm A}\frac{M_{\rm A}}{M_{\rm B}}\\
\label{eqs:ephempars2}
&\omega_{\rm B}=180^\circ+\omega_{\rm A},
\end{align}
where $T_{\odot}=(\mathcal{GM})^{\rm N}_\odot/c^3$ is the solar-mass parameter, expressed in units of time\footnote{See https://www.iau.org/static/resolutions/IAU2015\_English.pdf}; $n=2\pi/P_{\rm b}$ is the orbital frequency; $M_{\rm A}$, $M_{\rm B}$, and $M$ are the inertial masses of pulsar A, pulsar B, and the binary system, respectively; $x_{\rm A}$ is the projected semi-major axis of pulsar A's orbit; and $\omega_{\rm A}$ and $\omega_{\rm B}$ are the longitudes of periastron of pulsar A and pulsar B, respectively.

We must clarify here that, although we have assumed GR to calculate $x_{\rm B}$ from the precession of periastron and the Shapiro shape, the result actually covers all fully conservative gravity theories where the generalised Eddington parameters are sufficiently close to the values assumed by GR (cf.~Section III\,B in Damour \& Taylor 1992\nocite{dt92}, where we actually only require $\epsilon -\xi/2$ to be sufficiently close to the GR value). In particular, the modelling of the beam shape of pulsar B and the wind of pulsar A, which we present in Section~\ref{subsec:simemgeo}, is fairly independent of the choice of a gravity theory, as the precision we require for $x_{\rm B}$ in order to calculate orbital phases would be sufficiently good, even with the measured value of this parameter. Later on in this paper (Section~\ref{subsec:timimprov}), where we discuss the improvements on the precision measurements of $R$ that our modelling could bring about, the precisely calculated value of $x_{\rm B}$ becomes more important. Therefore, we must recognise that, because the timing performed in this paper relies upon the GR value of $x_{\rm B}$, our results cannot directly be used to test theories of gravity. However, since $x_{\rm B}$ is the only theory-dependent timing parameter in our analysis, and since as was mentioned above, the assumed values in our timing analysis are consistent across a range of gravity theories, so we are confident that our modelling and the conclusions that stem from it can contribute to future timing observations of pulsar B upon its reappearance (see discussion in Section~\ref{subsec:future}).

Besides the timing corrections of the above ephemeris, no further fits for any of the pulsar parameters were attempted. The residuals in each of the 14 100-day bins are shown in Fig.~\ref{fig:allresidsfits}. A visual inspection of the residuals reveals two types of delay as a function of orbital phase: (a) a slow, harmonic delay across the entire orbit, with an amplitude of $\sim 10$ ms, and (b) a fast and, to first-order, approximately linear delay across each of BP1, BP2, and the IP, of the same order of magnitude as (a). In addition, a secular increase of the amplitude of the above delays can be seen on a timescale of years, which is most likely caused by geodetic precession (e.g.~PMK+10\nocite{pmk+10}). 

One possible source of harmonic variations in pulsar timing is the beam aberration due to the pulsar rotation, which modifies the intrinsic direction of the emitted radiation, as seen by the inertial observer. The amplitude of this effect for pulsar B is $\sim (P_{\rm B}/P_{\rm b})x_{\rm B}\lesssim 0.5$\,ms, where $x_{\rm B}$ is the projected semi-major axis of pulsar B's orbit (Damour \& Taylor 1992\nocite{dt92}); therefore, it cannot account for the observed delays, which have larger amplitudes of 1--2 orders of magnitude. In Section~\ref{sec:simwindeff}, we propose that the presence of the harmonic delay is mainly due to the external action of pulsar A's wind, deflecting pulsar B's emission beam relative to our LOS, and thus modulating the observable part of the pulsar's emission. In this work, we mainly attempt to model that slow, harmonic modulation across the orbit. The fast profile changes across the BPs and the IP are briefly discussed in Section~\ref{sec:discussion}.

\subsection{Pulse profile analysis}
\label{sec:avgprofs}
\subsubsection{Average profiles}
The strong profile evolution of pulsar B across its orbit was already noted by Lyne et al.~(2004)\nocite{lbk+04}, for example. In Section~\ref{sec:simwindeff}, we try to draw conclusions about the structure of pulsar B's emission by modelling this evolution. However, the large number of profiles in the original data would render our modelling approach computationally very expensive. In order to reduce the complexity of the problem, we have decided to generate average profiles across each of the four orbital-phase windows, for every MJD bin.

The significant profile-shape modulation of pulsar B's emission as a function of orbital phase and MJD limits the integration time that can be used to obtain average profiles before averaging smears out any intrinsic features of the pulsed emission. The modelling performed in this paper relies on mapping the evolution of those features as a function of orbital phase and epoch. Therefore, we have tried to limit the amount of smearing caused by averaging over those two parameters. The original data profiles used in this work come from sub-integrations with a typical length of $t_{\rm sub}\approx 7P_{\rm B}$ ($\approx 0.002P_{\rm b}$), while only a small subset of profiles (before MJD\:53400) have $t_{\rm sub}\approx 65P_{\rm B}$ ($\approx 0.02P_{\rm b}$: see column 8 of Table~\ref{tab:paperdata}). Such short integrations are not enough to obtain a stable profile, as typically a few hundred to a few thousand pulses are required for this (Lorimer \& Kramer 2005\nocite{lk05}). In order to obtain more stable average profiles in each MJD bin, we have further averaged the original data in orbital-phase bins of $\Delta\phi_{\rm asc}/2\pi=0.02$. The number of profiles in the original data, contained in a given combination of MJD and orbital-phase bins, varied between 1 and 210. 

\begin{figure*}
	\includegraphics[width=1\textwidth]{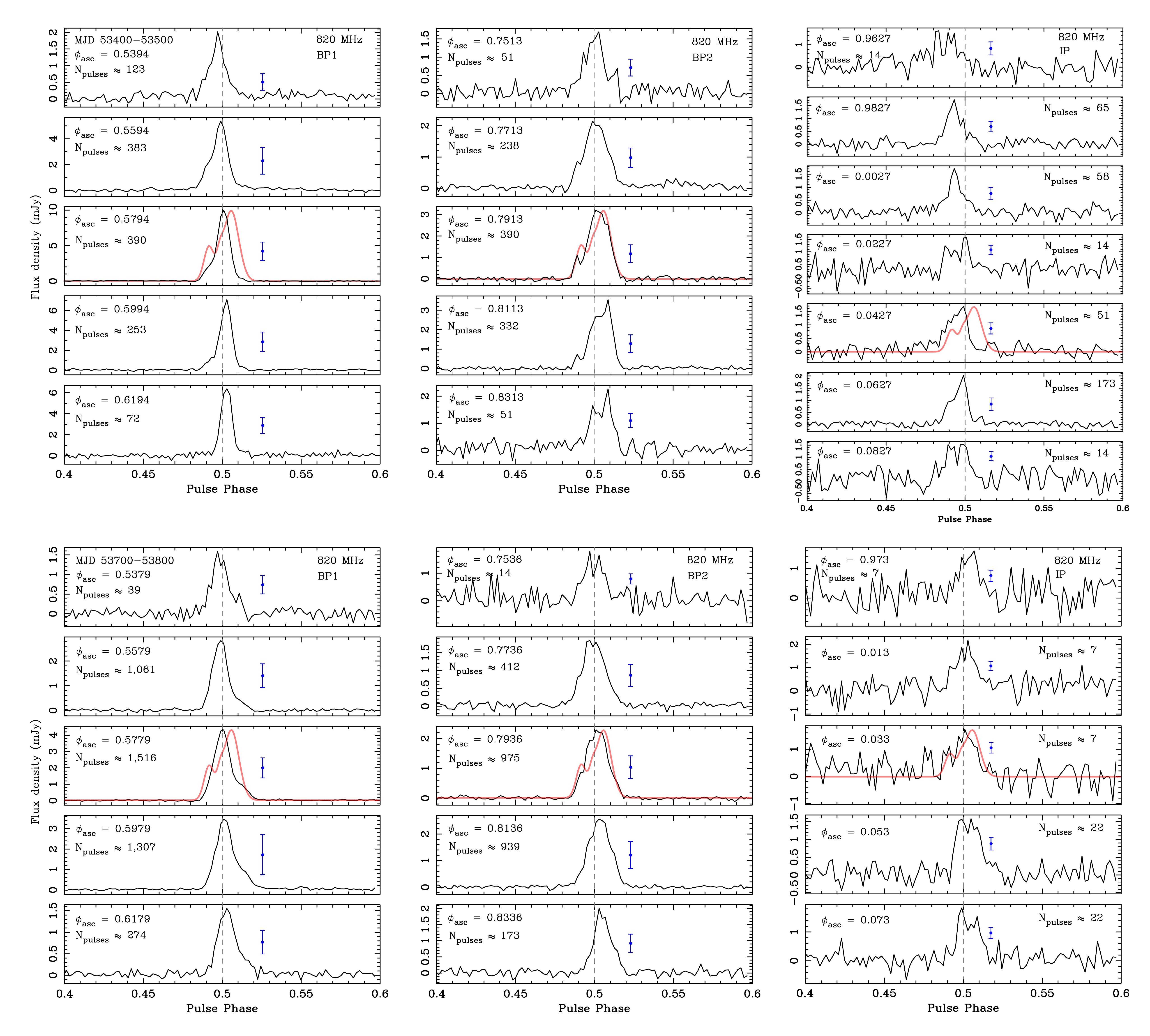}
    \caption{Profile evolution of pulsar B at 820\,MHz, during BP1, BP2, and the IP, corresponding to two MJD bins: (top half) MJD\:53400--53500 and (bottom half) MJD\:53700$-$53800. Each profile comes from averaging the original data in the corresponding MJD bin and an orbital-phase interval of $\Delta\phi_{\rm asc}/2\pi=0.02$, centred around the orbital phase, $\phi_{\rm asc}$, shown in each panel. The number of single pulses averaged to obtain each profile are also shown at the top-left corner of each panel. The grey, vertical dashed lines at pulse phase 0.5 are used here to guide the eye and to highlight the relative changes between profiles observed during most BPs. We note that the profile alignment in this figure can only be used to compare the profiles of a single MJD bin and orbital-phase window (see main text for details). The blue, vertical error bar next to each profile corresponds to the quadrature sum of the off-pulse RMS of the profile shown, $\sigma_{\rm stat}$, and the average RMS due to systematic pulse-to-pulse variations of the profiles that were summed to obtain the profile shown; see Section~\ref{subsubsec:profunc}). For comparison, we have overlaid the analytic WP template profile of Fig.~\ref{fig:wp_template} with the observed average profiles corresponding roughly to the middle of BP1, BP2, and the IP.}
    \label{fig:profileEvoPhasc}
\end{figure*}

In Fig.~\ref{fig:profileEvoPhasc}, we show the evolution of the average profiles across BP1, BP2, and the IP, for a number of orbital-phase bins and for two MJD bins: MJD\:53400--53500 (top panels) and MJD\:53700--53800 (bottom panels). All profiles shown come from folding and averaging the original data with the timing model presented in the previous section. Since the main purpose of this figure is to highlight the relative average profile changes across each of BP1, BP2, and the IP, each group of profiles, corresponding to a combination of MJD bin and an orbital-phase window, has been equally rotated in phase, such that the peak flux density of the centroid profile --- that is, the profile corresponding to the orbital-phase bin that includes the centroid phase, as is tabulated in Table~\ref{tab:bplocations} --- occurs at pulse phase 0.5 (vertical dashed lines). In that figure, the reader can also compare the analytic WP template of Fig.~\ref{fig:wp_template} (red profiles), which was constructed by fitting a three-Gaussian-component model to the average WP profile of pulsar B, with the observed average profiles. The corresponding number of single pulses averaged to obtain those profiles is also shown. In the original data, after the RFI excision and the application of our selection criteria, the fraction of sub-integrations, during the WP, which contained a significant signal ($>5\sigma$) amounts to less than 1\% of the total data set. In addition, during the WP, pulsar B shows little profile evolution. Therefore, for each MJD bin we averaged all available WP data to produce a single, average WP profile per MJD bin.

\subsubsection{Profile uncertainties}
\label{subsubsec:profunc}
The modelling we perform later in this paper is based on the minimisation of the $\chi^2$ between the model and the observed profiles. As such, our method depends on both the magnitude and the uncertainty of each profile's flux-density values. In particular, it is important to consider both the statistical uncertainties ( $\sigma_{\rm stat}$) arising from radiometer noise, as well as the systematic uncertainties ($\sigma_{\rm sys}$) borne from pulse-to-pulse variations within the integration length of our average profiles.

Although the off-pulse RMS in the average profiles ($=\sigma_{\rm stat}$) is a good indicator of the radiometer noise present in our observations, it does not contain any information about the pulse-to-pulse variability across the set of profiles that were averaged together to produce the final profile. Hence, we have decided to make an estimate of the systematic uncertainties corresponding to the flux density of each phase bin, $I_i$, in each average profile, by calculating the RMS of $I_i$ across the $n_{\rm prof}$ profiles that were averaged together. The final RMS of $I_i$, including both statistical ($\sigma_{\rm stat}$) and systematic ($\sigma_{{\rm sys}(i)}$) uncertainties is then given by
\begin{equation}
\label{eq:fluxsigma}
\sigma_i=\sqrt{\sigma_{\rm stat}^2+\sigma_{{\rm sys}(i)}^2}, 
\end{equation}
where 
\begin{equation}
\label{eq:fluxrms}
\sigma_{{\rm sys}(i)}=\sqrt{\frac{\sum_{m=1}^{n_{\rm prof}}(I_{im}-\langle I_i\rangle)^2}{n_{\rm prof}}}
\end{equation}
and $\langle I_i\rangle$ is the average flux density of the $i$th phase bin, across $n_{\rm prof}$ profiles.

Figure~\ref{fig:datprofs} shows a grid of the 41 observed average profiles (black lines), corresponding to the centroids of BP1, BP2, and the IP, and to the WP. The average, centroid profiles of BP1, BP2, and the IP were generated from all original-data profiles contained within the orbital-phase bins that included $\phi_{\rm asc}^{\rm BP1}$, $\phi_{\rm asc}^{\rm BP2}$, and $\phi_{\rm asc}^{\rm IP}$, respectively, according to Table~\ref{tab:bplocations}. The orbital-phase bins from which the above average profiles were calculated are indicated with blue horizontal error bars in Fig.~\ref{fig:allresidsfits}. Each column in the grid corresponds to BP1, BP2, the IP, or the WP, and each row corresponds to a 100-day MJD bin. For each profile, the figure also shows the $1\sigma_{\rm sys}$ flux-density envelope via grey lines. The $1\sigma_{\rm stat}$ value for each profile is shown with a blue vertical error bar.

\subsubsection{Profile templates}
\label{subsec:proftempl}
As a further step towards characterising the average centroid-profile evolution, as a function of orbital and precession phase, we fitted one- or two-component Gaussian templates, depending on the complexity of the profile, to the observed average profiles of Fig.~\ref{fig:datprofs}. We hence represented the noisy observed profiles with the smooth analytical versions shown in Fig.~\ref{fig:gaussdatprofs} (red lines), thus filtering out high-frequency noise and off-pulse artefacts. The description of pulsar B's profile evolution via noiseless templates is particularly advantageous in our modelling of Section~\ref{sec:simwindeff}, wherein we achieve model-parameter convergence by using such templates instead of the observed average profiles. The fitting of the templates was performed using PSRCHIVE, and, at this stage, considered only the statistical uncertainties ($\sigma_{\rm stat}$). Across all 41 profiles, the best fit Gaussian templates were a good match to the observed average profiles, with the total of the summed differences between them corresponding to $\chi^2_{\rm red}=1.26$. Indeed, the construction of those templates, and the decision, for example, of how many Gaussian components are needed to describe the data, may still have been influenced by the presence of non-Gaussian noise. However, based on the $\chi^2_{\rm red}$ of the individual fits, we estimate that for most profiles the contribution of such systematics is of the order of $|(\chi^2_{\rm red})^{1/2}-1|\sigma_{\rm stat}$, which is at least an order of magnitude smaller than the average $\sigma_{\rm sys}$ of the profiles. The exception is the IP profiles at MJD\:53600--53700 and MJD\:53700--53800 and the BP1 profile at MJD\:54300--54400, which come from a single profile in the original data and thus have zero systematic noise: these are profiles with very low S/N and it is difficult to be confident about the reliability of the fit.

\subsubsection{Profile evolution}
\begin{figure}
	\includegraphics[width=1\columnwidth]{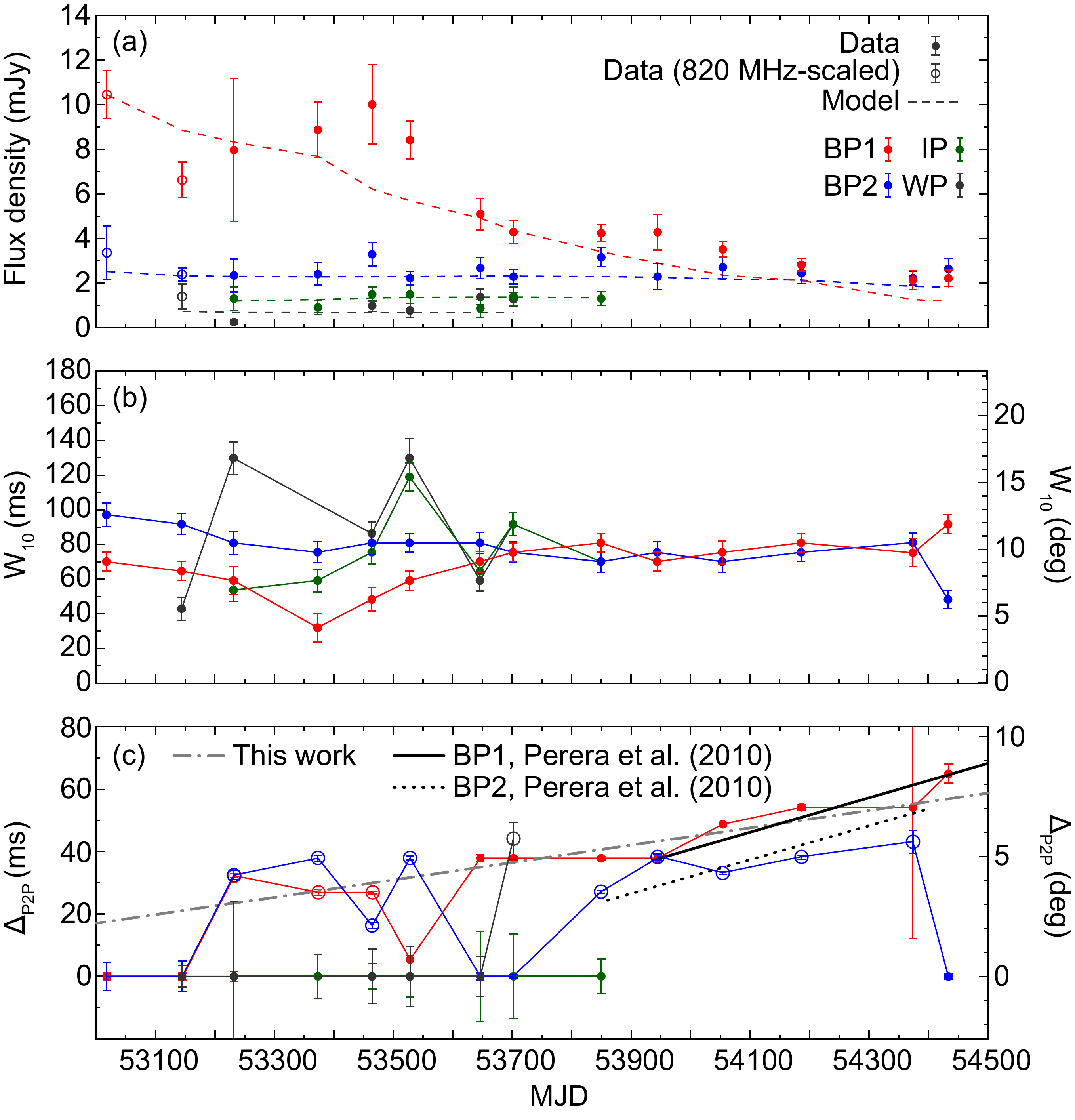}
    \caption{(a) Peak flux density of pulsar B at 820\,MHz (filled circles), as a function of MJD, calculated from the average pulse profile corresponding to the centre of BP1 (red), BP2 (blue), IP (green) and to the WP (grey). For the intervals MJD\:53000--53100 and MJD\:53100--53200, the flux densities (empty circles) were scaled to 820\,MHz from observations at 685\,MHz and 1400\,MHz, respectively, using the spectral indices reported in Section~\ref{subsec:spindx}. In the interval MJD\:54200--54300, there were no available data. The dashed lines show the best fit light curves of the peak flux density of the model. (b) Full pulse width at 10\% of the maximum ($W_{10}$) as a function of MJD, for the average profiles corresponding roughly to the centre of the orbital-phase range of BP1 (red), BP2 (blue), the IP (green), and the WP (grey). (c) Peak separation ($\Delta_{\rm P2P}$) between the leading and trailing component of the average profiles. Open circles indicate that the brightest component is trailing. The dash-dotted grey line is the best linear fit to the BP1 data from this work (see Section~\ref{sec:avgprofs}); the solid and dotted black lines are the best linear fits to the BP1 and BP2 data, respectively, across the corresponding ranges shown by Perera et al.~(2010). All values shown have been calculated from the best fit Gaussian templates of Fig.~\ref{fig:gaussdatprofs}.}
    \label{fig:peakfluxes}
\end{figure}

Using the Gaussian templates derived in the previous sub-section (Fig.~\ref{fig:gaussdatprofs}), we calculated the peak flux density, the pulse width at 10\% maximum, $W_{10}$, and the peak separation between the leading and trailing components ($\Delta_{\rm P2P}$) as a function of MJD, for each orbital-phase window. The evolution of these parameters across our data set is shown in Fig.~\ref{fig:peakfluxes}. It is interesting to note that only the average centroid-profiles of BP1 show a clear flux-density decrease with increasing MJD (see Fig.~\ref{fig:peakfluxes}a). The $W_{10}$ values for all orbital-phase windows do not exhibit any clear trends as a function of time over four years. More precisely, over that time interval, we have $\sigma_{\rm W10}/\langle W_{10}\rangle\approx 0.13-0.36$, where $\sigma_{\rm W10}$ is the RMS of the width over the data span and $\langle W_{10}\rangle$ is the mean value. Locally, the exception is the evolution of the BP1 profiles, during the interval MJD\:53300--53900. During that time, we observe an evolution from a profile with a weak leading component and bright trailing one (ca.~MJD\:53300--53600) --- and with a significant decrease of $\Delta_{\rm P2P}$ to roughly zero, during ca.~MJD\:53400--53600 --- to a profile with a bright leading component and weak trailing one, with $\Delta_{\rm P2P}$ monotonically increasing thereafter, for the remainder of the data span. This exchange of the relative position of the brightest component is represented in Fig.~\ref{fig:peakfluxes}c with a change in the sign of $\Delta_{\rm P2P}$.

Moreover, PMK+10\nocite{pmk+10} measured the average rate of change of $\Delta_{\rm P2P}$ for BP1 and BP2 during the interval MJD\:53900--54500. Their measured rates for BP1 and BP2 were $2\fdg6(1)$\,yr$^{-1}$ and $2\fdg6(2)$\,yr$^{-1}$, respectively. We can compare those rates with our measurements based on the average profiles at the centroids of BP1 and BP2 for a similar range of epochs. Our calculations considered only the absolute component separation ($|\Delta_{\rm P2P}|$) thus ignoring their relative intensity. We find that for BP1, for the above range of epochs, $|d\Delta_{\rm P2P}/dt|=2\fdg4(2)$\,yr$^{-1}$. For BP2, we could not detect a significant leading component beyond MJD\:54400; restricting the epoch range to MJD\:53800--54400, we estimated that $|d\Delta_{\rm P2P}/dt|=1\fdg4(5)$\,yr$^{-1}$. Furthermore, over the entire data span, we find that BP1 shows the most systematic change of $\Delta_{\rm P2P}$. In particular, if we exclude MJD\:53000--53100 and MJD\:53100--53200, where $\Delta_{\rm P2P}=0$, and MJD\:53500--53600, where the second component is marginally detected, a linear fit yields $|d\Delta_{\rm P2P}/dt|=1\fdg3(2)$\,yr$^{-1}$. The fits by PMK+10\nocite{pmk+10} and our fit to the BP1 values are shown in Fig.~\ref{fig:peakfluxes}c. The difference between our value of the global gradient of $\Delta_{\rm P2P}$, for BP1, being roughly half of that published by PMK+10, is most likely the result of the different MJD ranges considered.  

\subsubsection{Beam morphology}
According to the geometry of BKK+08\nocite{bkk+08}, the direction of geodetic precession leads to an increasing separation between our LOS and the visible magnetic pole of pulsar B over the span of our data. In other words, the absolute value of the impact angle ($|\beta|$) increases with time ($\beta$ itself becomes more negative). The approximate amount by which this happens is $|\Delta\beta|=|\beta({\rm MJD\:54500})-\beta({\rm MJD\:53000})|\approx |-15^\circ-(-2^\circ)|=13^\circ$ (see also Fig.~\ref{fig:bpsinefits}b). This fact, together with the evolution of the average BP1 and BP2 profiles, from mostly single-peaked profiles (before MJD\:53700) to predominantly double-peaked profiles (after MJD\:53800), suggests a convex emission beam, where the separation of the active regions increases with the distance from the magnetic axis. Therefore, at least based on these simple observations, the emission beam does not appear to be consistent with a concentric ring or a wedge, centred on the magnetic axis. In particular, it contradicts the proposed horseshoe model by LL14, that is, a concave wedge centred on the magnetic axis, which would result in a profile evolution from a double- to a single-peaked profile, based on the geometry of BKK+08\nocite{bkk+08}. Such a geometry would also lead to a decrease of the overall profile width with time. 

A simple calculation can be made in support of the above statements. Assuming that at the start of our data the LOS trace was tangent to an emission region with a circular-ring geometry of radius $R$ --- where the circular ring has a negligible thickness compared to $R$ --- then, according to Fig.~\ref{fig:peakfluxes}c, at the end of the data the peak separation would reach a maximum of $\Delta_{\rm P2P}\approx 8^\circ$. The radius $R$ is then simply given by $R=[0.25\Delta_{\rm P2P}^2+(\Delta\beta)^2]/(2|\Delta\beta|)\approx 7\fdg 1$, where $|\Delta\beta|$ is the absolute value of the change of $\beta$ across our data, as was defined above. For such a circular beam geometry, the rate of change of the pulse width with respect to time would be $d(\Delta_{\rm P2P})/dt=4(R-|\Delta\beta|)(d\beta/dt)/\Delta_{\rm P2P}$. The average rate is $\langle d(\Delta_{\rm P2P})/dt\rangle=4(R-|\Delta\beta|)(\Delta\beta/T)/\Delta_{\rm P2P}$, where $T\approx 4$\,yr is the length of our data. Using the values corresponding to the end of our data set, the rate is $\langle d(\Delta_{\rm P2P})/dt\rangle\approx 9\fdg6$\,yr$^{-1}$. Compared to the observed evolution of the peak-to-peak separation of BP1 across our data set, we find that this rate is over three times higher, suggesting that the emission region must indeed be elongated rather than circular. 
The exact shape of pulsar B's beam is investigated in Section~\ref{subsec:simemgeo}.

\begin{figure}
	\includegraphics[width=1.0\columnwidth]{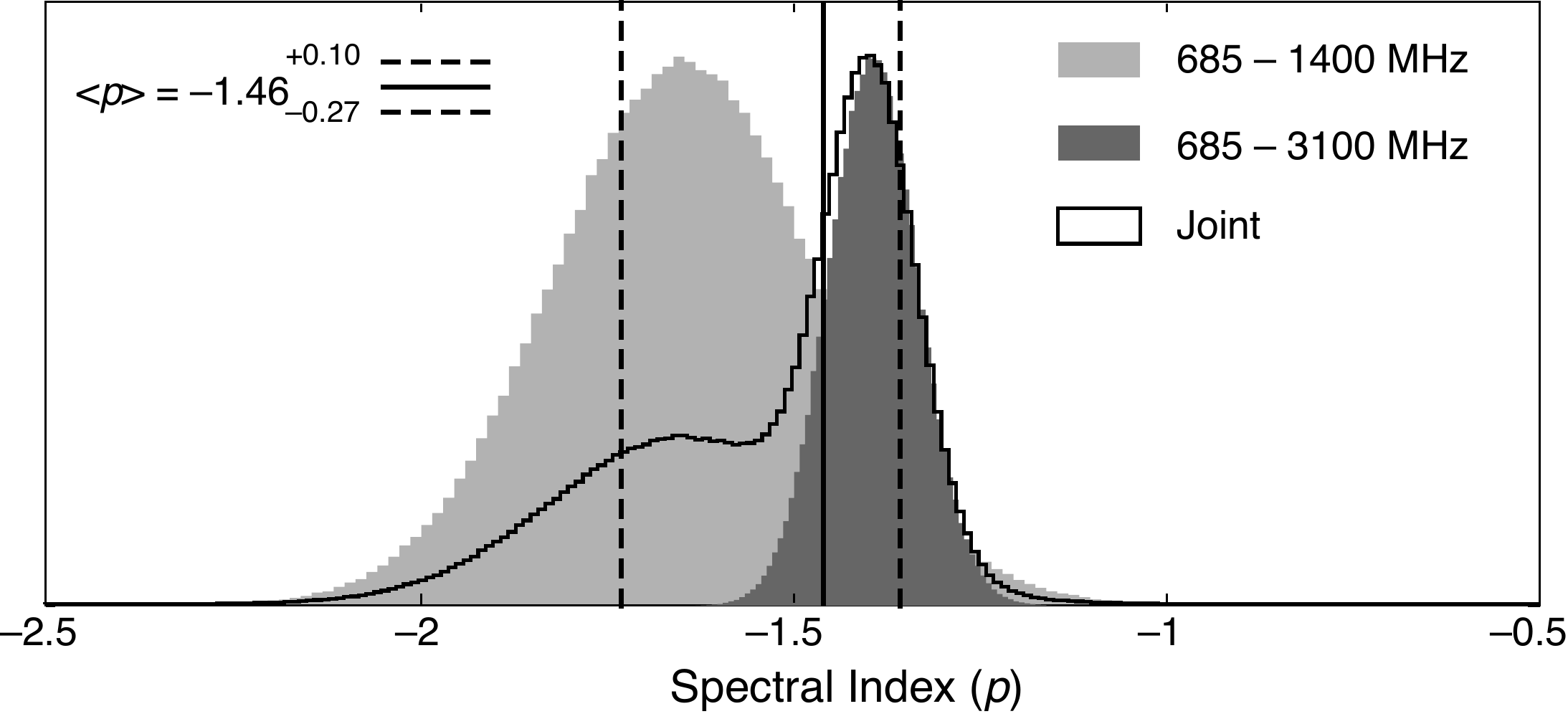}
    \caption{Probability distributions of the spectral index of pulsar B's radio emission, calculated assuming a single power-law between 685 and 1400\,MHz (light grey shade) and 685 and 3100 MHz (dark grey shade), from multi-frequency profiles corresponding to roughly the same orbital phase and separated in time by roughly 30 days. The joint probability distribution is shown with a solid black line, and the corresponding median spectral index and 68\% confidence interval, with solid and dashed vertical lines, respectively.}
    \label{fig:spindxdist}
\end{figure}

\subsection{Spectral index}
\label{subsec:spindx}
The modelling we perform in Section~\ref{sec:simwindeff} partly depends on the flux-density evolution of pulsar B's profile. Therefore, in order to have comparable flux densities across our data set, we must scale the flux density of our multi-frequency data set to a common reference frequency. To achieve that, we need to estimate the spectral index of pulsar B's flux-density spectrum. For this estimate, we have used the peak flux density of multi-frequency, flux-calibrated, average profiles at 685, 1400, and 3100\,MHz. We note that the 3100\,MHz data were only available during MJD\:53000--53100 and came from observations with Parkes and the WBCORR backend, using 768 MHz of bandwidth; these data were calibrated using $T_{\rm sys}=28$\,K and $G=0.62$\,K\,Jy$^{-1}$ (see KSM+06). As is shown in Fig.~\ref{fig:avgfluxes2}, the radio flux is significantly modulated both across the orbit and during the BPs, and, due to precession, it diminishes with increasing MJD. Hence, for the spectral-index fitting, we required an overlap in orbital phase and MJD. The optimal data combinations, for which we deemed this calculation to be reliable, contained two frequencies, had an orbital phase overlap with $\Delta\phi_{\rm asc}/2\pi<0.01$, and their epochs were separated by 30 days. We calculated the spectral index, $p$, assuming a single power law. 
The results were as follows: (a) from the combination of the 685\,MHz data at MJD\:53005 and $\phi_{\rm asc}/2\pi=0.6058$ with the 1400\,MHz data at MJD\:53033 and $\phi_{\rm asc}/2\pi=0.6068$, the spectral index was $p=-1.65(18)$; (b) from the combination of the 685\,MHz data at MJD\:53005 and $\phi_{\rm asc}/2\pi=0.5658$ with the 3100\,MHz data at MJD\:53034 and $\phi_{\rm asc}/2\pi=0.5721$, the spectral index was $p=-1.39(6)$. The probability distributions of $p$ from the individual data combinations and the joint probability distribution are shown in Fig.~\ref{fig:spindxdist}. The median value of the joint distribution was $\langle p\rangle=-1.46_{-0.27}^{+0.10}$ and was used to scale the average profiles at 685 and 1400\,MHz to their corresponding values at 820\,MHz. All profiles at those two frequencies presented herein have thus been scaled.

\section{Simulating the effects of pulsar A's wind}
\label{sec:simwindeff}

\subsection{Toy model description of the harmonic delays}
\label{subsec:toymodesc}
We explored simple linear relations between the observed harmonic delays in the timing residuals and orbital quantities that vary harmonically as a function of orbital phase. In particular, we tested functions such as the separation between the pulsars ($\propto [1+e\cos(\phi_{\rm asc}-\omega_{\rm B})]^{-1}$) and the radiation pressure of the dipole ($\propto [1+e\cos(\phi_{\rm asc}-\omega_{\rm B})]^2$). Such relations, although varying harmonically across the orbit like our data, are out of phase with the variations in the data. In Fig.~\ref{fig:allresidsfits}, it can be seen that for the epochs where the BPs, the IP, and the WP are detected, the maximum delay occurs near $\phi_{\rm asc}=n\pi$, where $n=0,1,2\ldots$ In contrast, the pulsar separation and the radiation pressure have maxima at $\phi_{\rm asc}=\omega_{\rm B}+n\pi$, with $\omega_{\rm B}$ increasing monotonically from $260^\circ$ to $326^\circ$ during the four-year span of our data due to the precession of periastron. This observation suggests that the harmonic delays are $\propto \cos\phi_{\rm asc}$. Such a dependence possibly implies a relation between the timing delays and the magnitude of the sky-projected component of an orbital property. 

\begin{figure}
    \centering
	\includegraphics[width=0.8\columnwidth]{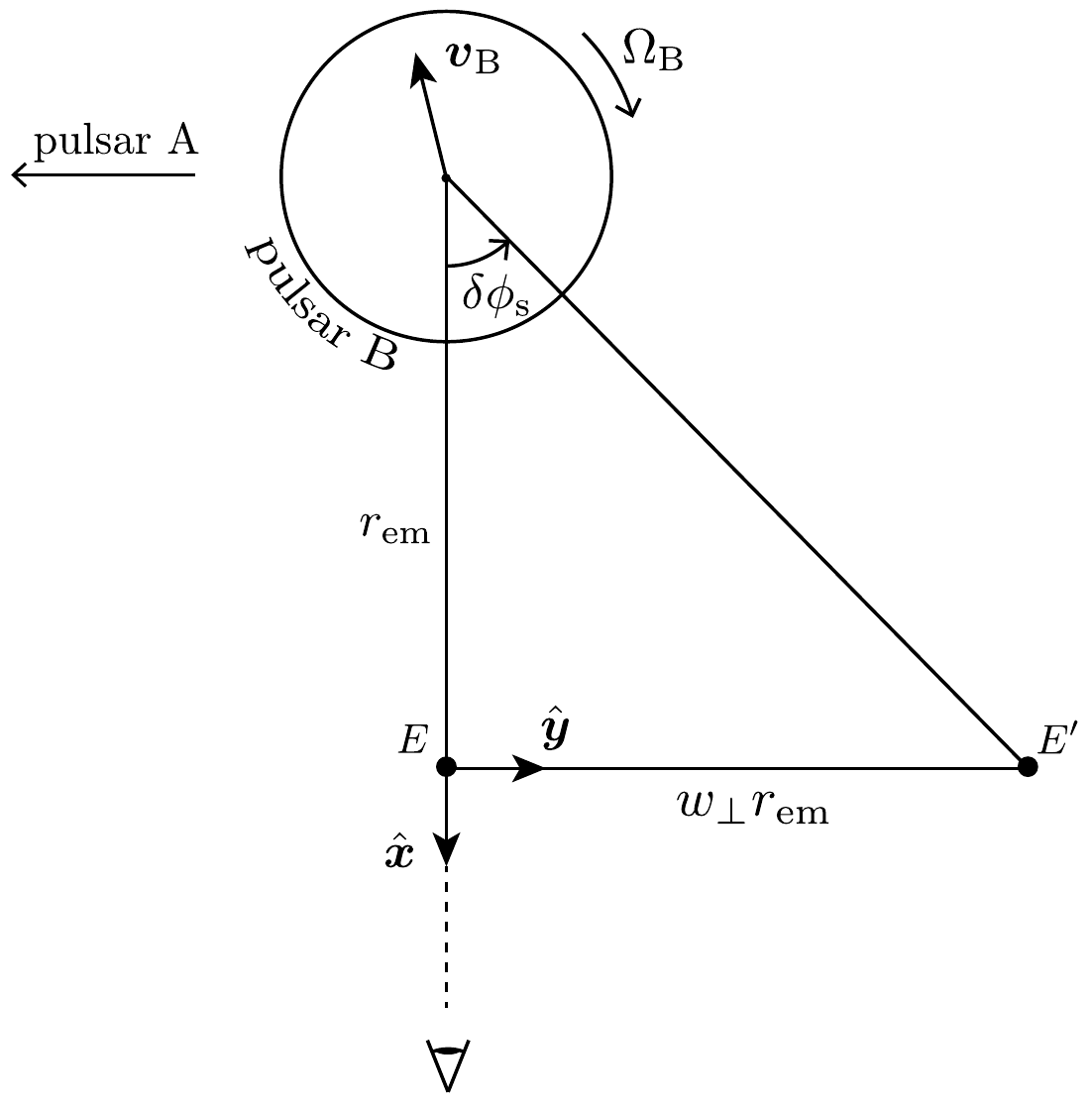}
    \caption{Geometry of the toy model used to fit the harmonic variation of the timing residuals in Fig.~\ref{fig:allresidsfits}. In this figure, pulsar B spins clockwise with a spin frequency of $\Omega_{\rm B}$, and it has a velocity of $\boldsymbol{v}_{\rm B}$, directed away from the LOS. The configuration shown corresponds to $\phi_{\rm asc}=0$, where the wind of pulsar A maximally deflects the emission beam of pulsar B, perpendicularly to the LOS (direction $\hat{\boldsymbol{y}}$) by angle $\delta\phi_{\rm s}$. The LOS direction is defined by $\hat{\boldsymbol{x}}$. In this model, the emission is assumed to be generated at a distance ($r_{\rm em}$) from the centre of the star. The wind of pulsar A, with magnitude equal to $w_\perp$ (in units of $r_{\rm em}$), deflects the emission site from position $E$ to $E^\prime$.}
    \label{fig:toymoddelay}
\end{figure}

In this work, we assume that a radial wind, directed from pulsar A to pulsar B, deflects the emission beam via its relativistic wind pressure. This assumption motivated previous studies of this system based on plasma physics (e.g.~Lyutikov 2005\nocite{lyu05}, Lomiashvili \& Lyutikov 2014\nocite{ll14}). Moreover, the harmonic delays in the residuals are caused by the component of the wind that is perpendicular to the LOS. That is to say,
\begin{equation}
\label{eq:windperp}
\boldsymbol{w}_\perp=w_0\left[1+e\cos(\phi_{\rm asc}-\omega_{\rm B})\right]^2\cos\phi_{\rm asc}\hat{\boldsymbol{y}},
\end{equation}
where we define $\mathbfcal{C}=\{\hat{\boldsymbol{x}},\hat{\boldsymbol{y}},\hat{\boldsymbol{z}}\}$ as the right-handed orthogonal Cartesian reference system, with $\{\hat{\boldsymbol{x}},\hat{\boldsymbol{y}}\}$ being co-planar with the orbital plane, $\hat{\boldsymbol{z}}$ being coincident with the orbital angular momentum vector, $\boldsymbol{L}$, and $\hat{\boldsymbol{x}}$ pointing towards the observer. The geometry assumed in this toy model is shown in Fig.~\ref{fig:toymoddelay}. In the above expression, $w_0$ is a dimensionless coefficient, equal to the orbit-averaged displacement perpendicular to the LOS (for $e\ll 1$), as a fraction of the radius of the emission site, $r_{\rm em}$. To first order, we expect that such a beam deflection introduces a phase delay ($\delta\phi_{\rm s}$), where $\phi_{\rm s}=\Omega_{\rm B}t$ is the spin phase of pulsar B, with $\Omega_{\rm B}$ being the instantaneous spin frequency accounting for spin-down. The amplitude of this delay varies harmonically across the orbit, thus reflecting what we observe in the residuals of pulsar B when a fixed template is used. The corresponding timing delay due to this deflection is
\begin{equation}
\label{eq:winddelay}
\Delta t=\Omega_{\rm B}^{-1}\tan^{-1}(w_\perp).
\end{equation}

We investigated the dependence of $w_0$, as a function of MJD, by fitting Eq.~(\ref{eq:winddelay}) to the residuals in each of the 14 MJD bins. Figure~\ref{fig:allresidsfits} shows the best fit functions and their $1\sigma$ confidence interval for each MJD bin, overlaid with the timing residuals. It can be seen that the amplitude of the function increases monotonically with time. Further analysis shows that the evolution of $w_0$ with time can be approximated well, within the considered MJD range, with a linear regression of the form $w_0=a(t-t_0)$ (see Fig.~\ref{fig:w0fit}). The slope of the regression was determined to be equal to $0.016(1)$\,yr$^{-1}$ and the epoch when $w_0=0$ was determined to be MJD\:52852(65). The MJD at which the amplitude of the harmonic variations of the residuals becomes zero, according to the best linear fit, corresponds to an impact angle of $\beta=0\fdg 0^{+0\fdg 3}_{-0\fdg 9}$; furthermore, $\beta<0$ corresponds to the configuration where the LOS at its minimum approach to the currently visible magnetic pole lies between said magnetic pole and the pulsar's north pole. In addition, we mapped the probability distribution of $a$ and $t_0$, given the entire data set of available residuals, using the Bayesian inference tool, MULTINEST (Feroz et al.~2009\nocite{fhb09}), with uniform parameter priors. For this step, we used the function of Eq.~(\ref{eq:winddelay}), but with a linear dependence of $w_0$ on time, as shown above, to determine the parameter values corresponding to the maximum likelihood. The 2D joint probability-density map for $a$ and $t_0$ is shown as an inset in Fig.~\ref{fig:w0fit}. The most likely values from that analysis were $a=0.015$\,yr$^{-1}$ and $t_0={\rm MJD\:}52820$; the latter epoch corresponds to $\beta\approx -0\fdg 048$. The small value of $\beta$ at $\approx {\rm MJD}\:52850$ implies that the discovery of pulsar B occurred when our LOS was tracing emission very near the magnetic pole.

\begin{figure}
	\includegraphics[width=\columnwidth]{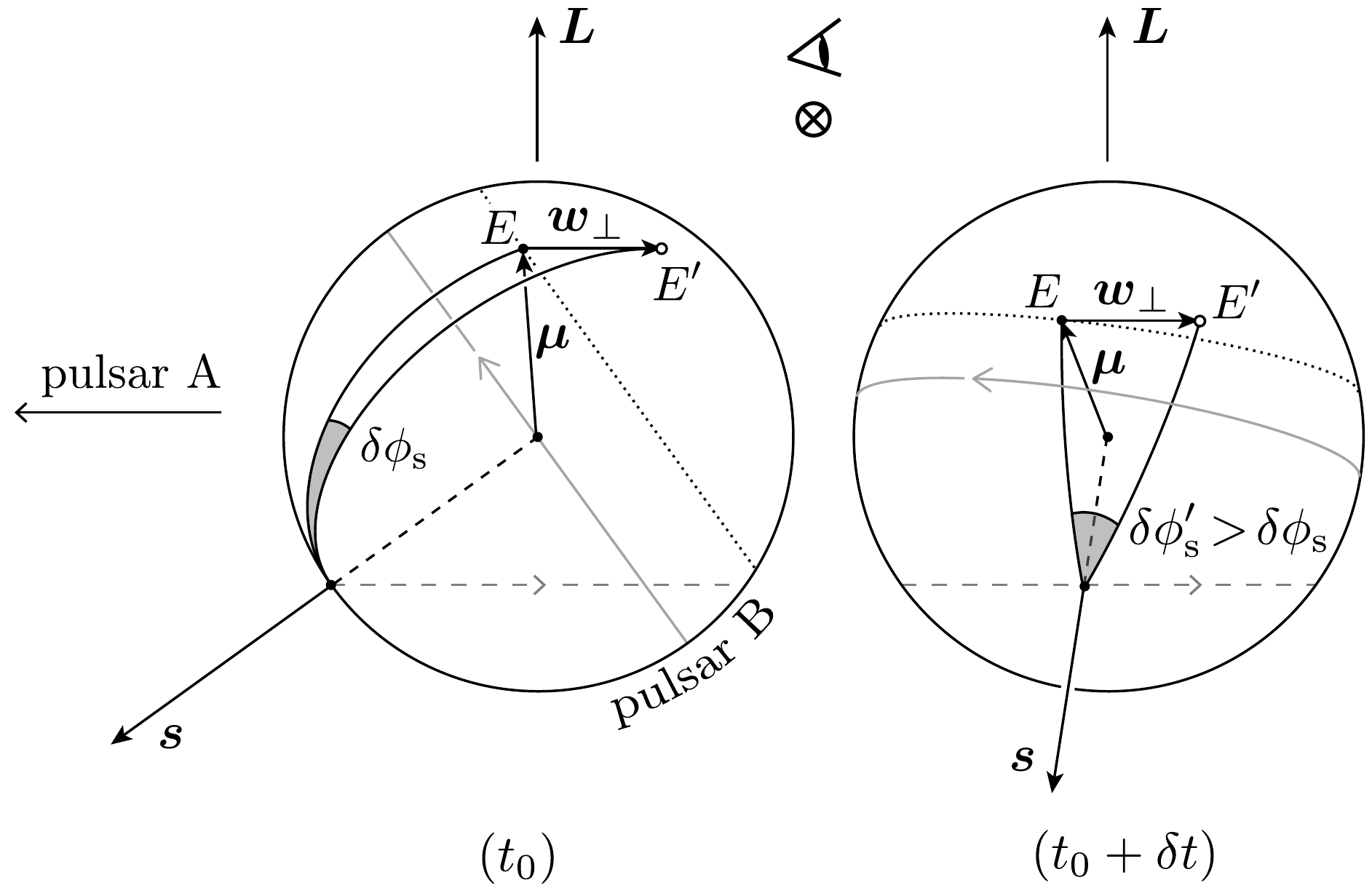}
    \caption{Cartoon representation of spin-phase delay ($\delta\phi_{\rm s}$) produced by the impact of the wind of pulsar A, with a perpendicular component to the LOS ($\boldsymbol{w}_{\perp}$) for two different phases of geodetic precession: at time $(t_0)$ and at a later time $(t_0+\delta t)$. In this representation, the observer's LOS is perpendicular to the plane of the figure, when looking down onto it; the plane of the orbit is perpendicular to the orbital angular momentum vector, $\boldsymbol{L}$; and pulsar A is located to the left of pulsar B. The precession of the spin axis, $\boldsymbol{s}$, about $\boldsymbol{L}$, proceeds along the path shown with a dashed, grey line, in the direction indicated by the arrow. The NS's equator is shown with a solid, grey line, with the grey arrow indicating the spin direction; the magnetic axis, $\boldsymbol{\mu}$, rotates about the spin axis in the same direction. In both instances of time, the action of $\boldsymbol{w}_{\perp}$ causes the deflection of the emission (assumed here to be at the magnetic pole) from position $E$ (solid black point) to $E^\prime$ (open circle). The equivalent spin-phase delay, $\delta\phi_{\rm s}$ at $(t_0)$ and $\delta\phi^\prime_{\rm s}$ at $(t_0+\delta t)$, is measured as the polar angle between $E$ and $E^\prime$. For the same wind magnitude ($w_{\perp}$), it can be seen that $\delta\phi^\prime_{\rm s}>\delta\phi_{\rm s}$.}
    \label{fig:toymoddelay3D}
\end{figure}

\begin{figure}
	\includegraphics[width=\columnwidth]{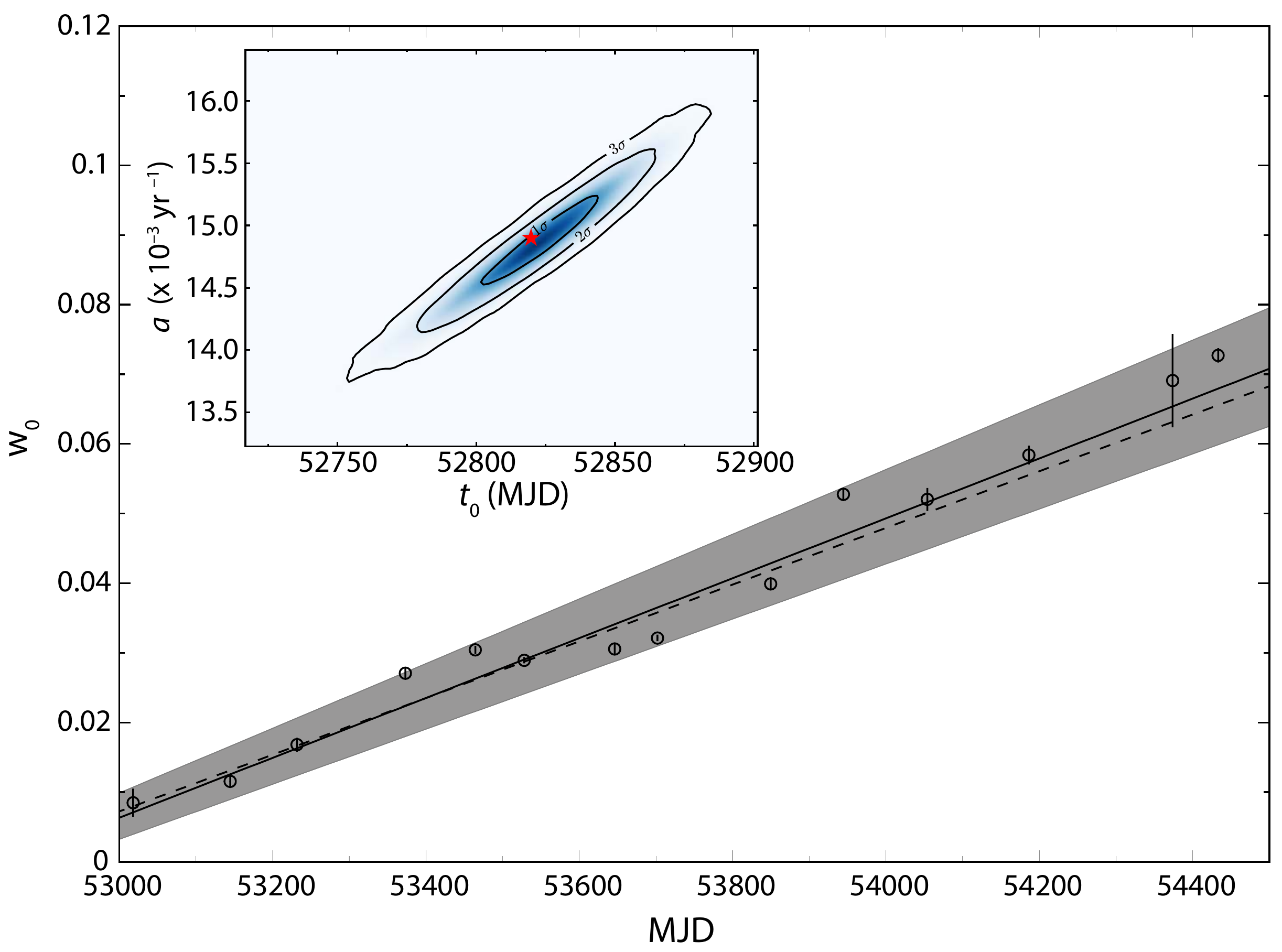}
    \caption{Orbit-averaged wind magnitude ($w_0$) as a function of MJD (data points) from fits of Eq.~(\ref{eq:winddelay}) to the residuals in each of the MJD bins shown in Fig.~\ref{fig:allresidsfits}. The solid grey line and the shaded area correspond to the linear fit to the data and the $1\sigma$ confidence interval of the best fit, respectively. The inset figure shows the 2D joint probability-density function of the rate of change of $w_0$ with time ($a$) and the reference MJD, $t_0$ (at which $w_0=0$), given the entire data set of residuals. The red star symbol corresponds to the location of the most likely values of $a$ and $t_0$. The dashed line corresponds to the most probable function that describes the evolution of $w_0$ with time.}
    \label{fig:w0fit}
\end{figure}

To first order, the above results imply that the impact of the wind displaces the emission region of pulsar B relative to our LOS, with an increasing orbit-averaged magnitude, across the span of our data. This conclusion is most likely the consequence of the simplified 2D model that we used here to fit the data, leading to Eq.~(\ref{eq:winddelay}), rather than that of a wind with a time-dependent orbit-averaged magnitude, $w_0(t)$. In reality, in contrast to the cartoon representation of Fig.~\ref{fig:toymoddelay}, the spin axis of pulsar B is non-orthogonal to the orbital plane, and geodetic precession changes the angles between the spin axis and the wind direction and the spin axis and our LOS, as a function of time. This results in a changing amount of spin-phase delay ($\delta\phi_{\rm s}$) as a function of time, as is depicted in the cartoon representation of Fig~\ref{fig:toymoddelay3D}. This effect is calculated more accurately, as part of the 3D modelling that we perform in the following sections. 

\subsection{Simulation of the emission geometry}
\label{subsec:simemgeo}
As a further step, we attempted to describe the profile evolution of pulsar B, as a function of orbital phase and epoch, with a geometric 3D model of the emission region. The intensity distribution of the emission of pulsar B, relative to the magnetic-axis direction ($\hat{\boldsymbol{\mu}}$) was parametrised with a two-component 2D Gaussian function:
\begin{equation}
\begin{aligned}
\label{eq:modelbeam}
&I(\gamma,\phi)=\\
=&\sum_{\ell=1}^{2}I_{0\ell}\exp[-[A_\ell \Phi_\ell^2(\gamma,\phi) + 2 B_\ell \Phi_\ell(\gamma,\phi) \Gamma_\ell(\gamma,\phi) + C_\ell \Gamma_\ell^2(\gamma,\phi)]], 
\end{aligned}
\end{equation}
where
\begin{equation}
\begin{aligned}
&\Phi_\ell(\gamma,\phi)=d_\ell(\gamma,\phi) \sin\psi_\ell(\gamma,\phi),\\
&\Gamma_\ell(\gamma,\phi)=d_\ell(\gamma,\phi) \cos\psi_\ell(\gamma,\phi),
\end{aligned}
\end{equation}
and
\begin{equation}
\begin{aligned}
&A_\ell=\frac{1}{2\sigma_{\Gamma\ell}^2}\left[\frac{\cos^2\zeta_\ell}{(1-f_\ell)^2}+\sin^2\zeta_\ell\right],\\
&B_\ell=-\frac{\sin(2\zeta_\ell)}{4\sigma_{\Gamma\ell}^2}\left[\frac{1}{(1-f_\ell)^2}-1\right],\\
&C_\ell=\frac{1}{2\sigma_{\Gamma\ell}^2}\left[\frac{\sin^2\zeta_\ell}{(1-f_\ell)^2}+\cos^2\zeta_\ell\right].
\end{aligned}
\end{equation}
The motivation for using such a general Gaussian-beam model, instead of a more specific `horseshoe' or fan-beam shape, came primarily from the high number of degrees of freedom that this model has, even allowing for beam shapes that are similar to those mentioned above. In addition, the LOS traces of such a beam produce one- or two-component Gaussian flux-density profiles, which is a close approximation of the observed profiles of pulsar B, as can be seen in Fig.~\ref{fig:gaussdatprofs}.

\begin{figure}
	\includegraphics[width=\columnwidth]{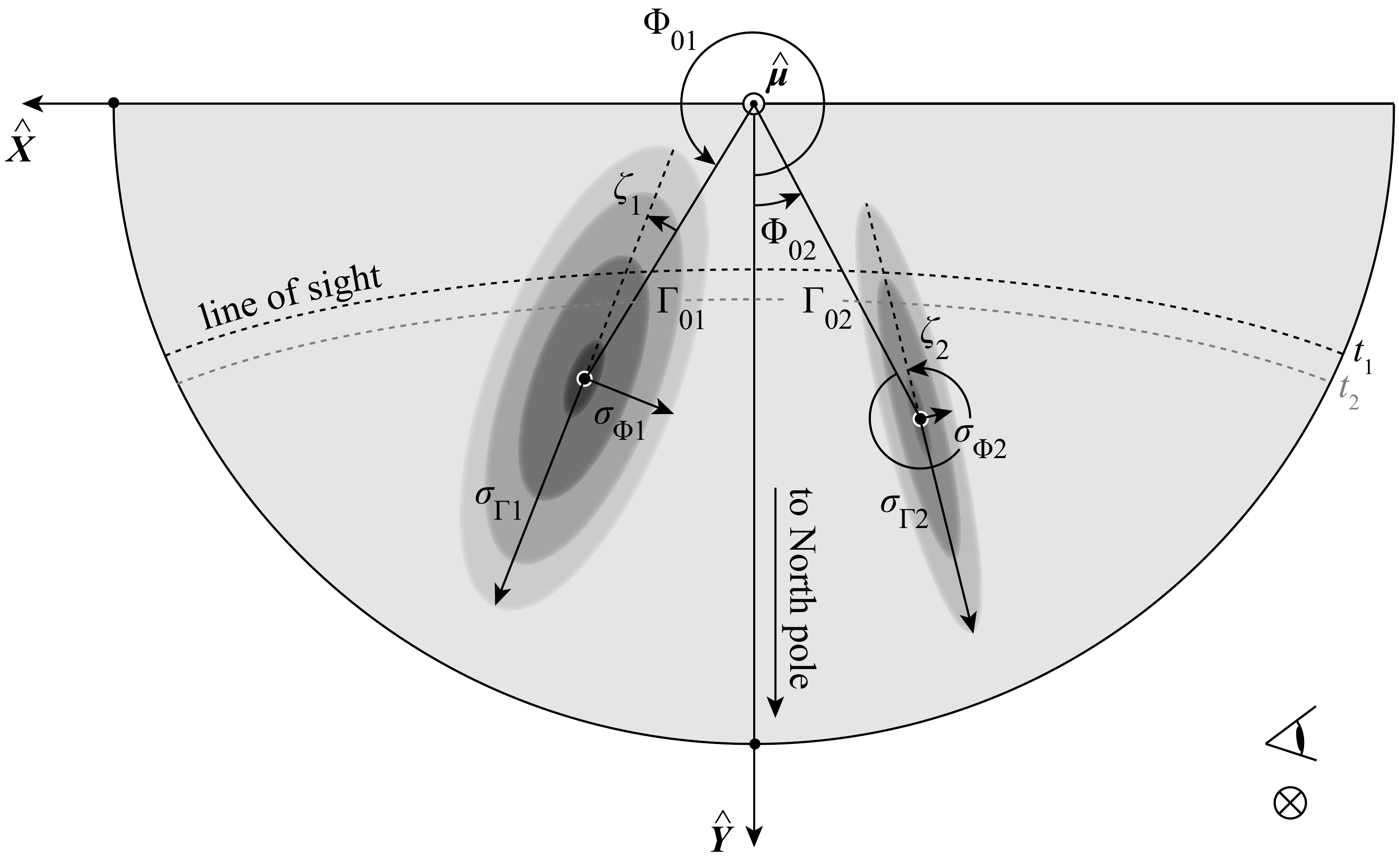}
    \caption{Schematic illustration of the emission beam of pulsar B, modelled in this work as a pair of Gaussian components, of which the intensity profile is represented here as greyscale contours. This figure shows a 2D projection of the emission beam on the plane of the sky and the definitions of the beam parameters that were used in our model to describe the location and shape of the emission (Eq.~(\ref{eq:modelbeam}); compare with Fig.~\ref{fig:simgeom}). The white circles at the centres of the elliptical contours show the locations of maximum beam intensity. In this representation, the observer's LOS is perpendicular to the plane of the paper (as denoted by the circled cross). The circled dot at the top of the figure shows the direction of the magnetic axis ($\hat{\boldsymbol{\mu}}$) in this projection pointing out of the plane of the paper, towards the observer. The black and grey dotted lines show two examples of the trace of the LOS across the beam, at epochs $t_1$ and $t_2(>t_1)$, respectively.}
    \label{fig:beammodelang}
\end{figure}

The definitions of the all the angles used in our parametrisation are depicted in Figs.~\ref{fig:simgeom}a,b and \ref{fig:beammodelang}. The polar coordinates (hereafter beam coordinates) in this parametrisation are $\gamma$, which is the beam colatitude, equal to the angle between the magnetic axis and the direction of emission; and $\phi$, which is the beam longitude, equal to the angle between the plane containing the pulsar's spin and magnetic axes (fiducial plane) and the plane containing the magnetic axis and the direction of the emission. The angle $\phi$ --- where $\phi\in[0,2\pi)$ --- is measured counter-clockwise on the plane of the sky, from the pulsar north through to the pulsar south. The parameter $I_{0\ell}$ corresponds to the peak intensity of each of the two Gaussian components; $\Gamma_{0\ell}$ and $\Phi_{0\ell}$ are the beam colatitude and longitude of the peak-intensity location of the Gaussian components, respectively; $\sigma_{\Gamma\ell}$ and $\sigma_{\Phi\ell}=(1-f_\ell)\sigma_{\Gamma\ell}$ are the half-beam widths of the Gaussian components, along the direction of increasing $\Gamma_{0\ell}$ and $\Phi_{0\ell}$, respectively (represented by the unit vectors, $\hat{\boldsymbol{\Gamma}}_{0\ell}$ and $\hat{\boldsymbol{\Phi}}_{0\ell}$) --- where $f_\ell$ is the flatness parameter, with $f_\ell=0$ corresponding to a circular Gaussian. Finally, $\zeta_{\ell}\in[0,\pi)$ is the angle of rotation about the direction of peak intensity of the Gaussian, measured anti-clockwise on the plane of the sky.

In addition, as can be seen in Fig.~\ref{fig:simgeom}b, $d_\ell$ is the angle between the location of maximum intensity of each of the Gaussians, with beam coordinates $(\Gamma_{0\ell},\Phi_{0\ell})$, and a given location on the sky, with coordinates $(\gamma,\phi)$:
\begin{equation}
\label{eq:dell}
d_\ell=\cos^{-1}\left[\cos\Gamma_{0\ell}\cos\gamma+\sin\Gamma_{0\ell}\sin\gamma\cos\left(\phi-\Phi_{0\ell}\right) \right]. \\
\end{equation}
$\psi_\ell$ is the angle between $\hat{\boldsymbol{\Gamma}}_{0\ell}$ and the direction of increasing $d_\ell$ (represented by the unit vector, $\hat{\boldsymbol{d}}_{\ell}$) measured anti-clockwise on the plane of the sky.

In order to parametrise the interaction of the wind with the pulsar beam, we have defined two orthogonal Cartesian reference systems, $\mathbfcal{C}=\{\hat{\boldsymbol{x}},\hat{\boldsymbol{y}},\hat{\boldsymbol{z}}\}$ and $\mathbfcal{C}^\Omega=\{\hat{\boldsymbol{x}}^\Omega,\hat{\boldsymbol{y}}^\Omega,\hat{\boldsymbol{z}}^\Omega\}$. $\mathbfcal{C}$ is defined such that $\hat{\boldsymbol{z}}$ coincides with the direction of the orbital angular momentum ($\boldsymbol{L}$) and $\hat{\boldsymbol{x}}$ points towards the observer. Here, we have approximated $(\boldsymbol{L}\wedge \hat{\boldsymbol{x}})=90^\circ$. In reality, the orbital inclination of the double pulsar system is $89\fdg 3(1)$ (Breton 2009\nocite{bre09})\footnote{Hereafter, the $\wedge$ operator is used to indicate angles between vectors.}. $\mathbfcal{C}^\Omega$ is defined such that $\hat{\boldsymbol{z}}^\Omega$ coincides with the spin axis of pulsar B ($\boldsymbol{\Omega}$) and $\hat{\boldsymbol{x}}^\Omega$ always lies in the plane defined by $\boldsymbol{\Omega}$ and $\hat{\boldsymbol{x}}$ (hereafter {\em fiducial plane}), while $(\hat{\boldsymbol{x}}\wedge \hat{\boldsymbol{x}}^\Omega)<90^\circ$. Hence, $\mathbfcal{C}^\Omega$ is inclined with respect to $\mathbfcal{C}$ by an angle $\delta$, meaning $\boldsymbol{L}\wedge \boldsymbol{\Omega}=\delta$. Figure~\ref{fig:simgeom} shows the full 3D geometry assumed in our model.

The calculation of the beam intensity towards the LOS (direction $\hat{\boldsymbol{x}}$) is defined with respect to the magnetic axis ($\gamma=\phi=0$); it requires the transformation of $\hat{\boldsymbol{\mu}}=\sin\alpha\cos\phi_{\rm s}\hat{\boldsymbol{x}}^\Omega+ \sin\alpha\sin\phi_{\rm s}\hat{\boldsymbol{y}}^\Omega+ \cos\alpha\hat{\boldsymbol{z}}^\Omega$, from $\mathbfcal{C}^\Omega$ to $\mathbfcal{C}$. The transformation from $\mathbfcal{C}$ to $\mathbfcal{C}^\Omega$ can be described as a rotation by $\phi_{\rm SO}$ around the $z$-axis, then a rotation by $\delta$ around the $y$ axis, and, finally, a rotation by $\pi-\phi_0$ about the $z$ axis, where $\phi_0=\tan^{-1}(\tan\phi_{\rm SO}/\cos\delta)$: the last rotation ensures that $\boldsymbol{\Omega}$, $\hat{\boldsymbol{x}}$ and $\hat{\boldsymbol{x}}^\Omega$ are co-planar, so that $\phi_{\rm s}=0$ at the closest approach of the observer's LOS to the magnetic pole. Hence, the magnetic axis in $\mathbfcal{C}$ is expressed as
\begin{equation}
\begin{aligned}
&\hat{\boldsymbol{\mu}}(x,y,z)=R_z(\phi_{\rm SO})\times R_y(\delta)\times R_z(\pi-\phi_0)\times\hat{\boldsymbol{\mu}}(x^\Omega,y^\Omega,z^\Omega)=\\
&=
\begin{bmatrix}
\cos\alpha \cos\phi_{\rm SO} \sin\delta - \sin\alpha \sin(\phi_0-\phi_{\rm s}) \sin\phi_{\rm SO} - \\
- \cos\delta \cos(\phi_0-\phi_{\rm s}) \cos\phi_{\rm SO} \sin\alpha \\
\\
\cos\alpha \sin\delta \sin\phi_{\rm SO} + \\
+ \cos\phi_{\rm s}\sin\alpha \left(\cos\phi_{\rm SO} \sin\phi_0-\cos\delta \cos\phi_0 \sin\phi_{\rm SO}\right)- \\
- \sin\phi_{\rm s}\sin\alpha \left(\cos\phi_0 \cos\phi_{\rm SO}+\cos\delta \sin\phi_0 \sin\phi_{\rm SO}\right) \\
\\
\cos\alpha \cos\delta + \cos(\phi_0-\phi_{\rm s}) \sin\alpha \sin\delta
\end{bmatrix},
\end{aligned}
\end{equation}
where $R_k(\xi)$ is the $3\times 3$ rotation matrix that rotates a 3D vector counter-clockwise by angle $\xi$, around direction $k$. The above operation is non-commutative. The precession phase, $\phi_{\rm SO}$, is a time-dependent quantity, meaning $\phi_{\rm SO}=\Omega_{\rm SO}(t-t_0)$, where $\Omega_{\rm SO}$ is the precession rate and $t_0$ is the reference epoch corresponding to $\phi_{\rm SO}=0$.

At any instant of time, the beam coordinates, $(\gamma_x,\phi_x)$, of the trace of the LOS on the unit sphere centred on pulsar B are given by
\begin{equation}
\begin{aligned}
\label{eq:gammaphi}
\gamma_x&=\cos^{-1}\mu_x, \\
\phi_x&=(\hat{\boldsymbol{\mu}}\cdot \hat{\boldsymbol{n}})\cos^{-1}(\hat{\boldsymbol{t}}_x\cdot\hat{\boldsymbol{Y}})=\\
&=(\hat{\boldsymbol{\mu}}\cdot \hat{\boldsymbol{n}})\cos^{-1}\left(\frac{\sin\delta\cos\phi_{\rm SO}-\mu_x\cos\alpha}{\sin\alpha\sin\gamma_x}\right).
\end{aligned}
\end{equation}
Here, we have defined the orthogonal Cartesian system $\mathbfcal{C}^\mu=\{\hat{\boldsymbol{X}},\hat{\boldsymbol{Y}},\hat{\boldsymbol{Z}}\}$, having its origin at the location of the magnetic pole; $\hat{\boldsymbol{X}}$ is the unit tangent vector pointing in the direction of the star's rotation (i.e.~eastwards), and $\hat{\boldsymbol{Y}}$ is the unit tangent vector pointing in the direction of the north pole (see Fig.~\ref{fig:simgeom}):
\begin{equation}
\label{eq:tvecs}
\hat{\boldsymbol{Y}}=\frac{\hat{\boldsymbol{\Omega}}-\cos\alpha\hat{\boldsymbol{\mu}}}{\sin\alpha}. \\
\end{equation}
Similarly, the unit tangent vector directed towards $\hat{\boldsymbol{x}}$ is given by
\begin{equation}
\hat{\boldsymbol{t}}_x=\frac{\hat{\boldsymbol{x}}-\cos\gamma_x\hat{\boldsymbol{\mu}}}{\sin\gamma_x}.
\end{equation}
The normal unit vector,
\begin{equation}
\hat{\boldsymbol{n}}=\frac{\hat{\boldsymbol{Y}}\times\hat{\boldsymbol{t}}_x}{|\hat{\boldsymbol{Y}}\times\hat{\boldsymbol{t}}_x|},
\end{equation}
at the position of the magnetic pole ensures the correct sense of $\phi_x$, as was defined at the beginning of this section.

The radial wind vector, with components 
\begin{equation}
w_x=-|\boldsymbol{w}|\sin{\phi_{\rm asc}}
\end{equation}
and
\begin{equation}
w_y=|\boldsymbol{w}|\cos{\phi_{\rm asc}}, 
\end{equation}
deflects the beam in the direction of $\hat{\boldsymbol{r}}$, such that the deflected emission forms an angle with $\hat{\boldsymbol{x}}$, which is given by
\begin{equation}
\delta\phi_x=(\hat{\boldsymbol{k}}\cdot\hat{\boldsymbol{z}})\cos^{-1}\left[\frac{1+w_x}{\sqrt{(1+w_x)^2+w_y^2}}\right],
\end{equation}
where
\begin{equation}
\hat{\boldsymbol{k}}=\frac{\cos\phi_{\rm asc}}{|\cos\phi_{\rm asc}|}\hat{\boldsymbol{z}}
\end{equation}
defines the direction of the deflection as anti-clockwise --- when viewed from the celestial north --- when $\phi_{\rm asc}\in [0,\pi/2)\cup[3\pi/2,2\pi)$, and clockwise elsewhere.
To calculate the pulse profile after the effect of the wind, it is sufficient to only rotate $\hat{\boldsymbol{x}}$ by $-\delta\phi_x$ (i.e.~in the opposite direction to the deflection) and recalculate the pulse profile from Eq.~(\ref{eq:modelbeam}), using the coordinates of Eqs.~(\ref{eq:gammaphi}). 

Ultimately, for a given set of beam-shape parameters, wind magnitude, and epoch, our simulation produces a flux-density profile: $F(t;I_0,\Gamma_0,\Phi_0,\sigma_\Gamma,f,\zeta,w_0)$. We stress here that our model does not account for the rapid profile evolution that is observed during each of the BPs, and, to a certain degree, during the IP.  

\subsection{Simulation setup}
We combined Eqs.~(\ref{eq:modelbeam}) and (\ref{eq:gammaphi}) to calculate the pulse profile of pulsar B, as function of $\phi_{\rm asc}$ and MJD, under the influence of pulsar A's wind, of average magnitude $w_0$. The orientation of pulsar B's spin and magnetic axes relative to the orbital plane, and its rate and phase of precession have been estimated with high precision in the work of BKK+08\nocite{bkk+08}, using a model of the eclipses of pulsar A. We used those values as fixed parameters in our simulation. BKK+08 calculated the above values in a global self-consistent fit. Although we are aware that the GR prediction for $\Omega_{\rm SO}$ (i.e.~$\Omega^{\rm GR}_{\rm SO}$) yields a much more precise value, to remain consistent with the global solution across all model parameters we decided to use the fitted value for $\Omega_{\rm SO}$. Specifically, we fixed the values\footnote{We warn the reader that the adoption of the values of BKK+08\nocite{bkk+08} renders our results dependent on the modelling performed in that work.} for $\alpha=70\fdg 92$, $\delta=49\fdg 98$, $\phi_{\rm SO}({\rm MJD}\:53857)=308\fdg 79$ and $\Omega_{\rm SO}=4\fdg 77$\,yr$^{-1}$. We note that we define $\boldsymbol{L}\parallel\hat{\boldsymbol{z}}$, which is in the opposite direction to that in BKK+08\nocite{bkk+08}; therefore, in our model, $\phi_{\rm SO}$ increases with time, and, accordingly, $\phi_{\rm SO}$ has the opposite value at the reference epoch. 

The free parameters of our simulation were $I_{\{01,02\}}$, $\Gamma_{\{01,02\}}$, $\Phi_{\{01,02\}}$, $\sigma_{\Gamma\{1,2\}}$, $f_{\{1,2\}}$, $\zeta_{\{1,2\}}$, $w_0$. In order to account for unmodelled physical effects that amplify or suppress the beam intensity during the four different orbital-phase windows, BP1, BP2, IP, and WP, we assigned separate free parameters for $I_{\{01,02\}}$, to each of those orbital-phase intervals: for example, $I_{\{01,02\}}^{\rm BP1}$, $I_{\{01,02\}}^{\rm BP2}$, $I_{\{01,02\}}^{\rm IP}$, $I_{\{01,02\}}^{\rm WP}$. For a beam model and wind magnitude that exactly describes the profile evolution of pulsar B, these 19 parameters should be constant and independent of the orbital phase or epoch of observation. 

In LL14, the authors constrained the beam-shape parameters by fitting the simulated intensity variations of pulsar B, as a function of epoch, to the observed peak-intensity maps of pulsar B, published by PMK+10\nocite{pmk+10}. The best fit parameters were determined via the computation of the correlation coefficient between the simulated and the observed intensity maps, across a multi-dimensional parameter grid with iteratively increasing resolution around the highest coefficient. Apart from the precession rate, which was fixed to $\Omega_{\rm SO}=4\fdg 8$\,yr$^{-1}$, LL14 assumed no prior knowledge of the pulsar's orientation and also fitted for $\alpha$ and $\theta$ $(=180^\circ-\delta)$, which they determined to be $56^\circ$ and $122^\circ$, respectively. An additional parameter that LL14 determined was the height of the emission, $r_{\rm em}=3,750R_{\rm NS}$, where $R_{\rm NS}\sim 10$\,km is the NS radius. In this work, we did not consider an emission height or any associated phase shift due to the lag between the rotating frame of the NS and that of an inertial observer. Indeed, for the slow-rotating pulsar B, such a phase shift only becomes significant if the emission is generated close to the light cylinder (Perera et al.~2012\nocite{plg+12}). We would like to note that the high value of $r_{\rm em}$ in the work of LL14 resulted in perpetual visibility for pulsar B, which contradicts observations. To mitigate this problem, the authors forced a cut-off on the intensity by means of a Gaussian filter function, allowing emission only from near the null-charge surface, meaning regions for which $\boldsymbol{\Omega}\cdot\boldsymbol{B}\sim 0$, where $\boldsymbol{B}$ is the local magnetic field at the region of interest. It is interesting to note that the horseshoe beam of LL14, when adapted to the geometry of BKK+08\nocite{bkk+08}, leads to no emission during the period from the pulsar's discovery to its disappearance. The possibility of emission with the beam of LL14 still exists, of course, but it requires the beam to be rotated by $\approx 180^\circ$ about the magnetic axis. In that case, precession would evolve the pulse profile from double-peaked to single-peaked, which is the opposite of what is observed. Notably, in LL14, the direction of precession is reversed (see Section 8.3 in that paper), leading to the desired pulse-shape evolution.

\begin{figure*}
\centering
	\includegraphics[width=1\textwidth]{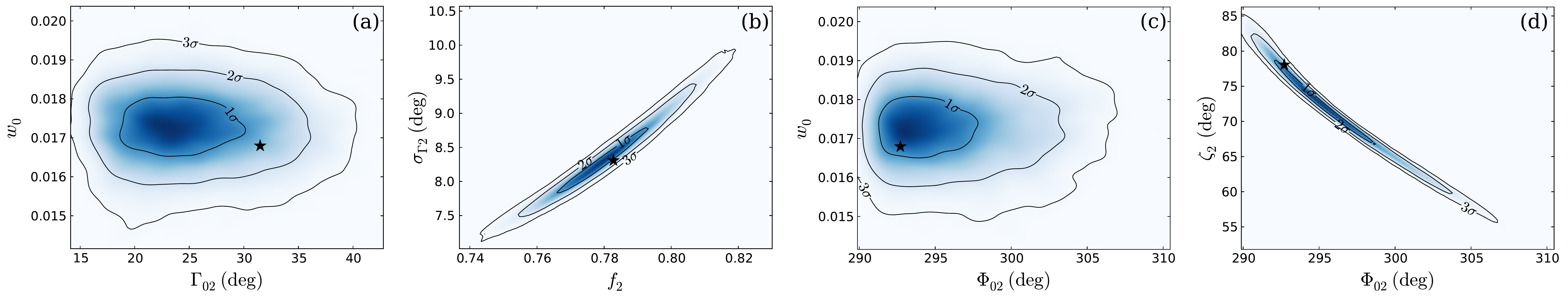}
    \caption{Probability density plots of four pairs of model parameters, corresponding to the second Gaussian beam component, from our analysis: (a) $\Gamma_{02}$--$w_0$, the beam colatitude of the peak intensity of the Gaussian component against the wind magnitude; (b) $f_2$--$\sigma_{\Gamma2}$, the flatness parameter against the half-beam width along the direction of increasing $\Gamma_{02}$; (c) $\Phi_{02}$--$w_0$, the beam longitude of the peak intensity of the Gaussian component against the wind magnitude; and (d) $\Phi_{02}$--$\zeta_2$, the beam longitude of the peak intensity of the Gaussian component against the angle of rotation about the direction of peak intensity. The star symbols indicate the values corresponding to the solution with the maximum global likelihood.}
    \label{fig:modcontours}
\end{figure*}

\subsection{Parameter estimation}
\label{subsec:modparest}
The significant profile-shape evolution of pulsar B across its orbit means that timing with a fixed profile template, as was done to derive the timing residuals shown in Fig.~\ref{fig:allresidsfits}, will introduce systematic offsets induced by the cross-correlation procedure: the latter tries to find the best phase offset that minimises the difference between the shape of the template and that of the data profile. If the shape of the observed profile changes significantly (see e.g.~Fig.~\ref{fig:profileEvoPhasc}), the phase offset does not only represent the relative shift between the template and the data but also a systematic offset that reflects the difference in pulse width, the number and the relative amplitude of the components, etc (see e.g.~Hassall et al.~2012\nocite{hsh+12}). As was mentioned in the introduction, to avoid this problem in timing pulsar B, KSM+06\nocite{ksm+06} generated a set of different templates for different epochs and orbital phases. Generated templates that are good approximations of the observed profile shapes would yield smaller systematic offsets between the model and data profiles, and the corresponding model parameters would be weighted higher. However, the templates of KSM+06\nocite{ksm+06} were not motivated by an underlying physical model but were constructed from the data. Our geometric model allows us to derive analytic templates for a given epoch and orbital phase, based on a coherent evolution of the viewing geometry, a parametrisation of the beam shape of pulsar B, and the impact of a radial wind. As is also discussed in Section~\ref{subsec:timimprov}, the systematic uncertainties of the cross-correlation procedure represent only a fraction of the `true' uncertainties, since they do not account for a potential plethora of alternative geometries that could provide an equally good or better fit to the observations.

We attempted to constrain our model's parameters, given the observed profiles of pulsar B, by mapping their probability distribution via the following likelihood function:
\begin{equation}
\label{eq:lnlike}
\ln\mathcal{L}\propto -\frac{1}{2}\sum_{k=1}^{N_{\rm epoch}}\sum_{j=1}^{N_{\rm phase}}\sum_{i=1}^{N_{\rm bin}}\left[\frac{\left[F_{ijk}^{\rm obs}-F_{ijk}^{\rm mod}(\boldsymbol{p}_J) \right]^2}{\sigma_{ijk}^2}+\ln\sigma_{ijk}^2\right],
\end{equation}
where $F_{ijk}^{\rm obs}$ and $F_{ijk}^{\rm mod}(\boldsymbol{p}_J)$ represent the flux density of the $i$-th pulse-phase bin of the observed and model profile, respectively, corresponding to orbital phase $\phi_{{\rm asc}(j)}$ and epoch $t_k$. The model profiles are calculated from Eqs.~(\ref{eq:modelbeam}) and (\ref{eq:gammaphi}), by calculating $I(\gamma_x,\phi_x)$ for every pulse-phase bin. The model parameters are represented here as a vector $\boldsymbol{p}_J=(I_{0\ell}^J,\Gamma_{0\ell},\Phi_{0\ell},\sigma_{\Gamma{\ell}},f_{\ell},\zeta_{\ell},w_0)$, where $J\in\{{\rm BP1,BP2,IP,WP}\}$, for each orbital-phase window, and $\ell\in\{1,2\}$, for each Gaussian component of the beam. Finally, $\sigma_{ijk}$ are the standard deviations of the observed profiles (calculated according to Eq.~\ref{eq:fluxsigma}).

Our aim was to determine the most likely parameters that describe the profile evolution both across the orbit, due to pulsar A's wind, and over the years, due to geodetic precession. For that reason, our model must be able to account for the profile changes that are observed across the entire span of our data. However, it is clear that the effect of the radial wind in our model can only produce a harmonic modulation of the profile shape (at the orbital period) via the periodic displacement of the beam as a function of orbital phase. As such, the fast evolution across each of the BPs and the IP cannot be accounted for, in our simple model. Therefore, we decided to only use the average profiles nearest to the centre of BP1, BP2, and the IP (see Fig.~\ref{fig:gaussdatprofs}). As was mentioned earlier, when the pulsar is detectable during the WP, the average profile was generated from all the available data in the corresponding orbital-phase window. Furthermore, as was explained in Section~\ref{sec:avgprofs}, to avoid biasing the parameter estimation due to off-pulse artefacts and other non-Gaussian noise unrelated to the pulsar's emission, we used the noiseless Gaussian templates of Fig.~\ref{fig:gaussdatprofs} as $F_{ijk}^{\rm obs}$ instead of the average data profiles. Our choice to use noiseless templates instead of the observed profiles leads to overestimated values of $\mathcal{L}$, because the magnitudes of the statistical and systematic noise, $\sigma_{ijk}$, which are considered in Eq.~(\ref{eq:lnlike}), are not reflected in the differences, $F_{ijk}^{\rm obs}-F_{ijk}^{\rm mod}(\boldsymbol{p}_J)$, leading to smaller values of the arguments being summed. In terms of the statistical noise, this decision has little impact, as $\sigma_{\rm stat}$ is roughly the same across our data set, with $\sigma_{\rm stat}=0.09_{-0.05}^{+0.17}$ mJy across the 41 profiles. More significant is the systematic noise, $\sigma_{\rm sys}$, which is both higher in the majority of cases and varies significantly between profiles: for example, by excluding all pulse phases outside the FWHM of the pulse profiles, and ignoring phase bins with $\sigma_{\rm sys}<\langle\sigma_{\rm stat}\rangle$, where $\langle\sigma_{\rm stat}\rangle$ is the median of the statistical noise, as shown above, we calculate $\sigma_{\rm sys}=0.63_{-0.33}^{+0.68}$ mJy. However, the contribution of the systematic noise could only be properly accounted for in Eq.~(\ref{eq:lnlike}), if all of the 1,467 original data profiles that were averaged to produce the final set of 41 averaged profiles were used in the calculation of $\mathcal{L}$. Unfortunately, this would have been prohibitively expensive in terms of computation.         

The sampling of the parameter space was done using POLYCHORD (Handley et al.~2015\nocite{hhl15}), which is a nested sampling algorithm that is tailored for problems with high dimensionality. For a given set of model-parameter values, the code calculates the likelihood of Eq.~(\ref{eq:lnlike}); it then converges towards the most likely region of the parameter space by means of sequentially sampling that space with 500 of so-called live points. These are updated sequentially towards increasingly constraining regions around the global maximum likelihood.

In addition to the 19 model parameters, we also constrained the global constant phase offset ($\delta\phi_{\rm s0}$) between the model profiles and the corresponding observed profiles: this was necessary because the reference phase of the observed profiles, which is typically defined by the start of the observations and the timing ephemeris, does not necessarily match the reference phase of the model. Moreover, we used a more conservative approach and introduced separate $\delta\phi_{\rm s0}$ parameters for the 685, 820, and 1400\,MHz profiles: $\delta\phi_{\rm s0}^{\rm 685MHz}$, $\delta\phi_{\rm s0}^{\rm 820MHz}$, and $\delta\phi_{\rm s0}^{\rm 1400MHz}$. This decision was motivated by the frequency-dependent phase offsets that can arise from differing instrumentation and/or observatories.

Finally, to account for the unknown amount of covariance between $\sigma_{\rm sys}$ and $\sigma_{\rm stat}$, for each of the 41 average profiles we introduced an additional error coefficient, $q_{kj}$, where $\sigma^{\prime}_{{\rm sys}(ijk)}=q_{kj}\sigma_{{\rm sys}(ijk)}$. In total, the number of free parameters that were simultaneously constrained given the 41 observed profiles was $N_{\rm par}=N_{\rm par}^{\rm beam}+N_{\rm par}^{\rm wind}+N_{\rm par}^{\rm offset}+N_{\rm par}^{\rm errc}=18+1+3+41=63$. For all parameters, we chose uniform priors: the corresponding ranges are shown in the last column of Table~\ref{tab:bestfitparamsNDK18}. For $q_{kj}$, the range of the priors was [0,10].

\begin{table*}
\centering
\caption{The mean and most likely values from our analysis, for the beam intensity, beam-shape, and average wind magnitude. The values in parentheses correspond to the uncertainty on the last significant digit.}
\label{tab:bestfitparamsNDK18}

\centering
\vspace{0pt}
\begin{tabular}{lrrcc} 
\hline
Parameter                    & Mean value  & Most likely value & Units   & Prior range \\
\hline
$\log I_{01}^{\rm BP1}$      & $1.18(4)$   & $1.16$   & mJy & $[-3,3]$      \\
$\log I_{02}^{\rm BP1}$      & $-0.119(8)$  & $-0.106$  & mJy & $[-3,3]$      \\
$\log I_{01}^{\rm BP2}$      & $0.36_{-0.04}^{+0.05}$   & $0.36$   & mJy & $[-3,3]$      \\
$\log I_{02}^{\rm BP2}$      & $0.365(4)$   & $0.365$   & mJy & $[-3,3]$      \\
$\log I_{01}^{\rm IP}$       & $-2.3_{-0.5}^{+0.6}$  & $-1.9$  & mJy & $[-3,3]$      \\
$\log I_{02}^{\rm IP}$       & $0.14(1)$   & $0.13$   & mJy & $[-3,3]$      \\
$\log I_{01}^{\rm WP}$       & $-0.17_{-0.07}^{+0.06}$  & $-0.14$  & mJy & $[-3,3]$      \\
$\log I_{02}^{\rm WP}$       & $-0.15(2)$  & $-0.15$  & mJy & $[-3,3]$      \\
$\Gamma_{01}$                & $25_{-5}^{+7}$  & $32$  & deg & $[0,180]$     \\
$\Gamma_{02}$                & $25_{-5}^{+6}$  & $31$  & deg & $[0,180]$     \\
$\Phi_{01}$                  & $256(5)$ & $263$ & deg & $[0,360]$    \\
$\Phi_{02}$                  & $295_{-3}^{+4}$ & $293$ & deg & $[0,360]$    \\
$\sigma_{\Gamma1}$           & $9.7(4)$ & $9.6$ & deg & $[0.5,90]$    \\
$\sigma_{\Gamma2}$           & $8.3_{-0.4}^{+0.5}$ & $8.3$   & deg & $[0.5,90]$    \\
$f_1$                        & $0.848_{-0.007}^{+0.006}$   & $0.846$   & --  & $[0,1]$   \\
$f_2$                        & $0.78(1)$   & $0.78$   & --  & $[0,1]$   \\
$\zeta_1$                    & $92(4)$  & $88$  & deg & $[0,180]$    \\
$\zeta_2$                    & $72_{-7}^{+6}$  & $78$  & deg & $[0,180]$    \\
$w_0$                        & $0.017(1)$   & $0.017$   & --  & $[0,0.5]$      \\
\hline
\end{tabular}
\end{table*}

\subsection{Results}
\label{subsec:results}
\subsubsection{Parameter distributions}
In Fig.~\ref{fig:modcontours}, we show a small subset of the PDFs that were generated by our POLYCHORD run, focusing on the shape and location parameters of the brightest Gaussian component of the BP1 beam, $\Gamma_{02}$, $\Phi_{02}$, $f_2$, and $\zeta_2$, as well as the wind parameter, $w_0$. The complete set of PDFs can be found in Figs.~\ref{fig:contab01} and~\ref{fig:contab02}. The values corresponding to the maximum global likelihood are indicated with a star symbol, in each of the panels. It is clear from these plots that certain parameter combinations, like $\zeta_2$ and $\Phi_{02}$, are highly covariant, whereas others, like $w_0$ and $\Phi_{02}$, are close to orthogonal. It can also be seen, most clearly in Fig.~\ref{fig:modcontours}a, that for certain distributions the most likely values lie outside the $1\sigma$ contour. The reason for this is that the non-Gaussianity of some of the distributions --- most notably, the one shown in Fig.~\ref{fig:modcontours}d, for example --- biases the joint likelihood of certain parameter combinations. Nevertheless, these values are still the most likely, when the likelihood over the entire multi-dimensional space is considered. The mean values, $1\sigma$ uncertainties, and the values corresponding to the solution with the highest global likelihood are shown in Table~\ref{tab:bestfitparamsNDK18}. It is worth noting that the constant phase offset, $\delta\phi_{\rm s0}$, was found to be identical for all three frequencies and equal to $\delta\phi_{\rm s0}/2\pi=0.428_{-0.020}^{+0.016}$. Apart from the three profiles corresponding to the IP at MJD\:53600--53700 and MJD\:53700--53800 and the BP1 at MJD\:54300--54400, which as mentioned earlier have $\sigma_{\rm sys}=0$ and for which $q_{kj}$ was unconstrained, the majority of the rest of the profiles have $q_{kj}<0.5$.

\subsubsection{Beam structure}
\begin{figure*}
\centering
	\includegraphics[height=0.65\textheight]{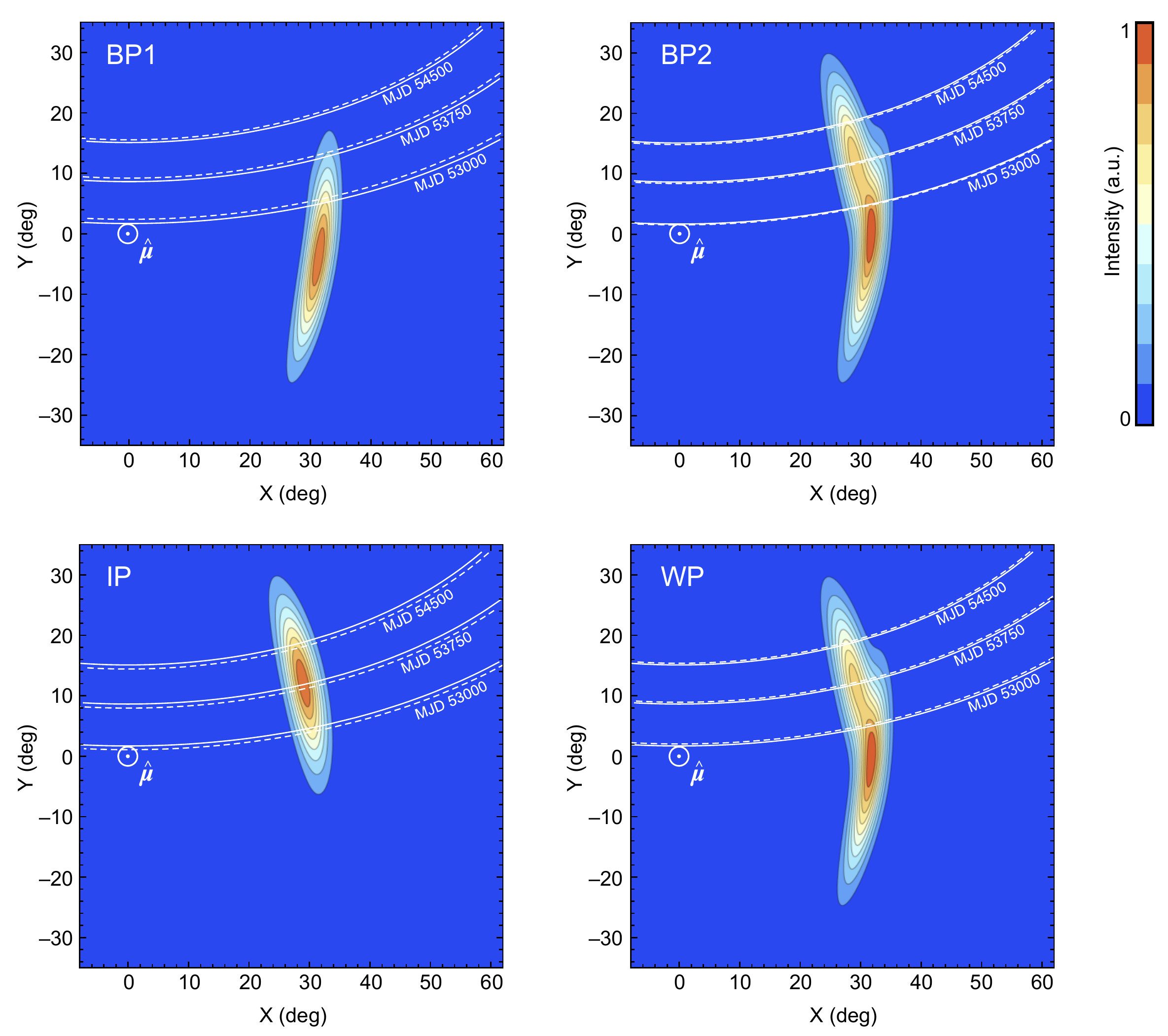}
    \caption{Two-dimensional projection on plane of the sky of most likely emission beams of pulsar B, corresponding to BP1, BP2, the IP, and the WP, as they were determined from our analysis. The Cartesian coordinates shown ($X,Y$) are defined in Section~\ref{subsec:simemgeo}. In each panel, the intensity has been colour-coded and normalised to a maximum value of 1 (see colour bar) to provide a clear representation of the beam shape; the relative intensity between the orbital phases is not shown. The solid white lines indicate the traces of the unperturbed LOS for MJD\:53000, MJD\:53750, and MJD\:54500; the dashed white lines show the same traces after applying the deflection caused by the most likely magnitude of the wind (i.e.~$w_0=0.017$). The circled-dot symbol indicates the position of the magnetic axis relative to the beam. A 3D rendering of these beams is provided in Fig.~\ref{fig:3dbeams}.}
    \label{fig:2dbeams}
\end{figure*}

We used the values corresponding to the most likely solution for the beam's parameters, to provide a visual representation of pulsar B's emission beam. Since our model parametrised the intensities of each component individually, for each orbital-phase window, the beam shape due to the different $I_{01}^J$ and $I_{02}^J$ values, for each $J\in\{{\rm BP1,BP2,IP,WP}\}$, also differs. Using the most likely values for the beam intensities, in Fig.~\ref{fig:2dbeams} we show the 2D projections of the beam of BP1, BP2, the IP, and the WP, where for clarity we have normalised the intensity to 1. Due to this normalisation, we stress that these plots show only the relative intensity between the Gaussian beam components in each orbital-phase window and should not be used to compare intensities between orbital phases. The trace of the observer's LOS at MJD\:53500, MJD\:53750, and MJD\:54500, for an unperturbed beam ($w_0=0$) and for $w_0\approx 0.017$, is shown with solid and dashed white lines, respectively. The beam intensity maps reveal that in BP1 and the IP, the most likely beam consists of a dominant primary component that, according to the values of $I_{0\ell}$, is roughly 20 and 100 times brighter, respectively, than the secondary; the absolute intensity of the dominant component in BP1 is roughly ten times brighter than in the IP. In contrast, in BP2 and the WP, the intensity of the primary and secondary components is roughly equal. Except for the IP, where the secondary component is practically invisible, in all other orbital-phase windows the two components are offset with respect to each other by roughly $15^\circ$ (combine $\Gamma_{0\ell}$ and $\Phi_{0\ell}$ with Eq.~(\ref{eq:dell})); also, the two components form an angle of $\approx 20^\circ$ (combine parameters $\Phi_{0\ell}$ and $\zeta_\ell$). Interestingly, although both the BP1 and IP beams comprise a clearly dominant component, for BP1 this corresponds to $\ell=1$, while for the IP it corresponds to $\ell=2$. In both cases, the secondary component is practically eclipsed by the brightness of the primary. For illustration purposes, Fig.~\ref{fig:3dbeams} shows a 3D representation of the most likely beam during BP1, BP2, the IP, and the WP, rendered on a sphere, where the viewing geometry corresponds to MJD\:53000, MJD\:53750, and MJD\:54500. 
 
\begin{sidewaysfigure*}
\centering
	\includegraphics[width=1\textwidth]{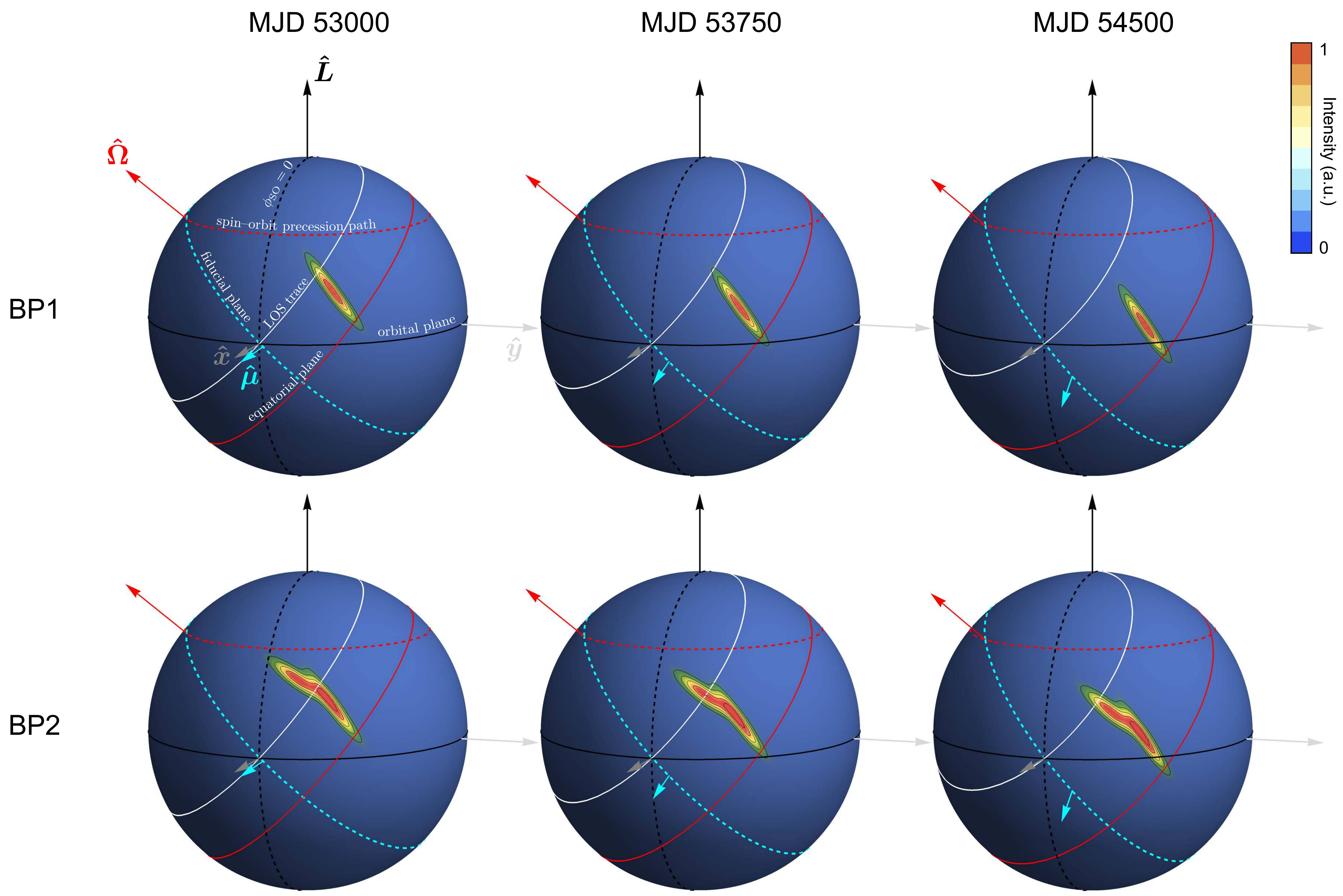}
    \caption{(a) 3D representation of most likely emission beams during BP1, BP2, the IP, and the WP, derived from this work, shown at the minimum approach of the magnetic axis ($\hat{\boldsymbol{\mu}}$) to the LOS ($\hat{\boldsymbol{x}}$). The figure shows three viewing configurations, corresponding to three geodetic-precession phases, at MJD\:53000, MJD\:53750, and MJD\:54500. The trace of the LOS is shown with a white line. The colour scale represents the beam intensity in arbitrary units, normalised between 0 (blue) and 1(red), in steps of 0.1.}
    \label{fig:3dbeams}
\end{sidewaysfigure*}

\addtocounter{figure}{-1}

\begin{sidewaysfigure*}
\centering
	\includegraphics[width=1\textwidth]{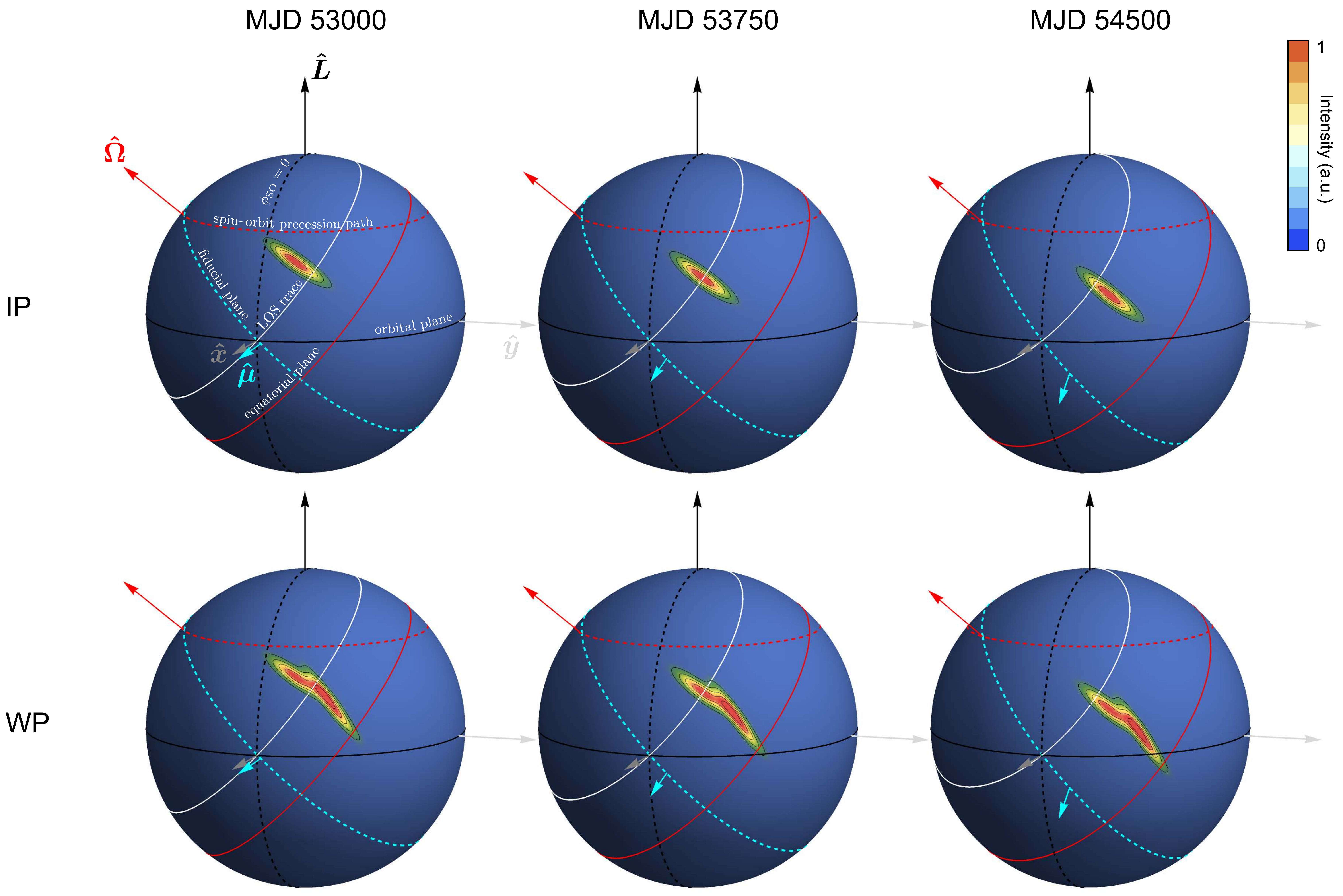}
    \caption{Continued.}
    \label{fig:3dbeams}
\end{sidewaysfigure*}

\subsubsection{Profile evolution}
An important aspect of any beam model is how closely it reproduces the observed flux-density profiles as a function of orbital and precessional phase. Using again the most likely beam and wind parameters, in Fig.~\ref{fig:moddatprofs} we show the model profiles (red lines) overlaid with the observed profiles (black lines). Similarly to Fig.~\ref{fig:datprofs}, the $1\sigma$ confidence interval of the systematic uncertainties, multiplied by the most likely error coefficient ($q_{kj}$), such as $\sigma^{\prime}_{{\rm sys}(ijk)}$, is shown for each profile, with grey lines. 

Qualitatively, the model tracks the precessional evolution of the profile well, beginning with mainly single-peaked profiles at the earliest epochs ($<$ MJD\:53500), and evolving to almost exclusively double-peaked profiles after MJD\:53900. Equally successful is the model's ability to track the phase drift of the profile as a function of orbital phase, although there are a few cases, almost exclusively in the IP (e.g.~MJD\:53400--53500 \& MJD\:53800--53900), where the peak of the model profile is shifted towards earlier phases, by a few phase bins, relative to the observed profile. The reason behind this discrepancy could be related to the following characteristics of the IP. Firstly, its orbital-phase extent is significantly larger than that of the BP1 and BP2 (see Table~\ref{tab:bplocations}). As a result, it is less certain that a single, average profile at the centroid of the IP would reflect the magnitude of the harmonic profile evolution at that orbital phase. Furthermore, the profile evolution during the IP, during the epochs where the largest discrepancies between the model and the data are seen, is more complex than that of BP1 and BP2. Indeed, the timing residuals during the IP, at MJD\:53400--53500 and MJD\:53500--53600, exhibit a more irregular drift pattern than the monotonic drifts observed during the BPs (see Fig.~\ref{fig:allresidsfits}). Moreover, at $\phi_{\rm asc}/2\pi\approx 0.02$, a discontinuity can be seen, accompanied with a decrease in S/N and the number of detections. These timing irregularities imply a more stochastic profile evolution across the IP compared to the other orbital-phase windows, a conclusion that is partly supported by the profiles shown in Fig.~\ref{fig:profileEvoPhasc}. Lastly, as can be seen in Fig.~\ref{fig:bp_fits}, the S/N during the IP is significantly lower than that of BP1 and BP2. As a result, fewer profiles were averaged together to create the average, centroid profiles for that orbital window, lending to less stable profiles that are possibly less representative of the average profile shape at that orbital phase. Despite those shortcomings, our work did not try to further characterise the complex structure of the IP and was restricted to considering it as a single interval, due to the limited statistics. Overall, the reduced chi-squared between the model and observed profiles was $\chi^2_{\rm red}=1.34$ (calculated from 20,992 data points and a model with 63 free parameters, i.e.~using 20,929 degrees of freedom). 

\subsubsection{Flux evolution}
As was mentioned earlier, our analysis has determined the most likely amplitudes of the Gaussian components, for each orbital-phase window. However, as well as changing the profile shape, geodetic precession also causes a change in the profiles' amplitudes, as our LOS traces different parts of the beam. It is therefore interesting to compare the peak flux-density evolution of the model profile to that observed. In Fig.~\ref{fig:peakfluxes}, together with the measured flux densities of the observed profiles (data points) we have plotted the peak flux-density of the model profiles with dashed lines, for each orbital-phase window. The roughly constant flux density of the profiles of BP2, the IP, and the WP is well reproduced by the model. More interestingly, the fading of the BP1 profiles with increasing MJD is, overall, tracked by the model. However, at around MJD\:53500 and after MJD\:53800 the model appears to systematically underestimate the BP1 flux density by as much as 2--3$\sigma$. It must be stressed, however, that this comparison does not consider the complex uncertainty of the model flux density, arising from the combination of the uncertainties of the model parameters (see Table~\ref{tab:bestfitparamsNDK18}). To first order, we can estimate that uncertainty, based on the $1\sigma$ uncertainty of $I_{01}^{\rm BP1}$, as being roughly 10\%. Although this alone does not eliminate the inconsistencies between the observed and model flux density, in particular around MJD\:53100, 53400, and 53800, it reduces them to within $2\sigma$.

\subsubsection{Sense of rotation}
Another important property that was determined by our parameter estimation is the sense of pulsar B's rotation. Such a determination in binary systems could provide useful information for studies of binary-star evolution and formation of such double-NS systems. Very recently, the sense of pulsar A's rotation was determined by Pol et al.~(2018)\nocite{pmk+18}, using the drifting features in the sub-pulse structure of pulsar B, caused by the electromagnetic radiation of pulsar A. The authors concluded that pulsar A spins in a prograde fashion relative to its orbital motion. 

As was mentioned in the introduction, the analysis of Breton (2009)\nocite{bre09} constrained the angle, $\delta$, between the orbital angular momentum and the spin axis of pulsar B. However, the solution was not unique and there was a degeneracy between $\delta$ and $180^\circ-\delta$, hence the sense of rotation was not uniquely determined. In our analysis, we tested both the retrograde and prograde model, while constraining the model parameters. In our model, reversing the spin of the pulsar affects the direction of the phase delay caused by the wind vector at a given orbital phase. For example, as is shown in Fig.~\ref{fig:toymoddelay}, at $\phi_{\rm asc}=0$ the radial wind deflects the pulsar beam counter to the pulsar's retrograde rotation, causing a phase delay in the pulse arrival time. This is of course what we observe in the residuals if a fixed template is used. However, in reality, the wind deflects the entire emission beam relative to the LOS, causing not only a translation but also a profile evolution, as our LOS traces different parts of the beam. Hence, depending on the complexity of the emission region, both retrograde and prograde configurations could result in the observed profile evolution and must be therefore tested.

A direct comparison could be made between the log-evidence values, $\log Z$, corresponding to the retrograde and prograde configurations, where $Z$ is the integral of the likelihood over the entire parameter space. For the retrograde case, POLYCHORD reported $\log Z_{\circlearrowleft}=19345.0(8)$, while for the prograde case the value was $\log Z_{\circlearrowright}=19923.7(8)$. Converting the log-evidence values to a probability of the retrograde solution compared to the prograde, we obtain $P=R/(1+R)\approx 10^{-578}\approx 0$, where $\log R=\log Z_{\circlearrowleft}-\log Z_{\circlearrowright}=-578$ is the logarithm of the Bayes factor. In addition, the most likely profiles in the retrograde case were exclusively single-component Gaussians, across the entire MJD range, and did not reproduce the observed profile evolution due to geodetic precession (see Fig.~\ref{fig:moddatprofs}). In particular, the total reduced chi-squared between the observed profiles and the retrograde model was $\chi^2_{\rm red}=2.16$ (cf.~$\chi^2_{\rm red}=1.34$, for the prograde case). Based on those facts, we have concluded that the most likely sense of rotation for pulsar B, relative to its orbital motion and relative to pulsar A, is prograde.

\section{Discussion}
\label{sec:discussion}

\subsection{Timing improvements}
\label{subsec:timimprov}
The model profiles of Fig.~\ref{fig:moddatprofs} can be used as analytic templates for timing pulsar B. We have generated TOAs for each of the original 4,115 observed profiles, using the model profiles corresponding to the orbital-phase and MJD range as timing templates. Here, we recall that the model templates we used for timing pulsar B represent only a small fraction of the pulsar's orbit and therefore we may not expect that they track the profile changes across each orbital-phase window. Nevertheless, we are interested in the model's performance with regards to the harmonic delays that are observed across the entire orbit. Ideally, a beam model must be able to match the observed profile shapes as well as the Gaussian templates of Fig.~\ref{fig:gaussdatprofs}. A visual representation of such an ideal case can be seen in Fig.~\ref{fig:paas_resids}, where we show the timing residuals derived from timing pulsar B with those Gaussian templates: compared to the fixed template, the total RMS of the residuals improves from ${\rm RMS}=9.33$\,ms to ${\rm RMS}=4$\,ms. This can be considered as the near-optimal case of correcting for the harmonic delays. However, such a heuristic method is not informative as to the physical processes behind those delays: our model attempts to interpret those delays in terms of the geometry of pulsar B's emission beam and how it is affected by pulsar A's wind pressure. 

The performance of our model can also be quantified in terms of the improvement on the RMS of the timing residuals, relative to the case where a fixed template was used to time pulsar B (see Fig.~\ref{fig:allresidsfits}). In Fig.~\ref{fig:ml_resids}, we show the residuals per MJD bin, generated from the original observed profiles and the most likely model profiles used as templates. The corresponding RMS was 6.6\,ms, this value being approximately mid-way between the na\"ive case of a fixed template and the ideal case of heuristic Gaussian templates. 

An interesting comparison can be made between the timing residuals using the fixed WP template of Fig.~\ref{fig:wp_template}, the Gaussian templates of Fig.~\ref{fig:gaussdatprofs}, and those using the model templates, for the MJD interval, MJD\:53900--54200. It is clear that the residuals corresponding to the BPs, based on the first two sets of templates, are clustered in two distinct sets separated by $\approx 0.01$ in pulse phase. This is roughly the separation of the two distinct peaks that the BP profiles develop during that interval. Since the fixed WP template cannot account for the complexity of the observed profile during those epochs, the cross-correlation procedure stochastically determines the phase delay based on the brightest of the two peaks, leading to two clusters. The heuristic Gaussian templates track the complexity of the BP profiles much better, leading to a less pronounced clustering (although some still remains, possibly owing to a small subset of single-peaked profiles, for which the double-peaked template is a bad fit). Finally, the model templates seem to track the profile shape well during BP1, resulting in practically no clustering of the residuals, in contrast to the residuals during BP2, where as expected from the fact that our model fails to reproduce the observed double-peaked profiles the clustering is more pronounced.

An important aspect of our efforts to improve the timing of pulsar B is the corresponding improvement on the precision of the orbital parameters that play a central role in tests of GR and alternative theories of gravity. One of these parameters is the ratio of the intrinsic semi-major axes, $R=x^{\rm int}_{\rm B}/x^{\rm int}_{\rm A}$, which is qualitatively different from the rest, as it provides theory-independent information, which can be used to constrain a large family of theories of gravity in a generic way. Although the precision of most PK parameters used in those tests improves significantly with time, solely by the continuing timing of pulsar A (Kramer et al.~{\em in preparation}), $R$ does not: its precision is dominated by the uncertainty of the observed semi-major axis of pulsar B, $x^{\rm obs}_{\rm B}=(1+\epsilon_{\rm aberr})x^{\rm int}_{\rm B}$, where $\epsilon_{\rm aberr}$ is the fractional change of the semi-major axis caused by beam aberration (Damour \& Taylor 1992\nocite{dt92}).

Crucially, as has been mentioned earlier, the timing of pulsar B presented here depends on the precise value of $x_{\rm B}$, which was calculated from the orbit of pulsar A, assuming GR. Therefore, the residuals between the GR timing model and the TOAs, whether the latter were derived using a fixed template or the templates of our model, are dependent on that assumption. It follows, then, that those residuals cannot directly be used in tests of GR and other theories of gravity. However, we would like to reiterate that our results are valid for a range of fully conservative gravity theories.

It should also be emphasised that there are a number of sources of systematic uncertainty, which we have not accounted for. Even in an ideal case where our model had been able to perfectly predict the observed profile shape as a function of orbital phase and MJD, it would still contain systematic uncertainties related to the assumption of theory- and model-dependent parameters used in this work, such as $x_{\rm B}$, $\Omega_{\rm SO}$, as well as all the parameters of pulsar B's geometry that were adopted from BKK+08\nocite{bkk+08}. In that hypothetical case, the RMS of the residuals would merely represent the statistical uncertainties between the observed data and the noiseless templates; the mean values of the timing parameters that one would derive from fitting a timing model to the TOAs would be consistent with those originally calculated from GR. Any additional red noise in the residuals, such as that seen using our imperfect model, would then be indication of the inability of our model to predict the exact phase and shape of the observed profiles. In this work, we have tried to account for part of that red noise using an analytical model based on geometry. As such, we cannot exclude the possibility that there are a number of alternative models that would perform equally well or better. This uncertainty mainly arises from the nature of such type of mathematical modelling, that is, one not based on physical principles derived from a deep knowledge of pulsar magnetospheric processes and how plasma winds interact with pulsar emission. Therefore, this systematic uncertainty associated with our model will always remain, even if our parametrisation is successfully applied in combination with future, independent timing of pulsar B to provide better constraints on $R$.

Apart from all the aforementioned shortcomings, it must also be stressed that the timing improvements that resulted from the application of our model do not automatically reflect corresponding improvements on the precision of the orbital parameters, as they do not account for the covariances between the parameters of the timing model and those of our model. To account for those covariances, a global fit including all parameters is necessary, which is beyond the scope of this paper. However, in the idealised case where there are no such covariances, we can estimate the maximal improvement expected on the precision of $x^{\rm obs}_{\rm B}$ relative to its previous estimate by KSM+06\nocite{ksm+06} using all the TOAs from our analysis in TEMPO2. More specifically, we performed a fit for $x^{\rm obs}_{\rm B}$, while keeping the rest of the parameters fixed. The precision from the analysis of KSM+06\nocite{ksm+06}, based on a smaller (507 TOAs) and shorter ($\approx 2.5 $\, y) data set than ours, was $\sigma_{x{\rm B}}=1.6$\,light-ms; using our wind model and beam, we were able to reduce this value by a factor of $\approx 2.6$, improving it to $\sigma_{x{\rm B}}=0.61$\,light-ms. This improvement also points towards an even more interesting prospect, the measurement of $\epsilon_{\rm aberr}$ for pulsar B for the first time. At the moment, however, the above precision falls slightly short of its predicted value, which is $\epsilon_{\rm aberr}x^{\rm int}_{\rm B}\approx 0.37$\,light-ms, $\approx 1.5\times$ larger by comparison. 

\begin{figure}
	\includegraphics[width=\columnwidth]{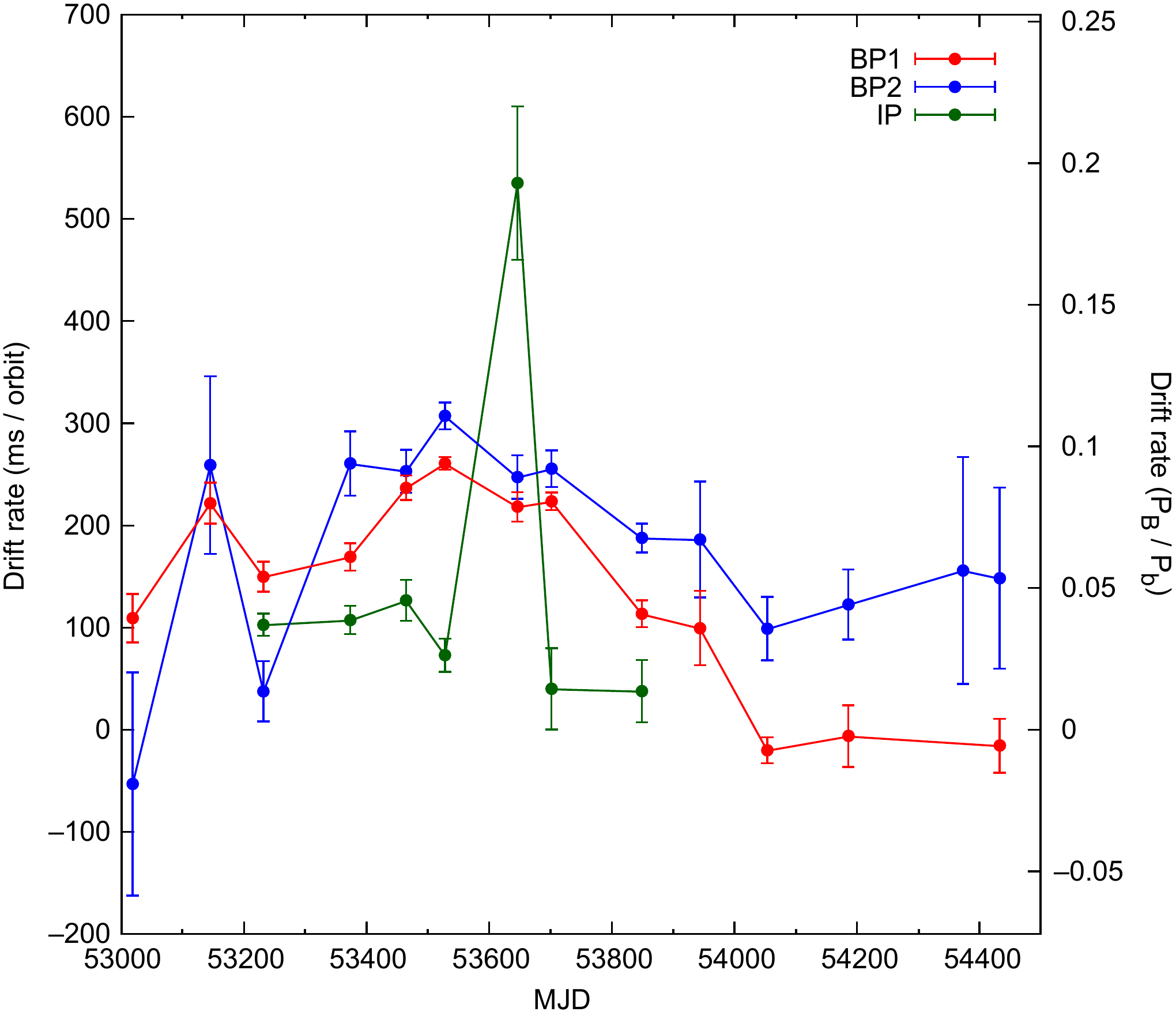}
    \caption{Average drift rate of timing residuals across BP1 (red), BP2 (blue), and the IP (green), as a function of MJD. The drift rates have been estimated from the linear fits shown in Fig.~\ref{fig:paas_resids} and are shown both in units of ms per orbit and as a fraction of pulsar B's spin phase per orbit.}
    \label{fig:lindriftrates}
\end{figure}

Finally, we briefly investigated the timing improvement that a future model could bring, if it is also able to account for the quasi-linear drifts observed across BP1, BP2, and the IP. After calculating the residuals based on the Gaussian templates, we estimated the average drift across each of BP1, BP2, and the IP, assuming a linear regression, via the fits shown in Fig.~\ref{fig:paas_resids}. The evolution of the slopes of the average drifts as a function of observation epoch is shown for each orbital-phase window in Fig.~\ref{fig:lindriftrates}. We also tested a quadratic model for the drifts, but found that on the whole it was not significantly better than the linear model, given the uncertainties of our timing residuals. After correcting for the linear drifts, using the best fit functions, we once again estimated the RMS of the timing residuals. The complete heuristic correction, based on the Gaussian templates and the best-fit linear drifts, resulted in ${\rm RMS}=2.5$\,ms. The timing residuals after those corrections can be seen in Fig.~\ref{fig:gausslinear_resids}.

\subsection{The migration of the bright phases}
\label{subsec:bpmigration}
An effect that is likely caused by geodetic precession is the migration of the locations of BP1 and BP2 as a function of time. However, a simple linear fit to the centroids of BP1 and BP2 shows that the average migration rate is significantly less than the precession rate (see Fig.~\ref{fig:bp_evol}). We note here that the pulse-profile evolution due to geodetic precession can only alter the phase of emission within a pulse period, that is, cause timing delays of $\lesssim P_{\rm B} \sim 10^{-4}P_{\rm b}$, and it cannot explain the shifts of the BPs by $\sim 10P_{\rm B}$\,yr$^{-1}$. Although our data span is short compared to the period of precession, it can be seen that the migration of the centroids of BP1 and BP2 deviates from being linear with time (see $\chi^2_{\rm red}$ values in Fig.~\ref{fig:bp_evol}) and that this deviation appears to progress in opposite directions. Indeed, towards the later epochs, the rate of change of the phase of BP1 increases, whereas at the same time that of BP2 decreases. To confirm this apparent complementarity between the migration rates, we plotted the average orbital-phase shift of BP1 and BP2 as a function of epoch: this was calculated using Table~\ref{tab:bplocations} as $\tfrac{1}{2}(\phi_{\rm asc}^{\rm BP1}+\phi_{\rm asc}^{\rm BP2})-\phi_{\rm asc(0)}$, where $\phi_{\rm asc(0)}=\tfrac{1}{2}[\phi_{\rm asc}^{\rm BP1}({\rm MJD\:53018.1})+\phi_{\rm asc}^{\rm BP2}({\rm MJD\:53018.1})]$ is the numerical average of the locations of BP1 and BP2, at the earliest epoch. As can be seen in the bottom plot of Fig.~\ref{fig:bp_evol}, the average shift can be fitted fairly well ($\chi^2_{\rm red}=2.6$) with a linear regression with a slope of $2\fdg4(1)$\,yr$^{-1}$, which confirms that the rates of change of the locations of BP1 and BP2 are complementary across the investigated range of epochs.

The above findings motivated us to further explore the connection between the migration rates of the BPs and the rate of geodetic precession. We would like to emphasise that the simple model employed here is entirely independent of the detailed 3D modelling of Section~\ref{sec:simwindeff}.

\begin{figure}
\centering
	\includegraphics[width=1\columnwidth]{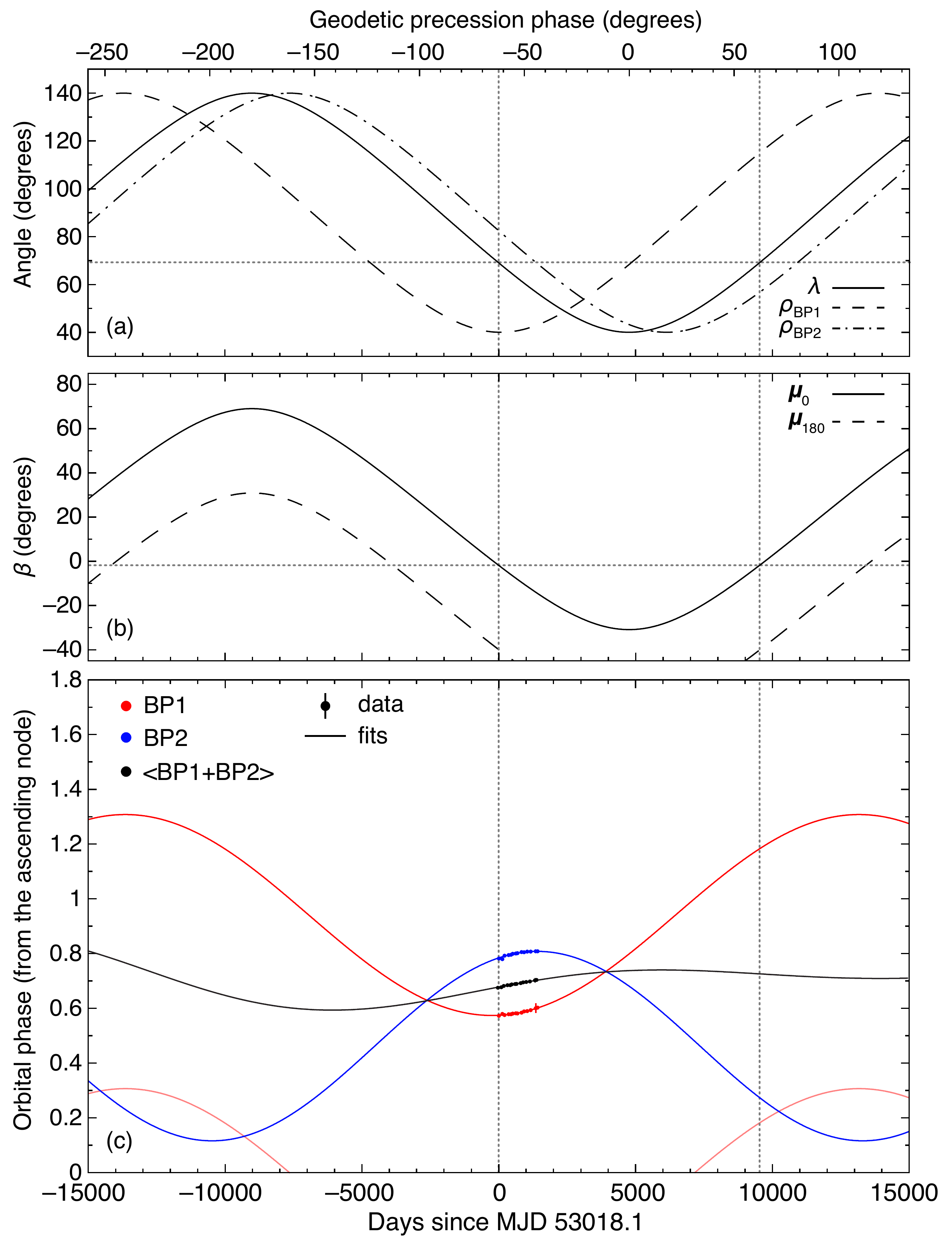}
    \caption{(a) Time evolution of $\lambda$, which is the angle between the spin axis of pulsar B and the LOS, and $\rho$, which is the angle between the spin axis of pulsar B and the radial wind direction, calculated at the centroids of BP1 and BP2, ca.~MJD\:53600. Time, along the bottom horizontal axis, is expressed as days since the epoch of the earliest MJD bin (MJD\:53018.1). Along the top horizontal axis, we show the corresponding phase of geodetic precession during the interval considered. (b) Time evolution of the impact angle, $\beta$, for the magnetic pole that was visible until ca.~2008 (solid curve), and the opposite magnetic pole (dashed curve). (c) Best fit sinusoidal functions of the location of BP1 (red curve and data points) and BP2 (blue curve and data points) as a function of time. The light red curve is the alias of the best fit function for BP1, modulo the orbital period. The solid black curve and data points show the average of the functions and data, respectively. Finally, the vertical dotted grey lines indicate the earliest observation epoch and the subsequent epoch at which the impact angle has the same value as that at the earliest epoch; in (a) and (b), the horizontal dotted grey lines indicate the values of $\lambda$ and $\beta$ at those two epochs.}
    \label{fig:bpsinefits}
\end{figure}

The evolution of $\phi_{\rm SO}$ is linear with time (to a high level of precision). However, as we saw above, the migration rates of BP1 and BP2 are not constant across the investigated range of MJDs. It is therefore possible that the locations of BP1 and BP2, hereafter $\phi_{\rm asc}^{\rm BP1}$ and $\phi_{\rm asc}^{\rm BP2}$, are not strictly proportional to $\phi_{\rm SO}$, but vary according to other defining angles of the system's geometry. Two such angles that can possibly influence the position of the BPs are the angle between pulsar A's wind and pulsar B's precessing spin axis, $\hat{\boldsymbol{r}}\wedge\boldsymbol{\Omega}=\rho$, and the angle between pulsar B's spin axis and the direction to the observer, $\hat{\boldsymbol{x}}\wedge\boldsymbol{\Omega}=\lambda$. These angles are related to the phase of geodetic precession, as follows:
\begin{align}
\lambda&=\cos^{-1}(\sin\delta\cos\phi_{\rm SO}) \\
\rho&=\cos^{-1}\left[\sin\delta\sin(\phi_{\rm SO}-\phi_{\rm asc})\right] 
\end{align}
As can be seen in Fig.~\ref{fig:bpsinefits}a, where we plot $\lambda$ and $\rho$ for $\phi_{\rm asc}/2\pi=0.58$ (BP1) and $\phi_{\rm asc}/2\pi=0.8$ (BP2), for a wide range of epochs, both angles are harmonic functions of time. It is thus reasonable to assume that if $\phi_{\rm asc}^{\rm BP1}$ and $\phi_{\rm asc}^{\rm BP2}$ are functions of $\lambda$ and $\rho$, they will also be harmonic functions of time. Based on that assumption, we have parametrised the orbital phases of BP1 and BP2 as
\begin{equation}
\begin{aligned}
\label{eq:sineeqs}
\phi_{\rm asc}^{\rm BP1}&=\phi_{{\rm asc}(0)}^{\rm BP1}+\delta\phi_{\rm asc}^{\rm BP1}\sin\left[\frac{2\pi(t-t_{01})}{P_{\rm BP1}}\right], \\
\phi_{\rm asc}^{\rm BP2}&=\phi_{{\rm asc}(0)}^{\rm BP2}+\delta\phi_{\rm asc}^{\rm BP2}\sin\left[\frac{2\pi(t-t_{02})}{P_{\rm BP2}}\right],
\end{aligned}
\end{equation}
where $P_{\rm BP1}$ and $P_{\rm BP2}$ are the periods of the harmonic movement of BP1 and BP2, respectively; $t_{01}$ and $t_{02}$ are the epochs when $\phi_{\rm asc}^{\rm BP1}=\phi_{{\rm asc}(0)}^{\rm BP1}$ and $\phi_{\rm asc}^{\rm BP2}=\phi_{{\rm asc}(0)}^{\rm BP2}$, respectively; finally, $\delta\phi_{\rm asc}^{\rm BP1}$ and $\delta\phi_{\rm asc}^{\rm BP2}$ are the respective amplitudes of the harmonic oscillation of $\phi_{\rm asc}^{\rm BP1}$ and $\phi_{\rm asc}^{\rm BP2}$. 

\begin{table}
\centering
\caption{The most likely values of the parameters used in Eqs.~(\ref{eq:sineeqs}).}
\label{tab:bestsinepars}
\begin{tabular}{lrrr} 
\hline
Parameter                        & Value          & Unit  &  Prior range\\
\hline
$\phi_{{\rm asc}(0)}^{\rm BP1}$  & 0.9(1)         & ${\rm rad}/2\pi$    &  $[0,1]$ \\
$\phi_{{\rm asc}(0)}^{\rm BP2}$  & 0.5(2)         & ${\rm rad}/2\pi$    &  $[0,1]$ \\
$\delta\phi_{\rm asc}^{\rm BP1}$ & 0.4(1)         & ${\rm rad}/2\pi$    &  $[0,1]$ \\
$\delta\phi_{\rm asc}^{\rm BP2}$ & 0.3(2)         & ${\rm rad}/2\pi$    &  $[0,1]$ \\
$t_{01}$                         & 6(2)            & $10^3$ days  &  $[0,2\pi/\Omega_{\rm SO}]$ \\
$t_{02}$                         & 19(5)          & $10^3$ days  &  $[0,2\pi/\Omega_{\rm SO}]$ \\
$P_{\rm BP1}$                    & 27(9)          & $10^3$ days  &  $[5,50]$ \\
$P_{\rm BP2}$                    & 24(6)          & $10^3$ days  &  $[5,50]$ \\
\hline
\end{tabular}
\end{table}

\begin{figure}
\centering
	\includegraphics[width=1.0\columnwidth]{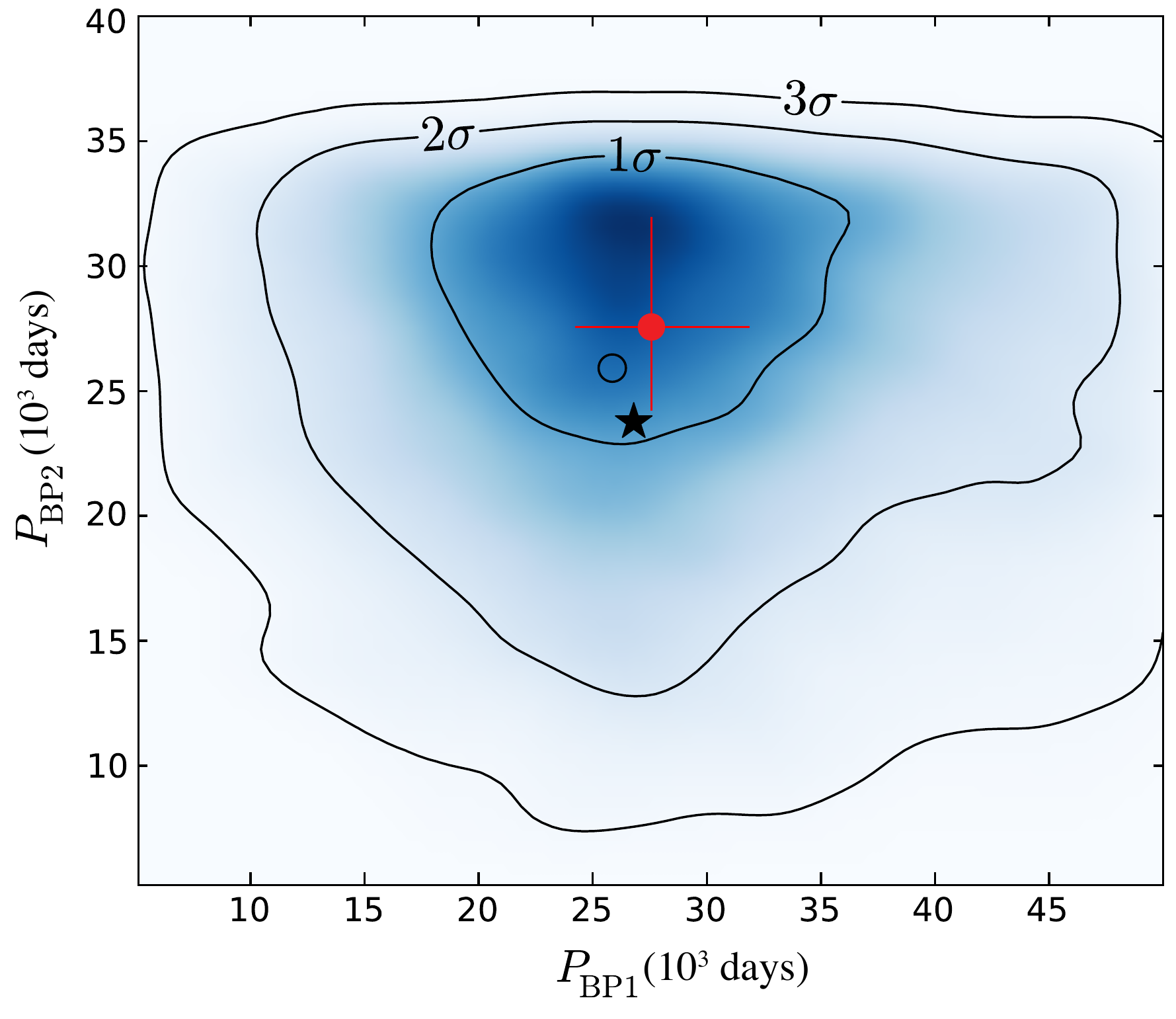}
    \caption{Joint 2D probability-density map of the period (in $10^3$ days) of the sinusoidal functions that were fit to the locations of BP1 and BP2 as a function of time (see Fig.~\ref{fig:bpsinefits}). The median and $1\sigma$ uncertainties of the period of geodetic precession period of pulsar B, published by Breton et al.~(2008), is shown with a filled red circle with error bars. The star symbol corresponds to the most likely values of the periods from our analysis. Finally, the empty circle corresponds to the period of geodetic precession predicted by GR.}
    \label{fig:sinepercont}
\end{figure}

We explored the above parameter space using nested sampling, as before, and determined the most likely values of the parameters by simultaneously maximising the likelihood of all eight parameters in Eqs.~(\ref{eq:sineeqs}). In Fig.~\ref{fig:sinepercont}, we show the joint probability density maps of $P_{\rm BP1}$ and $P_{\rm BP2}$ and of $t_{01}$ and $t_{02}$. The most likely values of $P_{\rm BP1}$ and $P_{\rm BP2}$ are shown with a star in that plot, and they are consistent within $1\sigma$ with the value for the period of geodetic precession, published by BKK+08\nocite{bkk+08}; red circle). It is noteworthy that the predicted value for the period of geodetic precession by GR (empty circle) is also consistent within $1\sigma$ with both the aforementioned values.

In Fig.~\ref{fig:bpsinefits}c, we plot Eqs.~(\ref{eq:sineeqs}) overlaid with the data, over 30,000\,days, centred at MJD\:53018.1 (the centre epoch of the earliest MJD bin), using the most likely parameter values (Table~\ref{tab:bestsinepars}). Our simple model of the movement of the BPs, as can be seen in that figure, has the following implications. Firstly, as shown with a solid black line and black data points, the average shift of BP1 and BP2 has a much flatter orbital-phase evolution with precession phase compared to those of BP1 and BP2. Consequently, over short data spans compared to the period of geodetic precession, such us ours, it can be approximated with a linear function. Secondly, according to Fig.~\ref{fig:bpsinefits}c, our model of the movement of the BPs suggests that when the pulsar returns to the same viewing-angle configuration as that of the earliest observations, the locations of BP1 and BP2 will be at orbital phase $\approx 0.17$ and $\approx 0.27$, respectively. These locations are roughly half an orbit away from where they where when the pulsar was originally observed. Also, the angles $\rho_{\rm BP1}$ and $\rho_{\rm BP2}$ will also be significantly different at that time. However, without a complete description of the effect of pulsar A's wind on pulsar B's emission region, we cannot make any predictions as to whether these orbital-phase intervals will still be where the pulsar will appear brightest. Certainly, irrespectively of where this happens in the orbit, our LOS will intersect the beam of pulsar B at the same latitude as when it was first seen, so during that time the pulsar will again be visible. The evolution of the impact angle, $\beta=\hat{\boldsymbol{x}}\wedge\hat{\boldsymbol{\mu}}$, is shown in Fig.~\ref{fig:bpsinefits}b, where we use grey dotted lines to indicate the epoch of the MJD bin corresponding to the earliest observations in our data and the subsequent epoch when the viewing geometry of the pulsar's beam will return to the same configuration (i.e.~to the same $\beta$; ca.~MJD\:62530).

\subsection{Emission height}
According to the most likely parameter values from our analysis, the magnitude of the wind corresponds to a maximum displacement of the emission region by $\approx 1.7-2\%$ of the emission height (Eq.~(\ref{eq:windperp}) with $w_0=0.017$). Equivalently, this corresponds to a maximum deflection by $\approx 1^\circ-1\fdg1$. In our model, we did not consider the emission height, $r_{\rm em}$, as a free parameter, but rather parametrised the wind magnitude as a fractional displacement. Nevertheless, based on the above amount of deflection and the simulations of Perera et al.~(2012)\nocite{plg+12}, we can estimate $r_{\rm em}$. The aforementioned work provides an analytical expression for the angular deflection of the emission, $\alpha_{\rm defl}$, as a function of the orbital phase and the emission height, $r_{\rm em}$, expressed in units of the stand-off radius, $r_{\rm s}$ (see their Eq.~(34)). The latter quantity corresponds to the distance from pulsar B, where its magnetic pressure, $B^2(r_{\rm s})/8\pi$, balances the dynamic pressure of pulsar A's wind, $\dot{E}_{\rm A}/4\pi c(a-r_{\rm s})^2$, where $B(r_{\rm s})$ is the magnetic field of pulsar B at $r_{\rm s}$, $\dot{E}_{\rm A}$ is the spin-down luminosity of pulsar A, and $a\approx x_{\rm A}+x_{\rm B}$ is the average distance between the two pulsars. In Perera et al.~(2012)\nocite{plg+12}, the stand-off radius was constrained to be $r_{\rm s}\sim 40,000$ km, which is roughly equal to 30\% of the undistorted light-cylinder radius, $R_{\rm LC}=cP_{\rm B}/2\pi$. In that work, the authors placed upper limits on $r_{\rm em}$ by postulating that the maximum deflection cannot exceed the width of the pulsar's beam, meaning $\alpha_{\rm defl}\lesssim 14\fdg 3$. Our analysis provided an estimate for $\alpha_{\rm defl}$, which can be directly converted to a value of $r_{\rm em}$. Using the centroid locations of BP1 and IP, which roughly correspond to the orbital phases where the maximum deflection occurs, and the maximum deflection of $\alpha_{\rm defl}= 1^\circ-1\fdg1$, we calculate $r_{\rm em}\approx 12,300-14,500$ km and $r_{\rm em}\approx 15,100-16,200$ km, respectively, across the span of our data. In other words, the emission site is located somewhere from 30--40\% of the stand-off radius above the star.

\section{Summary and future perspectives}
\label{sec:summaryfuture}
\subsection{Summary}
The work presented here has made use of roughly four years of archival pulse profiles of pulsar B (PSR J0737$-$3039B), from Parkes and GBT observations, covering nearly the entire range of epochs from the pulsar's discovery, to its disappearance in 2008. The main objective of this study was to make use of all available observations of this pulsar to construct a model that describes the systematic profile variations, as a function of orbital phase and observation epoch, and to investigate the level of improvement that such a model could bring about, in future tests of GR and alternative theories of gravity, with the double pulsar.

As has been noted in previous studies of the double pulsar, pulsar B is mainly detected during two narrow orbital-phase windows, `bright phase 1' (BP1; at $\approx 210^\circ$) and `bright phase 2' (BP2; at $\approx 280^\circ)$, whereas it is barely detectable elsewhere, during the so called weak phase (WP). Using available flux-calibrated multi-frequency profiles of pulsar B, observed in BP1 at nearly identical orbital phases, we estimated the spectral index of the radio emission to be $-1.46_{-0.27}^{+0.10}$. Furthermore, we determined the locations of BP1 and BP2, as a function of time, over the four-year data span. We find that the BP1 emission is confined within $\approx 180^\circ-235^\circ$, whereas the BP2 emission is within $\approx 260^\circ-308^\circ$. In addition, for half of our data span, we detected an orbital-phase window of intermediate brightness, between the BPs and the WP, which we call the intermediate phase (IP): we determined its bounds to be $\approx 318^\circ-360^\circ$ and $\approx 0^\circ-70^\circ$. Analogously to previous work, we also determined that BP1 and BP2 shift to later orbital phases as a function of time, over the span of the available observations. We measured the average rate of migration of BP1 and BP2 to be $2\fdg4(3)$\,yr$^{-1}$ and $1\fdg8(2)$\,yr$^{-1}$, respectively, but noted that (a) the migration rate deviates significantly from being constant across the data span, and that (b) this deviation progresses in opposite directions, for BP1 and BP2. Intriguingly, we found that the migration rates of BP1 and BP2 are complementary across the data span, which we showed by calculating the numerical average of the locations of BP1 and BP2, as a function of time, and modelling it with a linear regression. The linear fit to the average locations is a better description of the data, compared to the fits to each of the BP1 and BP2 locations separately; the slope of the best fit is equal to $2\fdg4(1)$\,yr$^{-1}$. In a further investigation, we explored the possibility that the migration rates of BP1 and BP2 are harmonic functions of time and found that the most likely periods of such functions were 74(25) and 66(16)\,yr, respectively, which are consistent with the published period of geodetic precession, $2\pi/\Omega_{\rm SO}=75^{+12}_{-9}$\,yr, and the period of geodetic precession predicted by GR, $2\pi/\Omega_{\rm SO}^{\rm GR}\approx 71$\,yr.

Pulsar B exhibits dramatic profile variations on different time scales due to the interaction with the electromagnetic wind of its companion and due to geodetic precession. To characterise these variations, we generated over 4,000 TOAs from the observed pulse profiles, using a synthetic template based on the pulsar's WP emission. After correcting for all known orbital and spin delays, assuming a model based on pulsar A timing and the validity of GR, we examined the systematic delays in the timing residuals, as a function of orbital phase and epoch of observation. The timing residuals exhibit quasi-linear drifts during each of BP1 and BP2, of the order of 10--20\,ms. In addition, across the entire orbit, the residuals exhibit on average a harmonic variation at the period of the orbit. Also, the amplitude of this variation slowly increases with time, over the four years of data, from $\approx 5$\,ms (ca.~MJD\:53000) to $\approx 30$\,ms (ca.~MJD\:54500). 

The present work focused on modelling solely the harmonic profile changes associated with the mean variation of the magnitude of the effect of pulsar A's wind on pulsar B, perpendicular to the observer's LOS. To achieve that, we employed a simple geometric model of a radial wind, directed from pulsar A to pulsar B, which deflects the emission beam of pulsar B to a varying degree, as a function of orbital phase. The assumed wind harmonically displaces pulsar B's beam direction, such that our LOS intersects pulsar B's emission at different magnetic latitudes, thus introducing profile variations across the orbit. 

In order to reproduce the observed profile variations, in conjunction with the effect of the wind we assumed a parametric 2D beam model, consisting of two 2D Gaussian components. The peak intensity of the Gaussian components was allowed to vary between the orbital-phase windows, to account for the significant brightness changes across the orbit, which we speculated are the result of the (herein unmodelled) interaction of pulsar A's wind with the magnetospheric plasma of pulsar B. 

On the whole, our model can be defined using 19 parameters, describing the beam shape and intensity, and the wind magnitude. It can provide an analytical approximation of the observed flux-density profiles, as a function of orbital phase and epoch of observation. It is important to stress the main inadequacies of the model. Since it assumes a harmonically varying wind as the source of profile variations, it cannot describe the fast profile variations across the BPs and the IP; for that reason, it was deemed sufficient to represent the profile changes across the orbit using only the average profiles corresponding to the centroids of BP1, BP2, the IP, and the WP (the orbital phases corresponding to maximum S/N). Consequently, our model is informed by only a small fraction of pulsar B's detectable emission, across its orbit. Moreover, our model cannot reproduce the lack of observed emission between the BPs, the IP and the WP. We can speculate that those gaps in the range of detectable emission are related to the interaction responsible for the fast profile variations during the BPs and the IP. For example, the timing drifts across each of these phases and the gradually diminishing S/N towards the phase bounds perhaps suggests that the beam also drifts across our LOS, becoming undetectable beyond the bounds of each phase. However, we cannot rule out that some other interaction between the wind of pulsar A and the magnetospheric plasma of pulsar B is responsible.

Given the observed average profiles corresponding to the centroids of BP1, BP2, the IP, and the WP between MJD\:53000 and MJD\:54500, we determined the most likely parameters of our model, using a nested-sampling Bayesian algorithm. 
Overall, the most likely template profiles from our model track the evolution of the average profile of BP1 well, from single-peaked to double-peaked ones, although towards later epochs the profile intensity is somewhat underestimated. Of course, it is important to take into consideration the original systematic uncertainties of the flux density (see Fig.~\ref{fig:datprofs}), that is, before the application of the error coefficients. In contrast, the evolution of the average BP2 profile, which clearly develops two distinct components during MJD\:53800--54200, is poorly tracked by our model --- although the profile-intensity evolution seems more consistent with the observations, compared to that of BP1. It is interesting to note, however, that during the above MJD range, the model appears to produce an additional very weak leading component, at approximately the pulse phase of the much brighter observed leading component of BP2. Lastly, the observed average profiles of the IP and the WP are much more weak and erratic, and our model was only able to approximate them with a single Gaussian --- although again an additional very weak component is generated by the model whenever the observed profiles are distinctly double peaked (e.g.~the WP profiles during MJD\:53600--53800). The inability of our model to fit the IP and WP profiles is also evident in the residuals of Fig.~\ref{fig:ml_resids}, where timing of those orbital phases with our model templates results in systematic offsets from zero.

An interesting property that was constrained by our modelling is the shape and intensity of pulsar B's emission beam. Our beam model was limited in that it only comprised a combination of two surface Gaussian distributions. However, the location, rotation, ellipticity, and peak intensity of those Gaussians was allowed to cover nearly all possible values of the respective parameters. We found that the most likely beam of BP1 and the IP comprises a primary component that is at least an order of magnitude brighter than the secondary Gaussian component. Interestingly, the brightness dominance of the primary and secondary Gaussian components alternates between BP1 and the IP, essentially shifting the phase of the pulse profile's peak. As those two orbital-phase windows represent the largest amount of timing delay between any two parts of the orbit where the pulsar is detected, the corresponding phase shift between the primary components is possibly, to a certain degree, covariant with the effect of the wind. On the other hand, in BP2 and the WP, the two components seen alternating between BP1 and IP are now both present with roughly equal intensity. In hindsight, our preliminary reasoning with regards to the beam shape (see Section~\ref{sec:avgprofs}) seems justified: we indeed find that the emission beam is consistent with an elongated, elliptical region, with components that diverge as we move away from the magnetic axis. 

Finally, our modelling provided tantalising evidence for a prograde rotation for pulsar B. Assuming that pulsar B spins in a prograde fashion with respect to its orbital motion yielded model profiles that were both a closer match to the observed profile evolution as a function of orbital phase and observation epoch, and at the same time demonstrated a prograde solution with significantly higher likelihood than its retrograde counterpart. If in future studies our conclusions are confirmed, then the prograde rotation of both pulsars in the double pulsar system will have interesting implications for the formation and evolutionary chain of double-NS systems. Moreover, in population syntheses of such systems, such information may also serve as an additional constraint when modelling the gravitational waves produced by double-NS mergers.

Our modelling is a first step towards improving the timing of pulsar B. We have estimated the improvement we can expect, if such a model is used to time pulsar B, by generating TOAs from the original 4,115 observed profiles and the model profiles as timing templates. The magnitude of the improvement is roughly half-way between the best case scenario, where the harmonic delays across the orbit are completely eliminated, and the uncorrected case. More importantly, the model offers a factor 2.6 improvement on the precision of the observed size of pulsar B's orbit, a parameter that is central in tests of GR with this system. Additionally, it brings us closer to measuring the amount of beam aberration for pulsar B for the first time, only falling short by a factor of 1.5. The timing of pulsar B presented in this work relied to some extent on the size of pulsar B's orbit, which was calculated assuming GR, from the orbit of pulsar A, in order to account for the orbital delays. Hence, the aforementioned improvement on the measurement precision of $x_{\rm B}$ is not independent of this assumption. However, the value of $x_{\rm B}$ is the only timing parameter in our work that significantly depends on the assumed theory of gravity and, moreover, its value is consistent across a range of fully conservative theories. Nevertheless, because of this dependence on a particular set of gravity theories, our timing results cannot directly be used to test GR and other, alternative theories of gravity. We must also caution that the above timing improvements account for neither the systematic uncertainties borne from the assumption that our simple model is the only one that correctly describes the observed profile evolution, nor the covariances between the model's parameters and those of the timing model used. As such, the above improvements can be considered as an idealised case, specific to our model. Ultimately, a more physically motivated model combined with a global fit over all the model and timing parameters is needed to estimate the true magnitude of the improvement.

Although our modelling presents a significant improvement over previous work, there are significant, unmodelled components in our timing, reflecting our lack of knowledge of pulsar B's emission geometry and its interaction with pulsar A. However, we are confident that, within its limitations, our model reflects real physical effects, and that it is not a phenomenological exercise in absorbing the observed timing systematics. This confidence is derived from the model's ability to reproduce long-term precessional effects, such as the amplitude of the observed harmonic delays, without introducing superfluous parametrisation, but with simple geometry.

\subsection{Future perspectives}
\label{subsec:future}
Future modelling of the profile variations observed in archival data of pulsar B, including the timing drifts observed during BP1, BP2, and the IP, will result in even higher timing precision. As an exercise of what can be expected, we combined the heuristic Gaussian templates with a linear model of the drifts during the orbital-phase windows, which resulted in roughly factor 4 improvement in the timing precision across four years. If made through physical modelling, such an improvement would reduce the uncertainty on the projected semi-major axis of pulsar B to levels comparable to or even lower than the expected value of beam aberration for this pulsar. The improved precision of $x_{\rm B}$ will yield an equally significant improvement on the precision of $R=x^{\rm int}_{\rm B}/x^{\rm int}_{\rm A}=M_{\rm A}/M_{\rm B}$, which will enable us to place stringent theory-independent constraints on the strong-field parameters of binary motion, as was done in Kramer \& Wex (2009)\nocite{kw09}.

Lastly, when pulsar B inevitably becomes visible again, which based on geometric arguments\footnote{Here, we are assuming that reappearance will occur when the impact angle ($\beta$) returns to the same value as that of the last epoch when the pulsar was still detectable (i.e.~ca.~2008). Of course, an earlier reappearance is still possible if additional beam components have become active in the meantime.} should be ca.~2024 at the latest (Breton 2009\nocite{bre09}), it will be possible to perform joint timing between both pulsars, A and B, and further increase the precision of the observed timing parameters. For pulsar B, increased timing precision can come from techniques that exploit the drifting sub-pulse features (Freire et al.~2009\nocite{fwk+09}; Liang et al.~2014\nocite{llw14}; Pol et al.~2018\nocite{pmk+18}). Upon continuing the timing of pulsar B, after its reappearance, having useful constraints on the phase of the pulsed emission will help to bridge the gap between the periods of visibility. To that effect, our model can be used to make predictions of the phase range we expect the emission to occur, at a given epoch. If coherent timing, either side of the pulsar's disappearance is achieved, it will provide the eclipse model of BKK+08 with a long timing baseline, which can be used in precise tests of GR via the $\Omega_{\rm SO}$ parameter. As discussed in Section~\ref{subsec:bpmigration}, when pulsar B returns, the relative orientation between the orbital angular momentum and pulsar B's spin, as well as the orientation of the orbit with respect to our LOS, will have changed due to geodetic and periastron precession, respectively. At that time, we expect the drift pattern in pulsar B’s pulsed emission, which is caused by the impact of pulsar A’s wind, to have also changed as a result. This future perspective will provide additional information with which we can further constrain our model's parameters.

\begin{acknowledgements}
This paper is partly based on data analysis that was performed on the {\sc Hercules} cluster of the Max Planck Computing \& Data Facility in Garching, Germany. AN and GD acknowledge financial support by the European Research Council (ERC) for the ERC Synergy Grant {\sf BlackHoleCam}. The authors would like to thank Drs.~David Champion, Ralph Eatough, Ewan Barr and Nataliya Porayko for invaluable discussions and advice, during the preparation of this manuscript. Additional thanks are given by the authors to Dr.~Alessandro Ridolfi for providing specialised software that greatly assisted the data-quality inspection and selection.
\end{acknowledgements}




\bibliographystyle{aa}
\bibliography{journals,modrefs,psrrefs}

\clearpage

\begin{appendix}
\section{Timing residuals}
The figures presented in this appendix come from the timing analysis performed in Sections~\ref{sec:timing} and \ref{subsec:timimprov}, as well as the modelling of Section~\ref{subsec:toymodesc}. All figures show the timing residuals after subtracting a timing model of the spin and orbital delays. For figures showing the residuals after the subtraction of supplementary timing models, describing the pulsar's profile evolution, please refer to the respective captions for details. The orbital phase is shown as a fraction of the orbit (measured from the ascending node) and it is aliased over two orbital periods for clarity. The error bars on the residuals correspond to $1\sigma$ uncertainty. The values of the weighted RMS and the corresponding reduced chi-squared ($\chi^2_{\rm red}$), shown for each data set, were calculated with TEMPO2 and refer to the total data span. For more details, please see the respective figure captions and the aforementioned sections.

\begin{figure*}[h]
	\includegraphics[width=\textwidth]{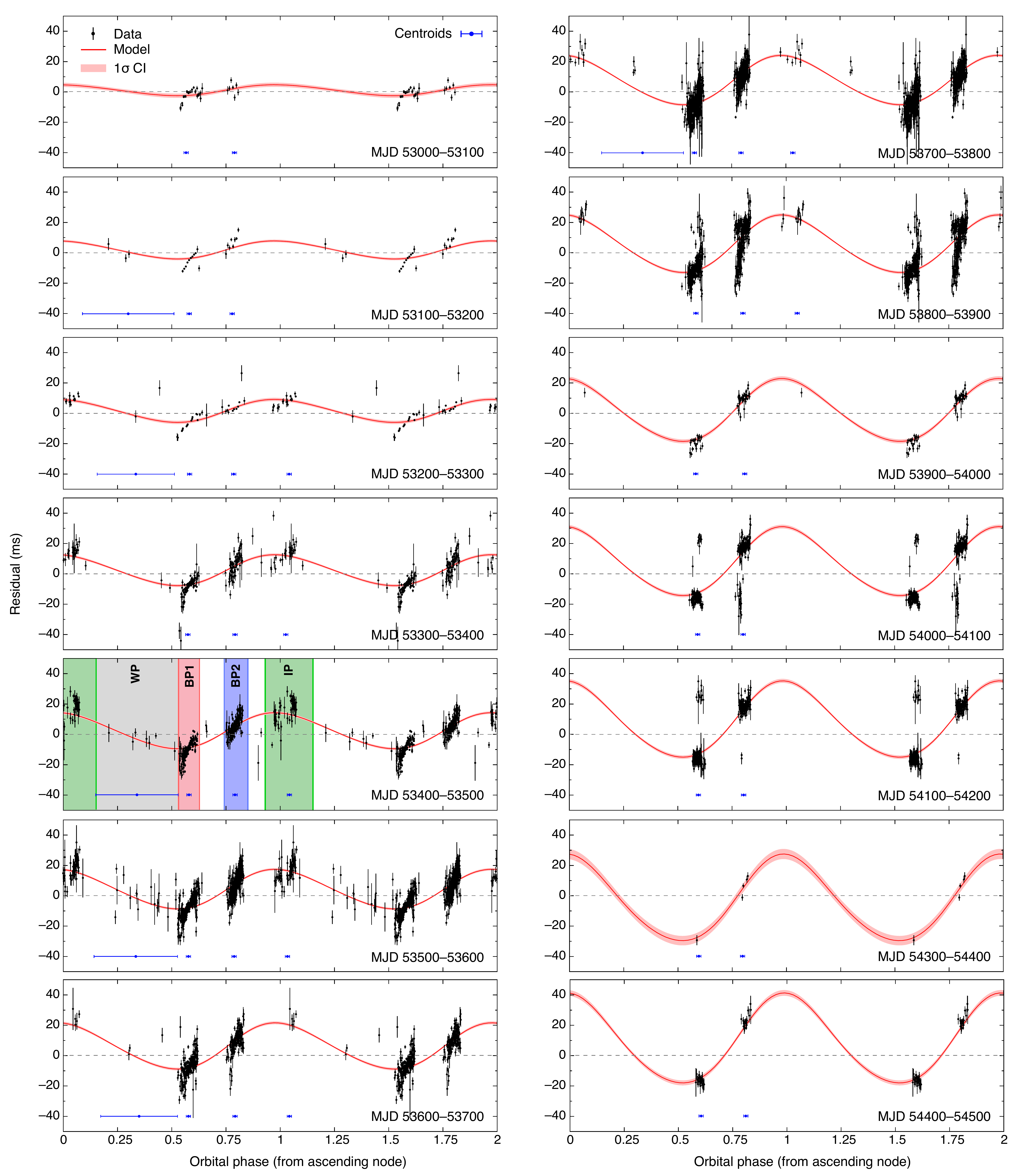}
    \caption{Residuals of pulsar B for 14 100-day intervals, calculated using the analytic WP template of Fig.~\ref{fig:wp_template}, constructed from fitting a three-Gaussian-component model to the average WP profile of pulsar B. The weighted RMS of the residuals, calculated over the entire data set shown in this figure, is 9.33\,ms; the corresponding $\chi^2_{\rm red}$ value is 75. The red lines correspond to the function of Eq.~(\ref{eq:winddelay}), using the best fit values for the parameter $w_0$. The shaded envelope corresponds to the $1\sigma$ confidence interval of the above function. The horizontal blue error bars indicate the orbital-phase intervals, at the location of peak brightness of BP1, BP2, and IP (centroids), that were used to calculate the average profiles of Fig.~\ref{fig:datprofs}; for the WP, the entire width of the orbital-phase window was used. Also, the panel corresponding to the interval MJD\:53400--54500 shows the extents of the orbital-phase regions, BP1, BP2, the IP, and the WP, for that interval, as red, blue, green, and grey shaded areas.}  
    \label{fig:allresidsfits}
\end{figure*}

\begin{figure*}[h]
	\includegraphics[width=\textwidth]{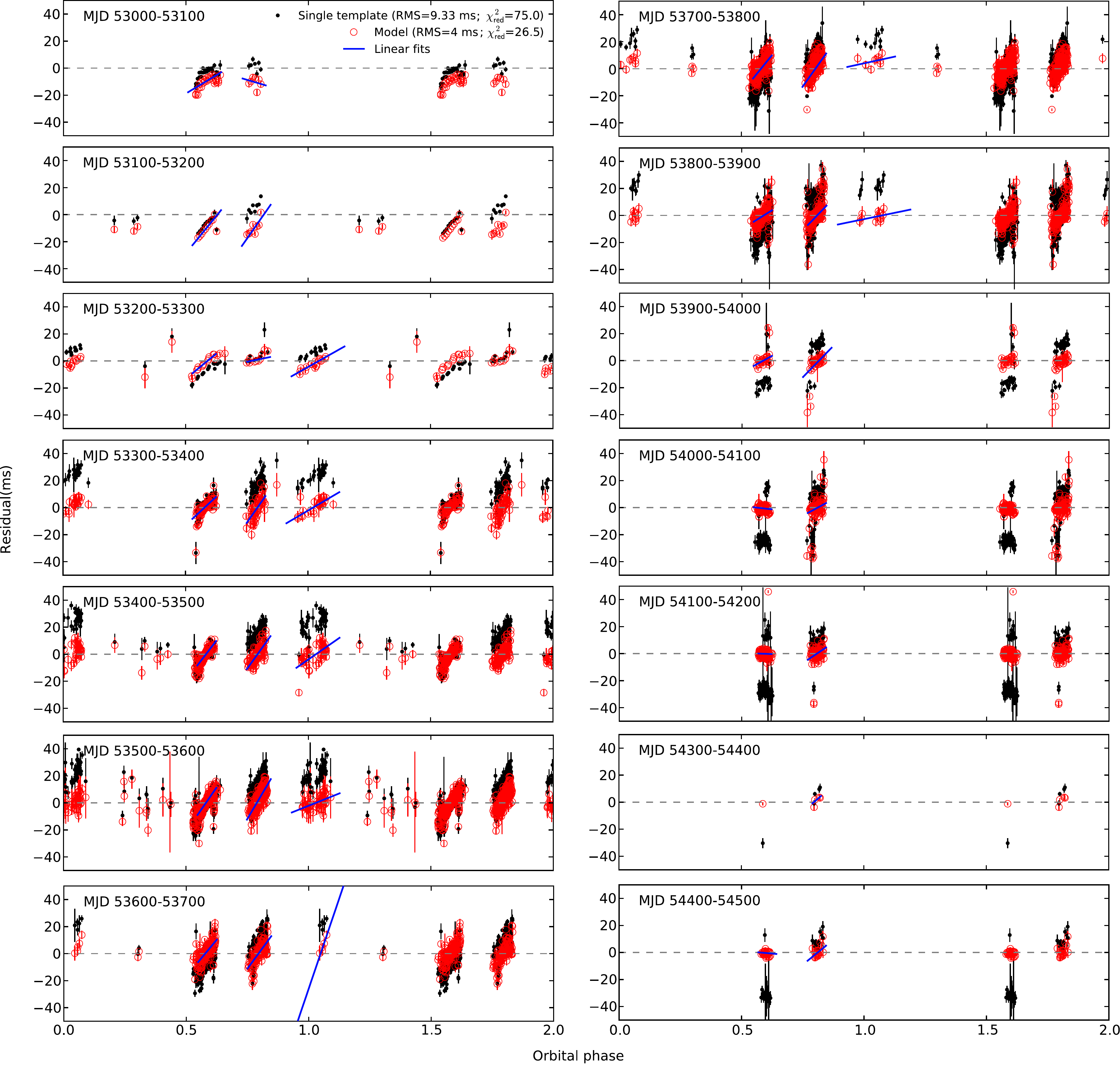}
    \caption{Timing residuals using fixed WP template profile of Fig.~\ref{fig:wp_template} (black points) and using the Gaussian templates of Fig.~\ref{fig:gaussdatprofs} (red open circles). For each data set, the weighted RMS and the corresponding $\chi^2_{\rm red}$, calculated over the entire data set shown in this figure, is shown in the top-right corner of the top-left sub-plot. The blue lines show the best linear fits to the residuals of BP1, BP2, and IP, which were used to correct for the linear drifts across those orbital-phase windows (see Fig.~\ref{fig:gausslinear_resids}). The orbital-phase extent of the lines is equal to the $W_{3\sigma}$ value of the corresponding orbital-phase window.}
    \label{fig:paas_resids}
\end{figure*}

\begin{figure*}[h]
	\includegraphics[width=\textwidth]{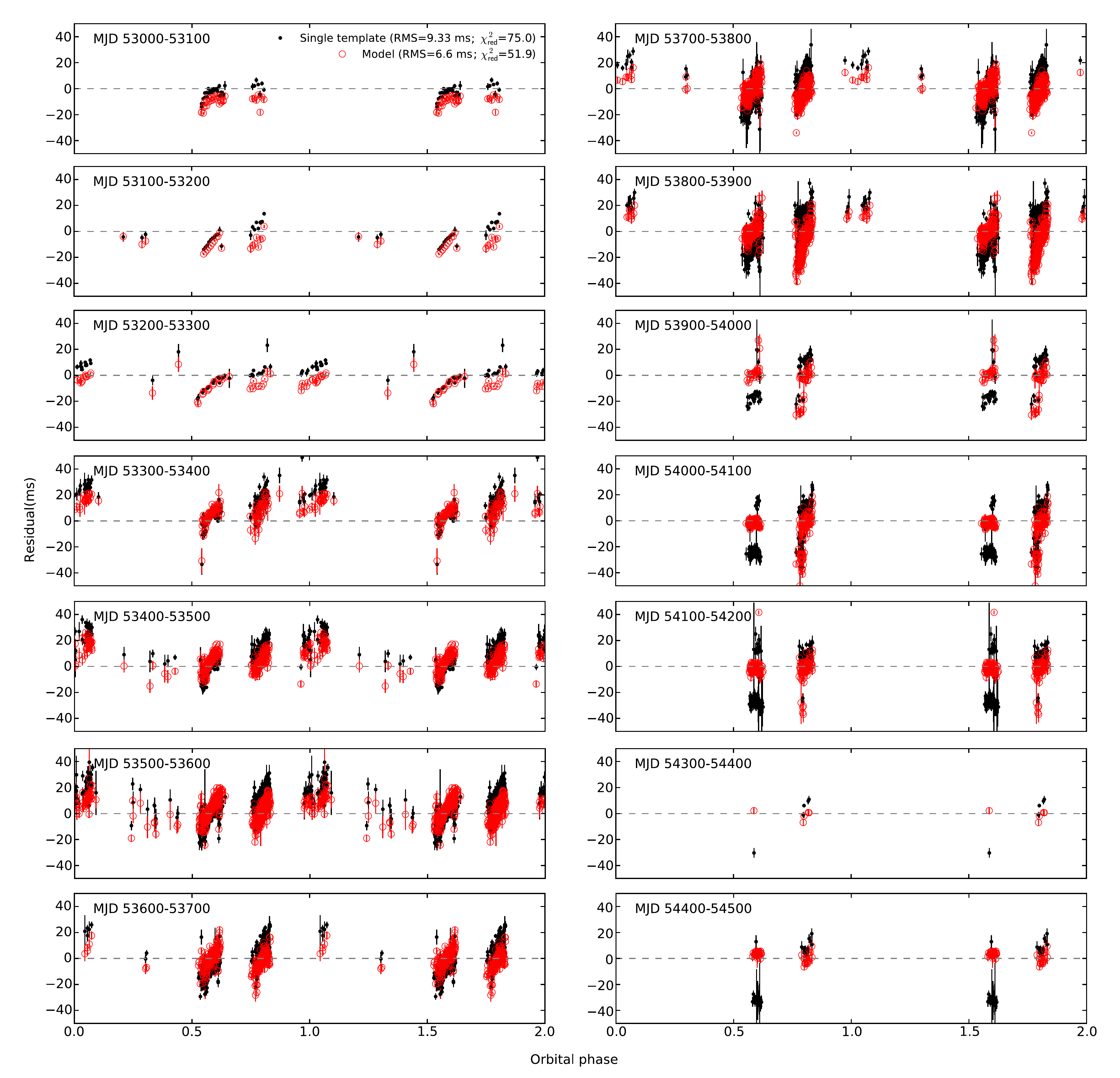}
    \caption{Timing residuals using fixed WP template profile of Fig.~\ref{fig:wp_template} (black points) and the model templates of Fig.~\ref{fig:moddatprofs} (red open circles). For each data set, the weighted RMS and the corresponding $\chi^2_{\rm red}$, calculated over the entire data set shown in this figure, is shown in the top-right corner of the top-left sub-plot.}
    \label{fig:ml_resids}
\end{figure*}

\begin{figure*}[h]
	\includegraphics[width=\textwidth]{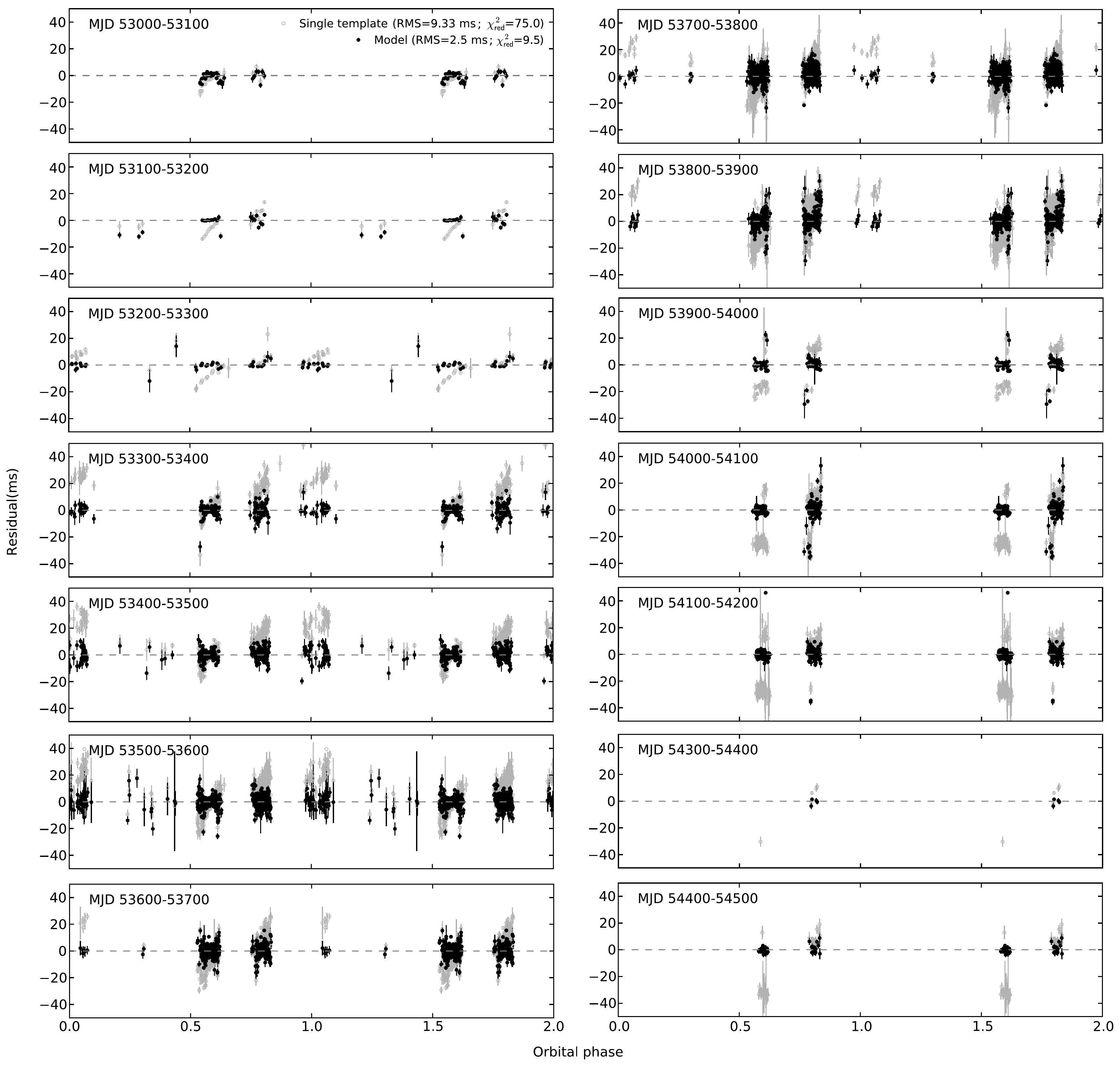}
    \caption{(black points) Timing residuals using Gaussian templates of Fig.~\ref{fig:gaussdatprofs} and after correcting for the linear drifts across each orbital-phase window, via the best fits shown in Fig.~\ref{fig:paas_resids}. For comparison, light grey circles show the timing residuals using the fixed WP template profile of Fig.~\ref{fig:wp_template}. For each data set, the weighted RMS and the corresponding $\chi^2_{\rm red}$, calculated over the entire data set shown in this figure, is shown in the top-right corner of the top-left sub-plot.}
    \label{fig:gausslinear_resids}
\end{figure*}

\clearpage

\section{Profile evolution}
The figures presented in this appendix show the positional and the average profile evolution of the BPs, the IP, and the WP of pulsar B, corresponding to the centroid of each orbital window (the orbital phase at peak brightness), as a function of time, binned in 14 100-day bins. The details of the corresponding analyses can be found in Sections~\ref{sec:dataredux}, \ref{subsec:simemgeo}, and \ref{subsec:modparest}.

\begin{figure*}[h]
	\includegraphics[width=1.0\textwidth]{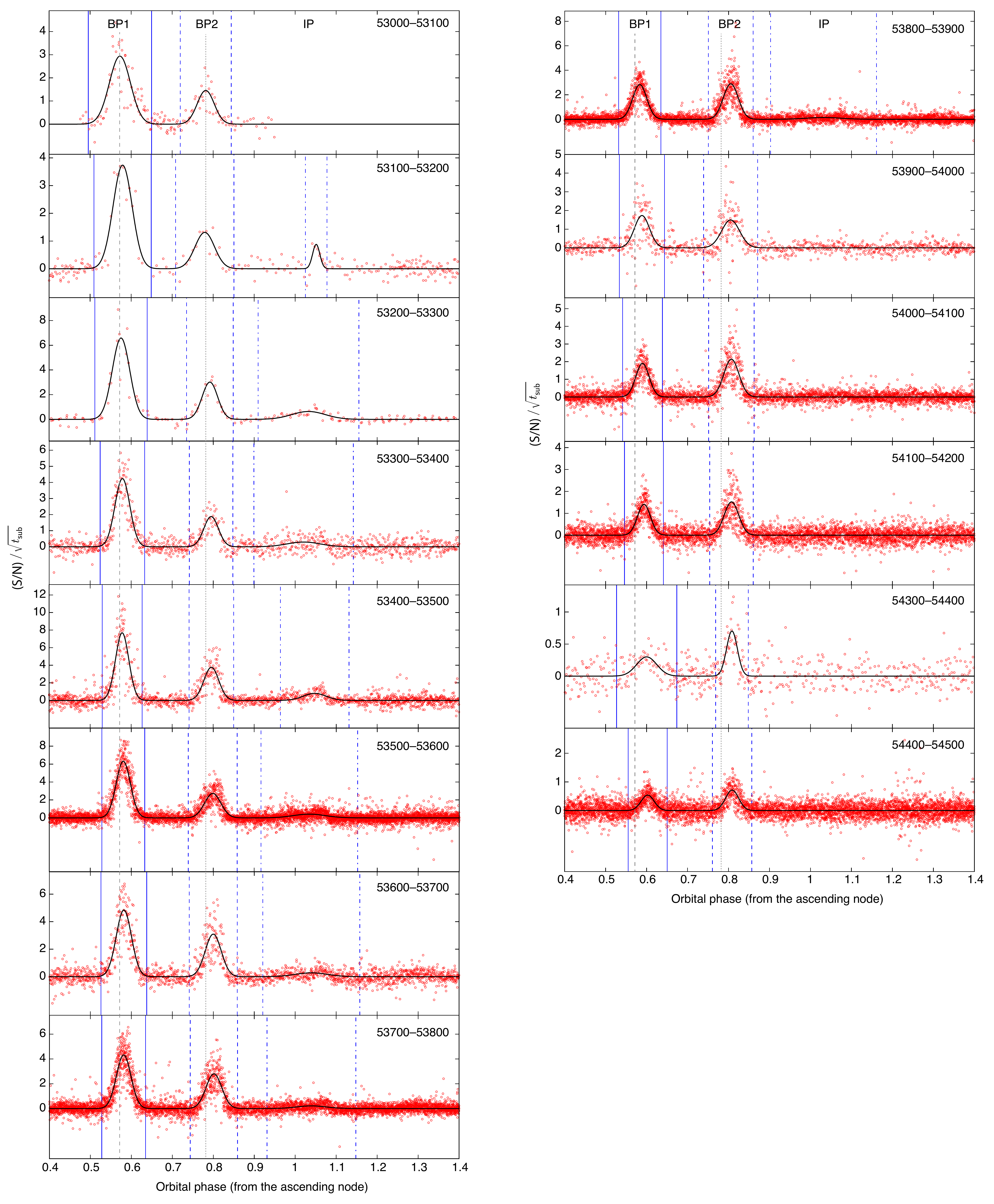}
    \caption{(data points) Distributions of S/N across the orbit of pulsar B, for each MJD bin. For this figure, the values of the S/N have been normalised by the square root of the integration time. The locations of the two BPs, BP1 and BP2, and where visible that of the IP have been determined using Gaussian fits to the data (black curves). The vertical, solid, dashed and dot--dashed blue lines show the 3$\sigma$ confidence intervals of the phase locations of BP1, BP2 and the IP, respectively. To emphasise the shift of the locations of BP1 and BP2, as a function of MJD, the dashed and dotted, vertical, grey lines mark the positions of the maxima of the fitted functions for BP1 and BP2, respectively, during MJD\:53000--53100.}
    \label{fig:bp_fits}
\end{figure*}

\begin{figure*}[h]
	\includegraphics[width=1\textwidth]{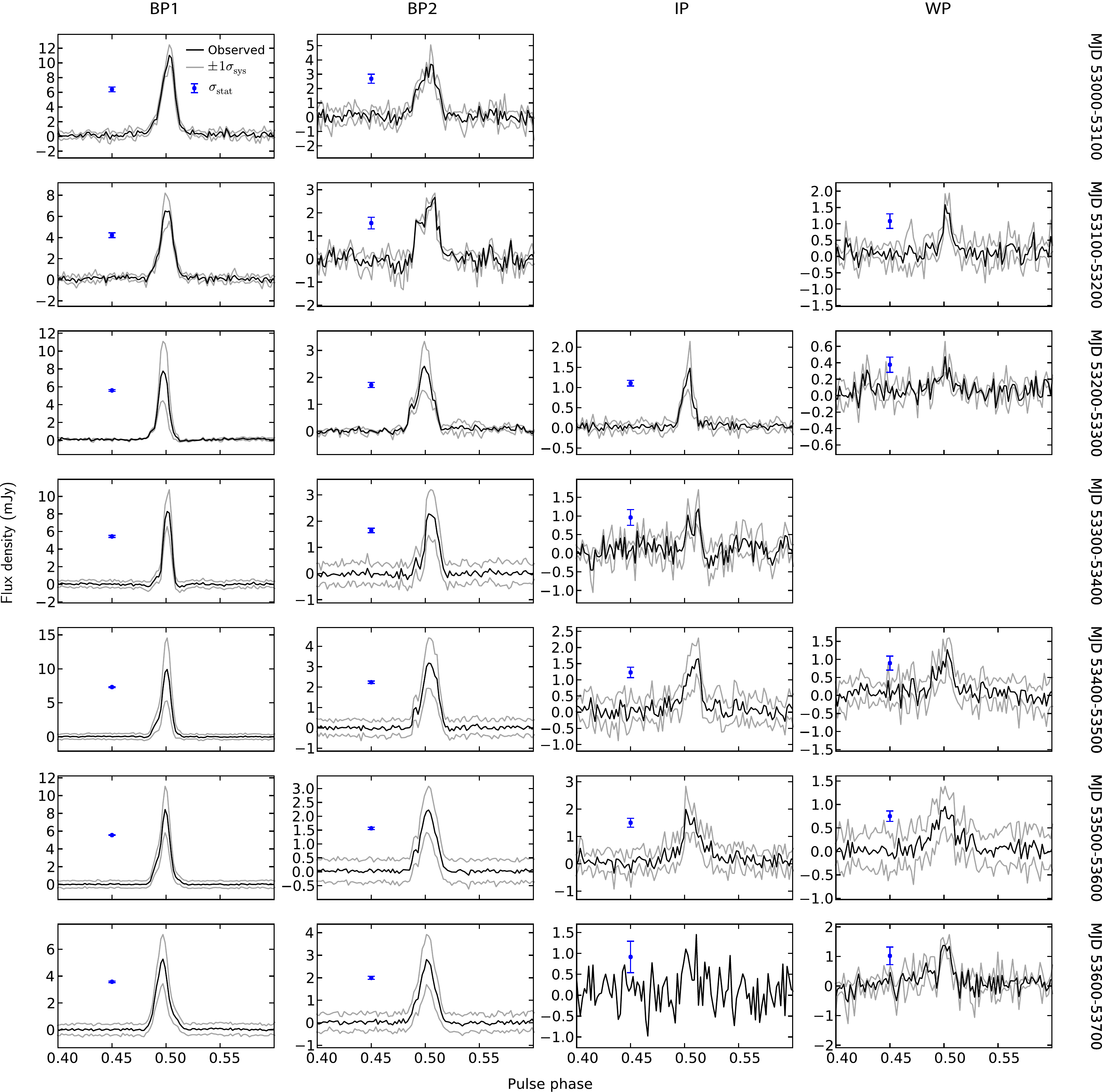}
    \caption{Observed average profiles of pulsar B (black lines), corresponding to the centroids of BP1, BP2, and the IP; the WP profiles were calculated using all the data in that orbital-phase window (see blue error bars in Fig.~\ref{fig:allresidsfits}). Each column corresponds to a different orbital-phase window (labelled along the top edge of the figure); each row corresponds to a different MJD bin (labelled along the right edge of the figure). The grey lines show the lower and upper $1\sigma_{\rm sys}$ confidence limits (Eq.~(\ref{eq:fluxrms})). The blue vertical error bar next to each profile corresponds to the off-pulse RMS, $\sigma_{\rm stat}$, of the average profile.}
    \label{fig:datprofs}
\end{figure*}

\addtocounter{figure}{-1}

\begin{figure*}[h]
	\includegraphics[width=1.0\textwidth]{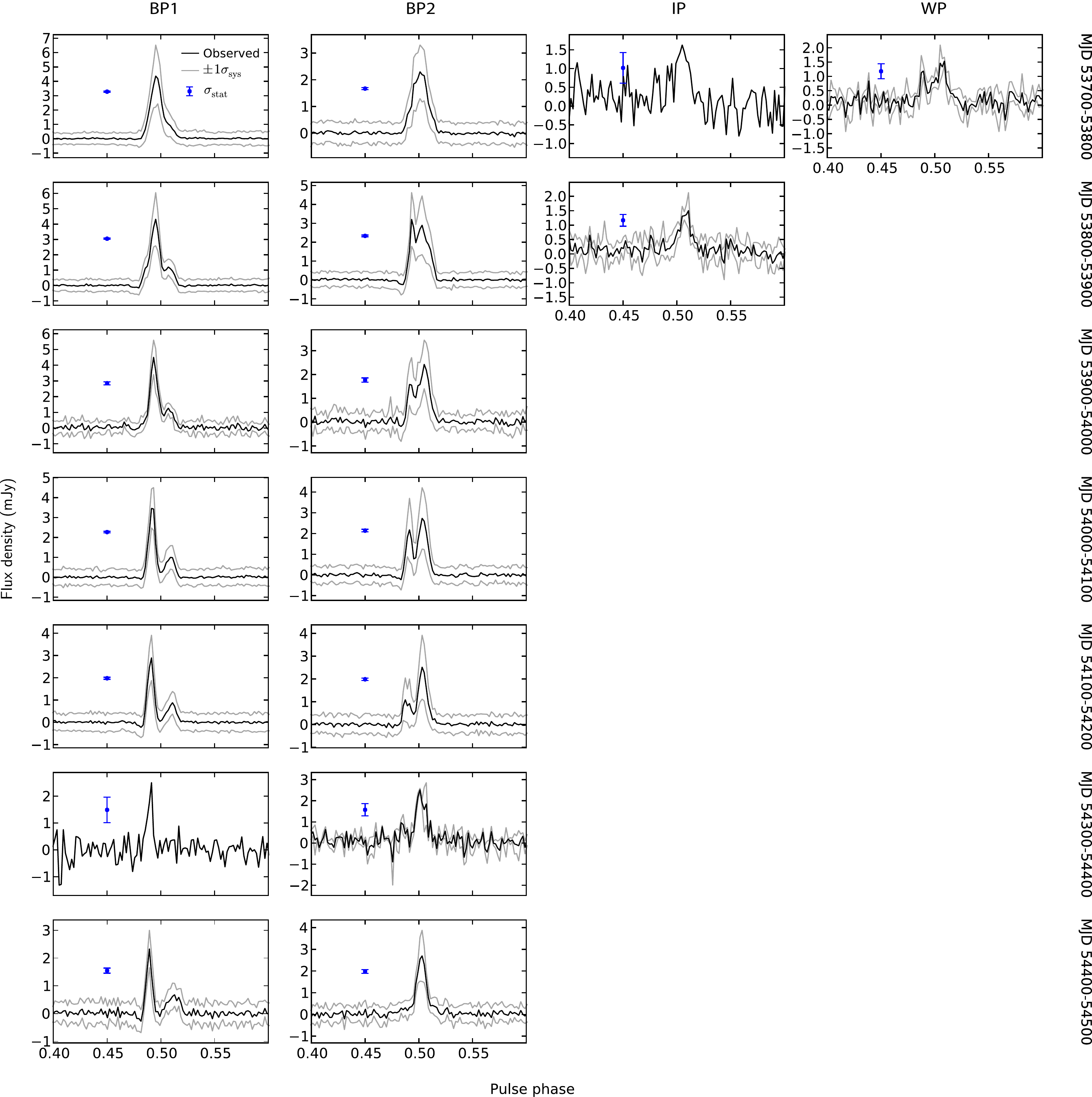}
    \caption{Continued.}
    \label{fig:datprofs}
\end{figure*}

\begin{figure*}[h]
	\includegraphics[width=1\textwidth]{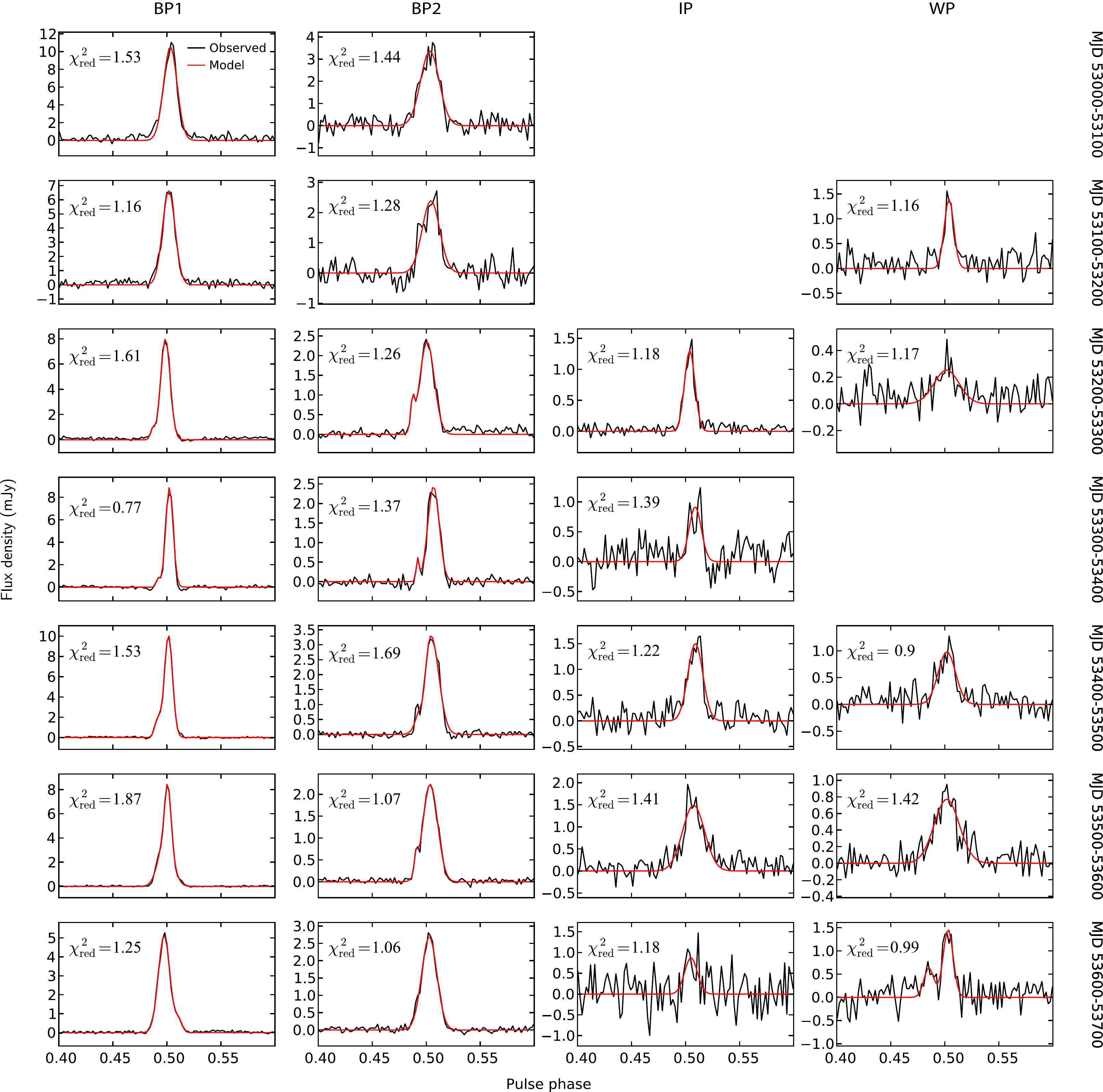}
    \caption{Comparison between observed average profiles of pulsar B (black) and the best fit Gaussian templates (red), as described in Section~\ref{subsec:proftempl}. The profiles shown have been tabulated according to orbital-phase window (columns): BP1, BP2, IP, and WP, and the MJD range (rows). The reduced chi-squared value shown for each plot has been calculated using only the off-pulse RMS ($\sigma_{\rm stat}$) as weights.}
    \label{fig:gaussdatprofs}
\end{figure*}

\addtocounter{figure}{-1}

\begin{figure*}[h]
	\includegraphics[width=1.0\textwidth]{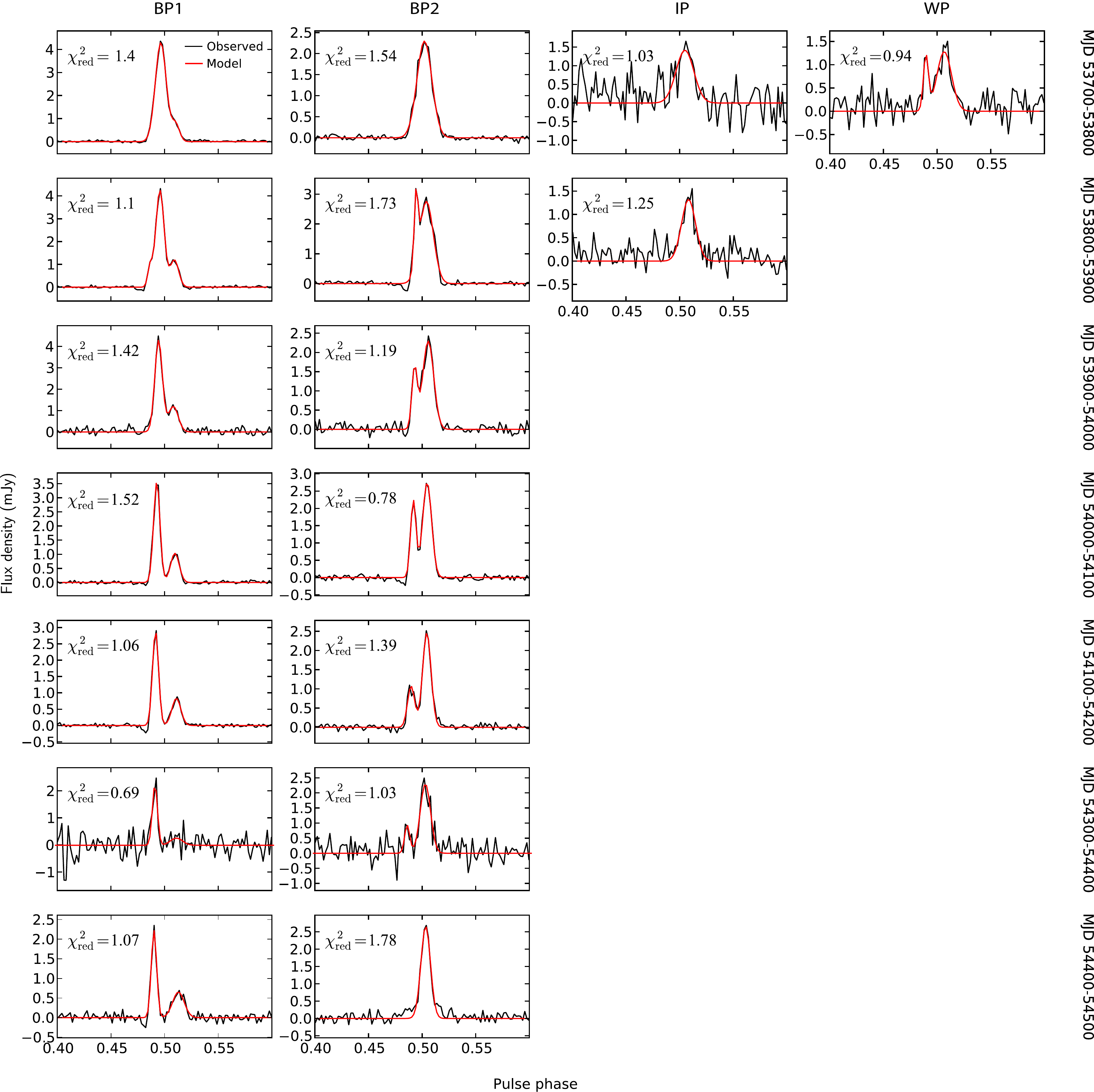}
    \caption{Continued.}
    \label{fig:gaussdatprofs}
\end{figure*}

\begin{figure*}[h]
	\includegraphics[width=1\textwidth]{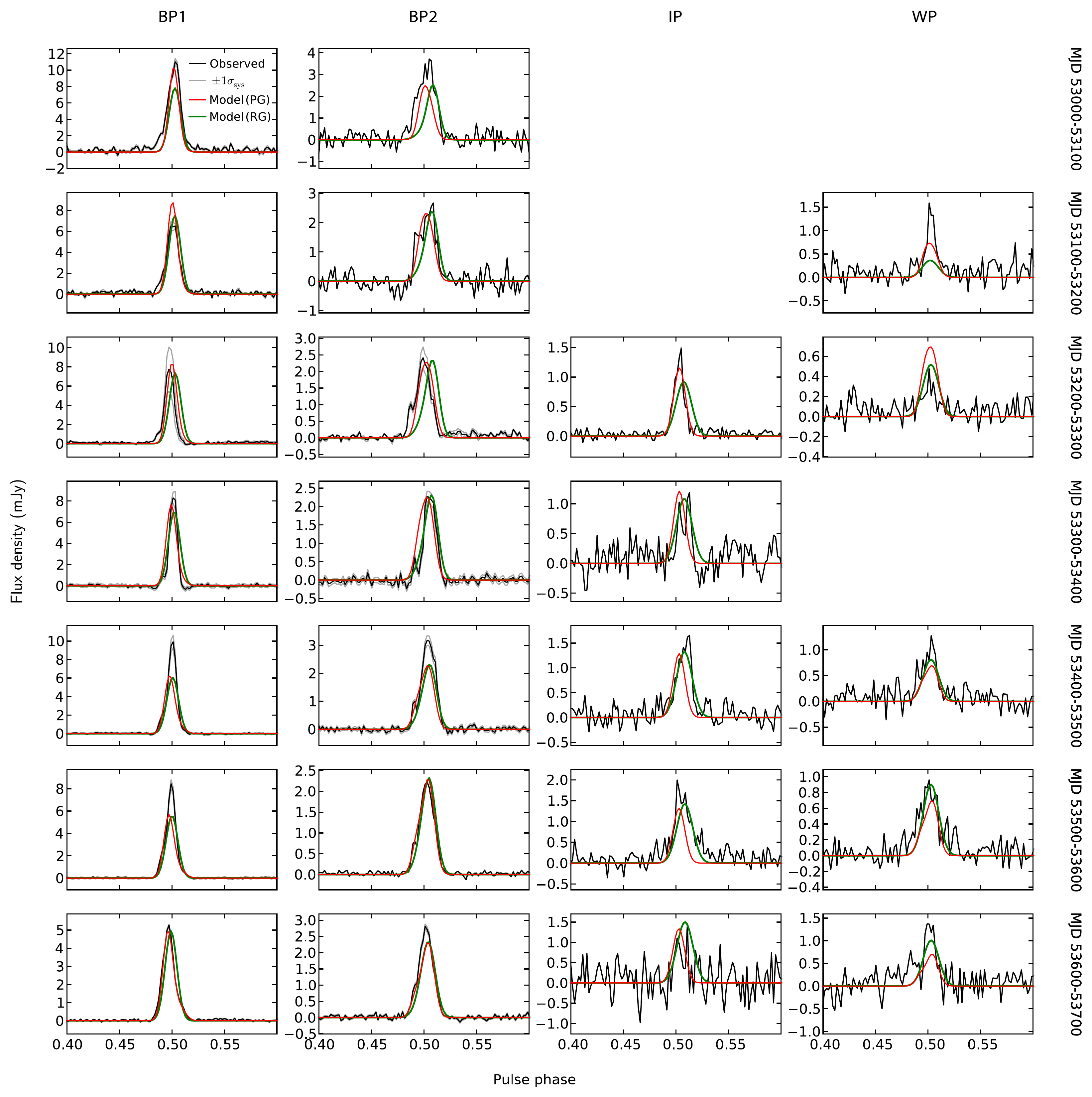}
    \caption{Comparison between observed profiles (black) and those produced by our simulation, as described in Section~\ref{sec:simwindeff}, using the model parameters with the highest likelihood and a prograde (PG) pulsar spin (red). Green lines show the most likely profiles of an alternative model with a retrograde (RG) pulsar spin. The profiles shown have been tabulated according to orbital phase window, that is, BP1, BP2, IP, and WP, and the MJD range to which they belong. The grey lines show the lower and upper 1$\sigma_{\rm sys}$ confidence limits of the observed flux density, multiplied by the most likely error coefficients ($q_{kj}$) that were derived from our analysis.}
    \label{fig:moddatprofs}
\end{figure*}

\addtocounter{figure}{-1}

\begin{figure*}[h]
	\includegraphics[width=1.0\textwidth]{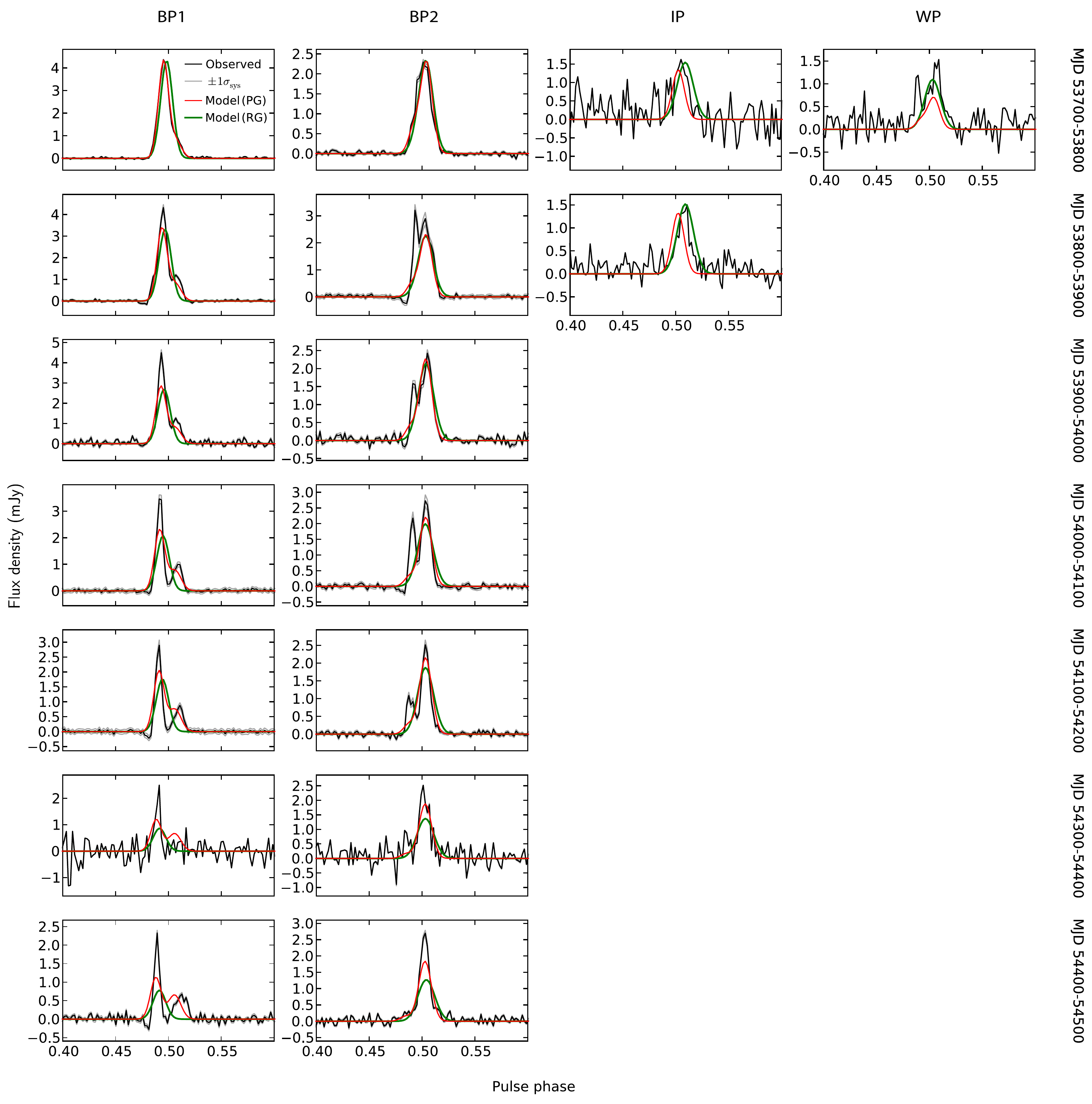}
    \caption{Continued.}
    \label{fig:moddatprofs}
\end{figure*}

\clearpage

\section{3D geometry}
The following figures accompany the description of our model's geometry, presented in Section~\ref{subsec:simemgeo}, as well as the description of the most likely beam shape of pulsar B, presented in Section~\ref{subsec:results}.

\begin{figure*}[h]
	\includegraphics[width=1\textwidth]{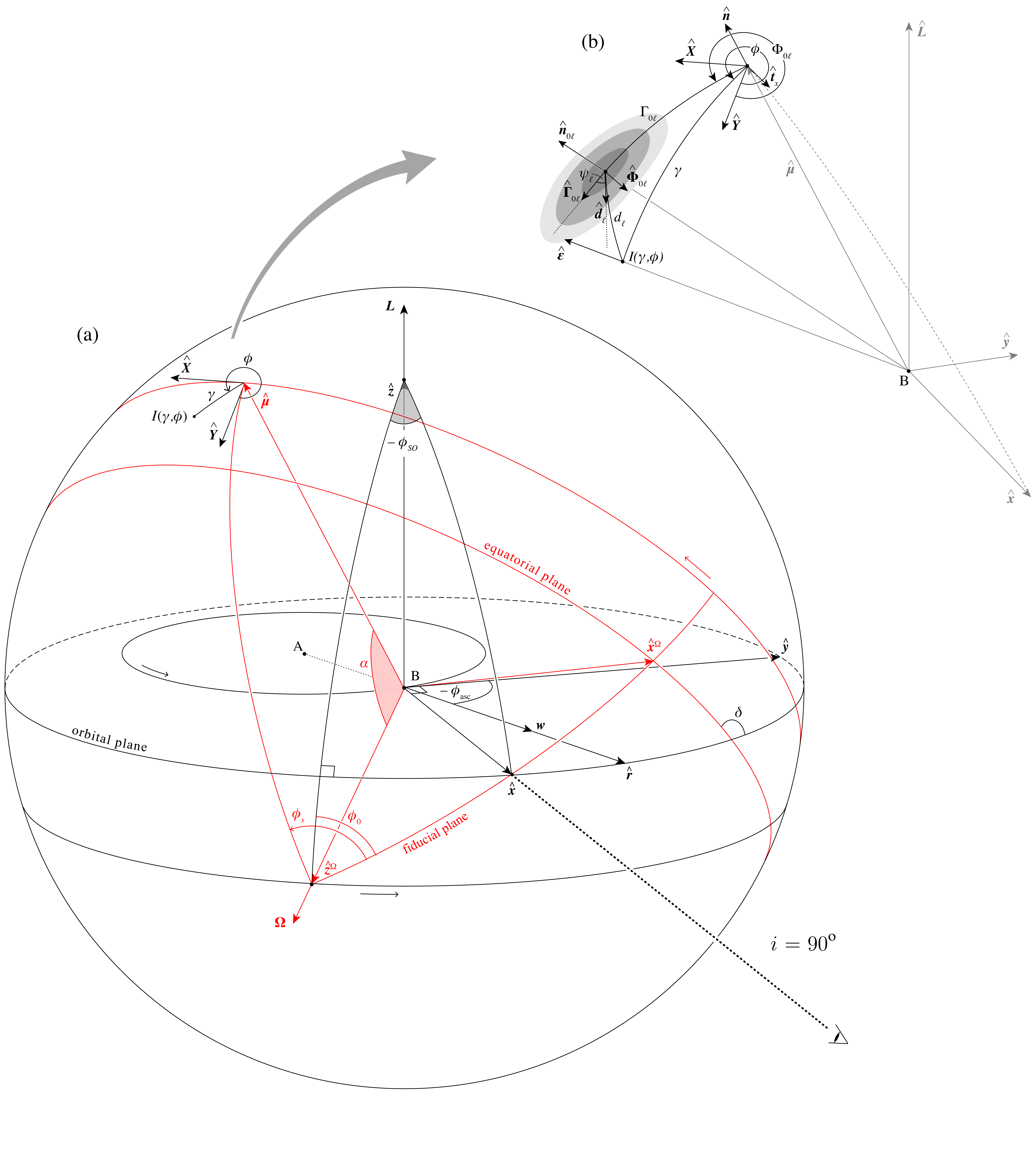}
    \caption{(a) Geometry of the model used in this work to describe the effect of the radial wind of pulsar A ($\boldsymbol{w}$) on the emission of pulsar B. In this figure, $\boldsymbol{\Omega}$ and $\hat{\boldsymbol{\mu}}$ are the spin and magnetic moment of pulsar B, respectively, and $\boldsymbol{L}$ is the orbital angular momentum of the binary system. For clarity, we separated the elements corresponding to the frame of the orbit ($\{\hat{\boldsymbol{x}},\hat{\boldsymbol{y}},\hat{\boldsymbol{z}}\}$) and those corresponding to the frame of the pulsar ($\{\hat{\boldsymbol{x}}^{\Omega},\hat{\boldsymbol{y}}^{\Omega},\hat{\boldsymbol{z}}^{\Omega}\}$) in black and red, respectively. The sense of pulsar B's orbit around pulsar A and that of $\boldsymbol{\mu}$ about $\boldsymbol{\Omega}$ are shown with black and red arrows, respectively. Also shown with a black arrow is the sense of the geodetic precession of $\boldsymbol{\Omega}$ about $\boldsymbol{L}$ (defined by an increasing $\phi_{\rm SO}$). We note that, compared to Breton et al.~(2008), we used the opposite direction for $\boldsymbol{L}$. The direction to the observer coincides with the unit vector $\hat{\boldsymbol{x}}$. In this work, we have assumed that the double pulsar system is viewed exactly edge-on: so, the orbital inclination, $i$, is exactly equal to 90$^\circ$. At the position of $\hat{\boldsymbol{\mu}}$, we show an example location in the beam, with beam coordinates $(\gamma,\phi)$, corresponding to intensity $I(\gamma,\phi)$ (see Eq.~(\ref{eq:modelbeam})). For the definition of the reference frames and angles shown, the reader is directed to Section~\ref{subsec:simemgeo}. (b) The definitions of the angles that are used in our parametrisation of pulsar B's emission beam location and intensity.}
    \label{fig:simgeom}
\end{figure*}

\begin{figure*}[h]
	\includegraphics[width=1\textwidth]{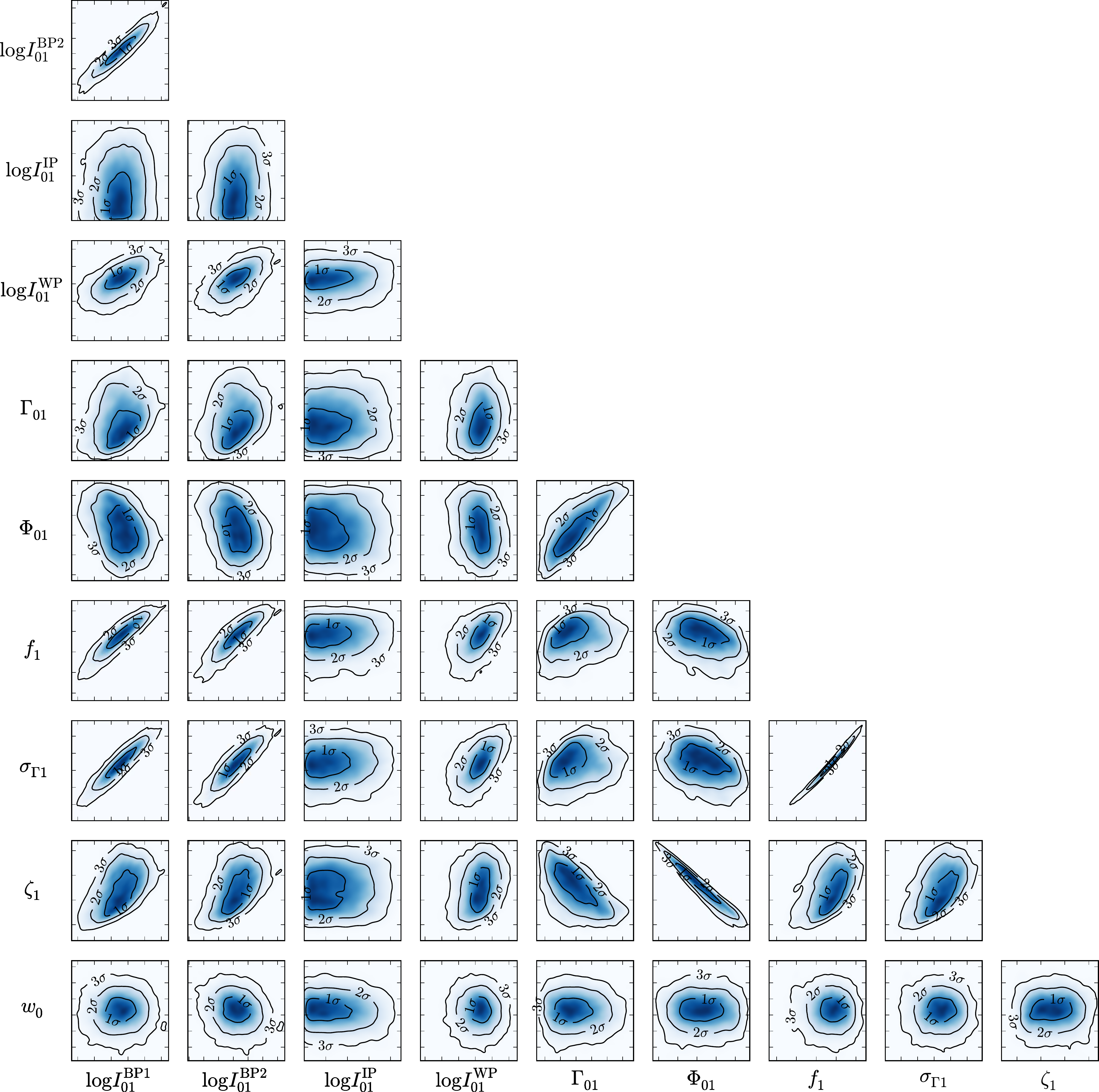}
    \caption{Matrix of probability density plots between the model parameters corresponding to the first Gaussian component in our beam model (indexed with `01' in Table~\ref{tab:bestfitparamsNDK18}), plus the wind magnitude. }
    \label{fig:contab01}
\end{figure*}

\begin{figure*}[h]
	\includegraphics[width=1\textwidth]{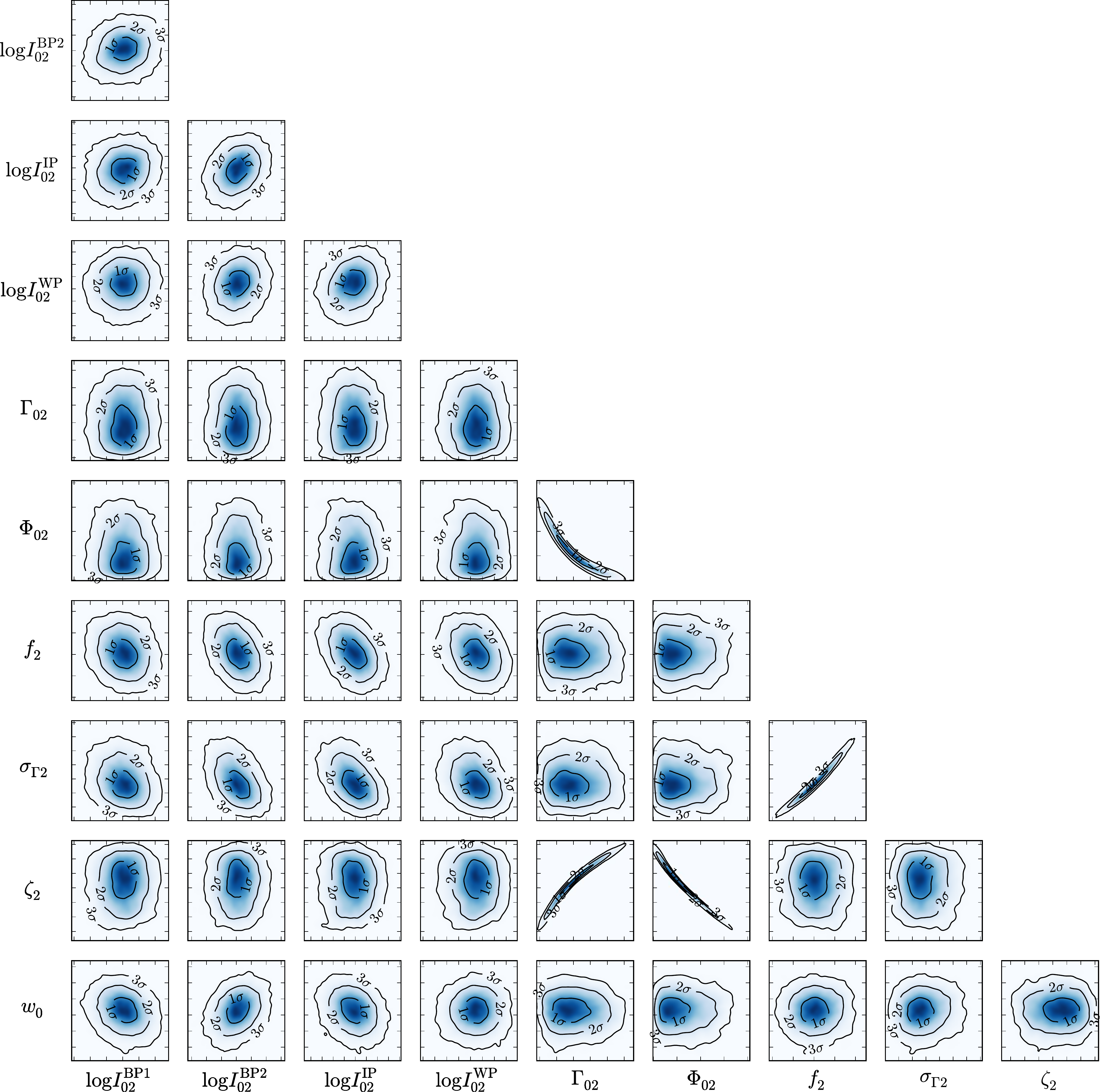}
    \caption{Same as in Fig.~\ref{fig:contab01}, for the second Gaussian component in our beam model (indexed with `02' in Table~\ref{tab:bestfitparamsNDK18}).}
    \label{fig:contab02}
\end{figure*}

\end{appendix}

\end{document}